\documentclass[letterpaper,12pt]{article}%
\usepackage{bm} %Letras griegas en negrita
\usepackage{amssymb,amsmath,amsthm}
\usepackage{graphicx}
\usepackage{subcaption}
\usepackage[height=9in,left=0.75in,right=0.75in,bottom=1in]{geometry}
\usepackage[breaklinks=true,bookmarksopen=true,colorlinks=true,citecolor=blue]{hyperref}
\usepackage[round]{natbib}
\usepackage{xcolor}
\usepackage{amsmath}%
\usepackage[title]{appendix} %titulo en los apendices
\usepackage{comment} %Comentar bloques
\usepackage{thmtools} %Para mejor manejo de los teoremas y poder continuarlos
\usepackage{bold-extra} %Small caps y bold al mismo tiempo
\usepackage{enumitem} %Enumeraciones con labels, etc.
\usepackage{graphicx} %Figures in a row
\usepackage{rotating} %Pagina apaisada
\usepackage{tikz} %Para los diagramas
\usepackage[font = small]{caption} %Modificar los captions de las figuras
\usepackage{booktabs}

\usetikzlibrary{shapes.geometric, arrows, positioning}
\RequirePackage{hypernat} %Hyperref y bibliografia

\DeclareCaptionLabelFormat{bsc}{\textbf{\textsc{#1}\ #2}}
\tikzstyle{space} = [rectangle, rounded corners, minimum width=3cm, minimum height=1cm,text centered, draw=white]

\tikzstyle{unblock} = [draw=white, rectangle, text centered, minimum height=8mm, node distance=0em, append after command={
	[very thick,shorten >=0.2bp, shorten <=0.2bp]
	(\tikzlastnode.south west) edge[-] (\tikzlastnode.south east)
}]

\tikzstyle{lblock} = [draw=white, rectangle, text centered, minimum height=8mm, node distance=0em, append after command={
	[very thick,shorten >=0.2bp, shorten <=0.2bp]
	(\tikzlastnode.south west) edge[-] (\tikzlastnode.south east)
	(\tikzlastnode.north west) edge[-] (\tikzlastnode.north east)
	(\tikzlastnode.south west) edge[-] (\tikzlastnode.north west)	
}]

\tikzstyle{cblock} = [draw=white, rectangle, text centered, minimum height=8mm, node distance=0em, append after command={
	[very thick,shorten >=0.2bp, shorten <=0.2bp]
	(\tikzlastnode.south west) edge[-] (\tikzlastnode.south east)
	(\tikzlastnode.north west) edge[-] (\tikzlastnode.north east)	
}]

\tikzstyle{rblock} = [draw=white, rectangle, text centered, minimum height=8mm, node distance=0em, append after command={
	[very thick,shorten >=0.2bp, shorten <=0.2bp]
	(\tikzlastnode.south west) edge[-] (\tikzlastnode.south east)
	(\tikzlastnode.north west) edge[-] (\tikzlastnode.north east)
	(\tikzlastnode.south east) edge[-] (\tikzlastnode.north east)	
}]

\newcommand{\argmin}{\operatornamewithlimits{argmin}}
\newcommand*{\E}{\mathbb{E}}
\newcommand*{\difTau}{\frac{d}{d\tau}}
\newcommand{\norm}[1]{\lVert#1\rVert}
\newcommand{\func}[1]{\operatorname{#1}}

\declaretheoremstyle[
spaceabove=10pt, spacebelow=10pt,
headfont=\normalfont\scshape\bfseries,
notefont=\normalfont\scshape,
notebraces={(}{)},
bodyfont=\itshape,
headpunct=,
postheadspace=1em
]{THMS}

\declaretheoremstyle[
spaceabove=6pt, spacebelow=6pt,
headfont=\normalfont\scshape\bfseries,
notefont=\normalfont\scshape,
notebraces={(}{)},
bodyfont=\normalfont,
headpunct=,
postheadspace=1em
]{DEF}

\declaretheoremstyle[
spaceabove=10pt, spacebelow=6pt,
headfont=\normalfont\scshape\bfseries,
notefont=\normalfont\scshape,
notebraces={(}{)},
bodyfont=\normalfont,
headpunct=,
postheadspace=1em,
qed=$\blacksquare$
]{EX}

\declaretheoremstyle[
spaceabove=12pt, spacebelow=12pt,
headfont=\normalfont\scshape\bfseries,
notefont=\normalfont\scshape\bfseries,
notebraces={}{},
bodyfont=\normalfont,
headpunct={:},
qed=$\blacksquare$,
postheadspace=1em
]{PROOF}
\declaretheorem[style=THMS, numberwithin=section, name=Theorem]{thm}
\declaretheorem[style=THMS, numberwithin=section, name=Proposition]{prop}
\declaretheorem[style=THMS, numberwithin=section, name=Lemma]{lma}

\declaretheorem[style=DEF, numberwithin=section, name=Assumption]{ass}

\declaretheorem[style=EX, name=Example]{ex}
\declaretheorem[style=EX, numberwithin=section, name=Remark]{rem}

\declaretheorem[style=PROOF, numbered=no, name=Proof of ]{proofc}

\numberwithin{equation}{section}

\begin{document}
% Old Automatic Debiased Estimation with Machine Learning-Generated Regressors
\title{Automatic Locally Robust GMM with Machine-Learning-Generated Regressors\thanks{ First version: 25 Jan 2023 (arXiv:2301.10643v1). Research supported by MICIN/AEI/10.13039/501100011033, grant
CEX2021-001181-M, Comunidad de Madrid, grants EPUC3M11 (V PRICIT) and
H2019/HUM-5891, and grant PID2021-127794NB-I00 (MCI/AEI/FEDER, UE), Programa Primas y Problemas de la Fundaci\'on BBVA 2023. }}
\author{Juan Carlos Escanciano \\
%EndAName
\textit{Universidad Carlos III de Madrid} \and Telmo P\'{e}rez-Izquierdo
\\
%EndAName
\textit{University of the Basque Country} }
\date{\today}
\maketitle

\begin{abstract}
Machine-learning (ML) methods now routinely generate regressors used in subsequent econometric analyses—for example, estimated propensity scores, control-function residuals, imputed covariates, learned proxies, or low-dimensional embeddings of high-dimensional data. As these ML-generated regressors become ubiquitous, the lack of general inference methods for models that use them has become a critical limitation. Standard plug-in and Double ML procedures ignore how generated regressors enter later stages, leading to large biases and invalid inference. We develop a three-step locally robust GMM framework for inference with ML generated regressors. A key new insight is \textit{downstream local robustness}: by a functional chain rule, moment functions that are constructed to be orthogonal to the second step eliminate the complicated indirect (conditioning) effects from the ML-generated regressors. We show how to implement this automatically by estimating the associated Riesz representers through cross-fitted auxiliary regressions, allowing for generic non-Donsker ML in both early steps. In leading treatment-effect and counterfactual settings, simulations demonstrate severe bias in existing methods and reductions of 85–95\% using our procedures.

\vspace{1cm}


\vspace{0.5cm}

\end{abstract}
\newpage

%%%%%%%%%%%%%%%%%%%%%%%%%%%%%%%%%%%%%%%%%%%%%%%%%%%%%%%%%%%%%%%%%%%%%%%%%%%%%%%%%%%%%%%%%%%%%%%%%%%%%%
%%%%%%%%%%%%%%%%%%%%%%%%%%%%%%%%%%%%%%%%%%%%%%%%%%%%%%%%%%%%%%%%%%%%%%%%%%%%%%%%%%%%%%%%%%%%%%%%%%%%%%
\section{Introduction}

Many parameters of interest depend on predicted or generated regressors. Leading examples include structural parameters in models with endogenous variables estimated by control functions \citep[see, e.g.,][]{stock1989nonparametric,stock1991nonparametric,blundell2004endogeneity,imbens2009identification}, average partial effects in sample selection models \citep{ahn1993semiparametric,das2003nonparametric,newey2009two}, propensity score matching \citep{heckman1998matching,abadie2006large}, and marginal treatment effects \citep{heckman2005structural}. More recently, machine learning (ML) is routinely used to generate regressors for imputing missing covariates \citep{fongtyler}, dimension reduction \citep{sorzano2014survey}, learned proxies, confounders, and treatments \citep{knox2022testing}, and feature engineering with unstructured data such as text, images, or audio \citep{feder2022causal}, among many others.\footnote{\cite{knox2022testing} estimate that about two thirds of recent computational work in political science uses predictions of unobserved concepts as regressors in their analyses.} In all of these settings, the parameter of interest depends on a regressor that is itself estimated in a preliminary step, often by flexible or high-dimensional methods.

Despite the prevalence of such problems in modern empirical work, there is currently no general inference framework that remains valid when regressors are generated by flexible or high-dimensional ML methods and then used again in downstream estimation. A common practice is to treat the generated regressors as if they were known, and to apply standard Generalized Method of Moments (GMM) or double/debiased machine-learning (DML) methods as if one were in a two-step setting. This practice typically yields invalid inference: the influence functions and asymptotic variances of such plug-in estimators have complicated analytic forms \citep{hahn2013asymptotic}, and ignoring first-step estimation (i.e., the estimation of the generated regressor) generally leads to distorted standard errors and large regularization or model-selection bias in the final estimates.\footnote{For other biases induced by the repeated use of ML-generated data, see \cite{shumailov2024ai}.} These difficulties are exacerbated in ML settings, where preliminary estimators are high-dimensional, nonparametric, and often non-Donsker \citep{chernozhukov2018double}.

This paper develops automatic locally robust/debiased estimation and inference for structural parameters in three-step models with ML-generated regressors, generalizing the two-step setting of \citet{chernozhukov2022locally}. A new idea is \textit{downstream local robustness}: valid inference must neutralize not only the direct impact of estimating generated regressors, but also the indirect downstream effects that arise because these regressors are themselves inputs to later nuisance functions. Indirect effects are annihilated by making the moment robust to the second step. This paper shows how to automatically achieve downstream local robustness.

A simple toy example may help to fix ideas. Let $V(g)$ denote a ML-generated regressor produced by a first-step $g$ with true value $g_0$, and let the second-step nuisance $h(g)$ denote the optimal linear predictor of $Y$ on $V(g)$, with slope coefficient $\beta_g$, i.e., $h(g)(v)=\beta_g v$. The direct effect of the first step is related to the mapping $g\mapsto \beta_0 V(g)$, where $\beta_0=\beta_{g_0}$, while the indirect effect operates through the second step via the mapping $g\mapsto \beta_g v$. The indirect effect is more complex than the direct effect, as can be seen from $\beta_g=\E[YV(g)]/\E[V(g)^2]$. Let the parameter of interest be a functional of $(g,h)$, say $\theta_0=\theta(g_0,h_0)$ with $h_0=h(g_0)$. Downstream local robustness means that, by the functional chain rule, if $\partial\theta/\partial h (g_0,h_0)=0$, then
\[
\left. \frac{\partial}{\partial g} \theta(g_0,h(g)) \right|_{g=g_0}
=\frac{\partial\theta}{\partial h}(g_0,h_0)\cdot \frac{\partial h}{\partial g}(g_0)
=0,
\]
so orthogonality with respect to the second step removes the indirect effect of $g$. The direct effect $\partial\theta/\partial g(g_0,h_0)$ still remains. This paper provides an automatic construction of functionals that delivers zero derivatives for both direct and indirect effects.

Automatic debiased estimators with generated regressors are useful for two main reasons. First, debiased estimators deliver downstream local robustness and correct for the large regularization and model-selection biases that arise when ML-generated regressors are plugged into subsequent stages. In our simulations, with a moderately large sample ($n=1000$), they reduce the bias of the DML estimator by up to 95\%. Second, in three-step procedures, the analytic form of influence functions and asymptotic variances becomes complex and hard to derive (cf. \citealp{hahn2013asymptotic}). Our estimators and tests are automatic in the sense that these objects are estimated directly from data and identifying moments, without requiring analytic derivations or bootstrap approximations whose theoretical justification is delicate in the presence of ML-generated regressors.

A key feature of these problems is that the use of generated regressors induces a natural three-step structure. We therefore generalize the existing debiasing literature from a two-step to our three-step framework. In the first step, some regressors are predicted (for example, via imputation, ML-estimated propensity scores, or control functions with high-dimensional covariates). In the second step, a nuisance function is constructed as a (potentially high-dimensional) least-squares projection using the generated regressors and possibly other covariates. In the third step, the parameter of interest is identified by a GMM criterion involving the first two steps and the data. Existing debiasing methods could be applied by treating either the generated regressor or the second-step nuisance as known, effectively reducing the problem to two steps, but this generally leads to invalid inference. Additionally, the three-step structure induces a constrained, non-product parameter space in which the second-step nuisance depends on first-step generated regressors, thereby invalidating standard local-robustness arguments that rely on product-space perturbations in the two-step literature (see Remark~\ref{twostep}).

We now summarize our main contributions.

First, we develop a general three-step locally robust GMM framework for models with generated regressors. We fully and separately account for the first and second steps and characterize their contributions to the parameter's influence function, including an indirect effect of the first step that operates through the generated regressors as conditioning variables in the second step. We show that when the second-step effect is zero, this indirect effect is also zero, extending a remark in \citet{hahn2013asymptotic} to a more general class of three-step procedures that include leading ML methods. This establishes the broader orthogonality principle of \textit{downstream local robustness}: by the functional chain rule, moment functions constructed to be orthogonal to the second step eliminate the indirect effects of generated regressors (see Proposition~\ref{prop:zero_indirect}).

Second, we provide automatic estimation of influence functions and asymptotic variances for models with generated regressors. Under a linearization assumption \citep[as in, e.g.,][]{newey1994asymptotic,ichimura2022influence}, we show how the Riesz representers in the first- and second-step influence functions can be identified and estimated separately without knowing their analytic form. This is achieved via cross-fitted auxiliary regressions that remain valid for generic non-Donsker ML methods in the first and second steps; see, e.g., \citet{chernozhukov2021automatic,chernozhukov2022locally,chernozhukov2023automatic,chernozhukov2022automatic}. Automatic estimation is particularly well motivated for generated regressors, where the Riesz representers typically have complex forms \citep[see, for instance,][]{hahn2013asymptotic,mammen2016semiparametric,escanciano2014uniform}. Together with the first contribution, we establish feasible standard errors and valid asymptotically normal inference for debiased estimators with ML-generated regressors. Relative to the DML literature, the presence of generated regressors makes the asymptotic analysis---and, in particular, the control of higher-order terms in the asymptotic expansions---more delicate, and we address this issue.

Third, we propose novel automatic three-step debiased estimators for leading applications such as high-dimensional propensity score (Hd-PS) regression adjustment, treatment effects with learned confounders (autoencoders), nonparametric Average Treatment Effect (ATE) estimation on a boosted propensity score, and the nonparametric Counterfactual Average Structural Function (CASF). In these settings, the generated regressors arise, for example, from a control-function approach using Lasso, Random Forest, or Deep Learning; from Logit-Lasso or Boosting Hd-PS; or from learned confounders via autoencoders \citep{bengio2013representation}. The nonparametric ATE estimator with a Hd-PS generalizes \cite{heckman1998matching}, \cite{hahn2013asymptotic}, and \cite{mammen2016semiparametric} to a ML setup with debiasing and automatic inference, reducing regularization bias from both first and second steps. The application to the CASF with a control-function approach appears to be novel even in low dimensions, and it is related to the literature on domain adaptation, transfer learning, and covariate shift. Relative to that literature, we allow for endogeneity and a flexible non-separable structural model, which is important in applications where counterfactuals involve endogenous variables such as prices.

Our work builds on two strands of the literature. The first is the classical literature on semiparametric estimation with generated regressors \citep[see, among many others,][]{ichimura1991semiparametric,ahn1993semiparametric,heckman1998matching,newey1999nonparametric,li2002semiparametric,rothe2009semiparametric,imbens2009identification,escanciano2010testing,song2012smoothness}. In an important work, \citet{hahn2013asymptotic} derive the influence function of three-step estimators that are averages of evaluation functionals of nonparametric regressions with generated regressors. We build on these influence-function calculations by considering a more general class of first, second, and third steps, including high-dimensional regressions (e.g., Logit-Lasso) and targets that may depend on the entire second step (not only evaluation functionals, as in, the CASF example). For estimation, \citet{mammen2012nonparametric,mammen2016semiparametric} and \citet{escanciano2014uniform} study the asymptotic properties of (non--locally robust) estimators using empirical process methods. These existing results are formulated for nonparametric first and second steps in Donsker classes and are generally not applicable to ML estimators, which often fall outside Donsker classes \citep[see][]{chernozhukov2018double}. We contribute to this literature by providing automatic debiased GMM estimators that explicitly account for ML-generated regressors and reduce regularization and model-selection biases, and by proving their asymptotic properties accounting for ML-generated regressors.
%\footnote{Our results do not encompass the specific three-stage and sieve settings studied in \cite{hahn2019three} and \cite{hahn2018nonparametric}. These papers, however, focus on plug-in and do not consider locally robust inference, which is the main goal of our paper.}

The second strand is the literature on locally robust/debiased estimators \citep[e.g.,][]{chernozhukov2018double,chernozhukov2022locally,chernozhukov2022automatic}. With the exception of \citet{sasaki2021estimation}, the DML literature prior to our work has not considered or accounted for generated regressors in inference. Our results complement \citet{sasaki2021estimation} by providing a general three-step framework and automatic estimation of adjustment terms for a broad class of models with generated regressors, including empirically relevant settings such as the partially linear model with ML-generated regressors. Relative to the Automatic DML literature, we innovate by (i) working in a three-step setting where the second step depends on the generated regressor and the product-space structure of \citet{chernozhukov2022locally} fails; (ii) exploiting novel partial and downstream local robustness results that allow separate identification and automatic estimation of individual Riesz representers; and (iii) accounting for generated regressors in the estimation of Riesz representers and the bounds for higher-order terms in functional derivatives with respect to the high-dimensional generated regressors.

The rest of the paper is organized as follows. Section~\ref{Setting} introduces the setting and examples. Section~\ref{sec:debiased_est} describes the debiased moment functions and defines the debiased GMM estimator in the presence of ML-generated regressors, illustrating its performance in two Monte Carlo experiments. Section~\ref{sec:Automatic} gives the separate identification and automatic estimation of the Riesz representers. Debiased automatic estimators for the examples are presented in Section~\ref{sec:Examples}. The asymptotic theory is developed in Section~\ref{sec:asymptotic}. Section~\ref{sec:conclusion} concludes. Appendix~\ref{sec:algorithm} summarizes the estimation steps. Appendix~\ref{sec:AdditionalExamples} provides a further application to a ML implementation of the nonparametric ATE estimator of \cite{heckman1998matching}. Appendix~\ref{sec:MC_details} contains details about the Monte Carlo simulations. Appendix~\ref{sec:app_inclusion} discusses regularity conditions, and Appendix~\ref{sec:Proofs} gathers the proofs of the main results.
%%%%%%%%%%%%%%%%%%%%%%%%%%%%%%%%%%%%%%%%%%%%%%%%%%%%%%%%%%%%%%%%%%%%%%%%%%%%%%%%%%%%%%%%%%%%%%%
\section{Setting and examples}
\label{Setting}
%%%%%%%%%%%%%%%%%%%%%%%%%%%%%%%%%%%%%%%%%%%%%%%%%%%%%%%%%%%%%%%%%%%%%%%%%%%%%%%%%%%%%%%%%%%%%%%%%%%%%%
\subsection{Three-step setting}

We observe data $W=(Y,D,Z)$ from a cumulative distribution function (cdf) $F_{0}$. We describe our three-step setting as follows: \bigskip

\textbf{First step}. There is a first-step nuisance function $g_{0}(Z)$ satisfying the moment restrictions
\begin{equation}
\mathbb{E}[\delta _{1}(Z)\epsilon (W,g_{0})]=0\text{ for all }\delta _{1}\in
\Delta _{1},  \label{orth1}
\end{equation}%
where $\epsilon (W,g_{0})$ is a generalized error depending on the data $W$ and the nuisance $g_{0}\in \Delta _{1}$, where $\Delta _{1}$ is a linear and closed subspace of $L_{2}(Z)$. Henceforth, for a generic random
variable $U$, we denote by $L_{2}(U)$ the Hilbert space of square-integrable
functions of $U$, i.e., $g\in L_{2}(U)$ iff $\mathbb{E}[g^2(U)]<\infty$. 

This setting covers a wide variety of semiparametric and nonparametric first steps. For example, when $\epsilon
(W,g_{0})=D-g_{0}(Z)$ and $\Delta _{1}=L_{2}(Z)$, we have $g_{0}(Z)=\mathbb{E}
[D|Z]$, as in \cite{hahn2013asymptotic}. However, if $\func{dim}(Z)$ is high, fully nonparametric first steps may not be feasible to implement. We could then consider a high-dimensional additive regression model with the same error but with $\Delta _{1}=\sum_{j =1}^{\func{dim}(Z)}\Delta _{1,j}$, where $\Delta _{1,j}$ is a subset of $L_{2}(Z_{j})$ for the $j$-th component of $Z$ \citep[see Chapter 7 in][]{wainwright2019high}. When $\Delta _{1}$ is the mean-square limit of linear combinations
$\sum_{k=1}^{K}\beta _{0k}c_{k}(Z)$ for $K\in\mathbb{N}$, a sequence of real numbers $(\beta
_{0k})_{k=1}^{\infty }$, a dictionary $(c_{k})_{k=1}^{\infty }$ of
functions in $L_{2}(Z)$, and $\epsilon (W,g_{0})=D-\Lambda (g_{0}(Z))$ for
the logistic cdf $\Lambda$, this setting covers high-dimensional
logistic regression (Logit-Lasso), which is commonly used for propensity-score and classification modeling in high dimensions. These ML-generated regressors complement the fully nonparametric mean-regression first steps in \cite{hahn2013asymptotic}. For numerous other examples of $\epsilon (W,g_{0})$, including
quantile regression, see Section 3 of \citet{ichimura2022influence}. For
general parametric first steps, see Remark \ref{Parfirststep}; and for other first steps not covered by our setting, see Remark \ref{OtherFS}. 
%%%%%%%%%%%%%%

The first-step nuisance $g_{0}$ in (\ref{orth1}) is used to construct the population generated
regressors
\begin{equation*}
V\equiv \varphi (D,Z,g_{0}),
\end{equation*}%
where $\varphi$ is a known function of observed variables $(D,Z)$ and the
unknown function $g_{0}.$ Note the simplified notation $V\equiv V(g_{0}).$
A high-dimensional extension of propensity score matching in \cite{heckman1998matching} has $V=\Lambda (g_{0}(Z))$; some dimension-reduction methods have $\varphi (D,Z,g_{0})=g_{0}(Z)$, as in \cite{hahn2013asymptotic}, or some components of $g_{0}$ (as with autoencoders); imputation for conditionally missing-at-random regressors has $\varphi
(D,Z,g_{0})=Z_{1}D+(1-Z_{1})g_{0}(Z_{2})$, where $Z_{1}$ is a ``not missing"
indicator for the covariate $D$ and $g_{0}(Z_{2})=\mathbb{E}
[D|Z_{1}=1,Z_{2}]$ for observed covariates $Z_{2}$; and control-function methods often
lead to $\varphi (D,Z,g_{0})=D-g_{0}(Z)$, for an endogenous variable $D$
and exogenous variables $Z.$ Our setting covers these and other generated
regressors.\bigskip

\textbf{Second step}. Let $S$ and $X$ denote some components (or all) of $(Y,D)$
and $(D,Z)$, respectively. The second step links $S$ with $X$ and the
generated regressor $V$ through the moment restrictions
\begin{equation}
\mathbb{E}[\delta _{2}(X,V)(S-h_{0}(X,V))]=0\text{ for all }\delta _{2}\in
\Delta _{2}(g_{0}),  \label{orth2}
\end{equation}%
where $\Delta _{2}(g_{0})$ is a linear, closed subspace of $L_{2}(X,V)$
(note $\Delta _{2}$ depends on $g_{0}$ because $V$ depends on $g_{0}$). When $S$ (and hence $h_{0}$) has dimension $\func{dim}(S)>1$, we understand (\ref{orth2}) as being applied to each component of $S.$ This dependence of the parameter space $\Delta _{2}(g_{0})$ on the first step $g_{0}$ is a point of departure from the existing debiasing literature in, e.g., \cite{chernozhukov2022locally}. \citet{hahn2013asymptotic} and \citet{mammen2016semiparametric}
consider cases where the second step $h_{0}$ is a nonparametric regression
of $Y$ on $(X,V)$, corresponding to $\Delta _{2}(g_{0})=L_{2}(X,V)$. In contrast, we also allow $\Delta _{2}(g_{0})$ to be a strict subset of $L_{2}(X,V)$ (e.g., with sparse or sieve restrictions).  
\bigskip

\textbf{Third step}. Let $\Theta \subseteq \mathbb{R}^{p}$ denote the
parameter space where the parameter of interest lies. Consider
the moment function $m\colon \mathbb{R}^{\func{dim}(W)}\times L_{2}(Z)\times
L_{2}(X,V)^{\func{dim}(S)}\times \Theta \rightarrow \mathbb{R}^{q}$, $q\geq p.$ The
parameter of interest $\theta _{0}$ is identified in a third step by a GMM
moment condition
\begin{equation*}
\mathbb{E}[m(W,g_{0},h_{0},\theta _{0})]=0.
\end{equation*}%
Here we assume that $\theta _{0}$ is identified by these moments, i.e., that
$\theta _{0}$ is the unique solution to $\mathbb{E}[m(W,g_{0},h_{0},\theta
)]=0$ over $\theta \in \Theta $. 

%%%%%%%%%%%%%%%%%%%%%%%%%%%%%%%%%%%%%%%%%%%%%%%%%%%%%%%%%%%%%%%%%%%%%%%%%%%%%%%%%%%%%%
\subsection{Examples}

The following examples are used to illustrate the main results of this paper.

\begin{ex}[Partially linear model with ML-generated regressors]
\label{ex:PLM_ML}

We first consider a general partially linear model with a generated regressor.
We observe $W\equiv (Y, D, Z)$, where $Y$ is an outcome, $D$ is a (possibly
vector-valued) treatment or regressor of interest, and $Z$ is a (potentially)
high-dimensional covariate. The starting point is the partially linear model
\begin{equation*}
    Y = \theta_0' D + \kappa_0(V) + \varepsilon
    \quad \text{with } \E[\varepsilon\mid D,V] = 0,
\end{equation*}
where $\theta_0$ is the parameter of interest, $V \equiv \varphi(D,Z,g_0)$
is a generated regressor constructed from a first-step nuisance $g_0$ and the
covariates $(D,Z)$, and $\kappa_0(\cdot)$ is an unknown nuisance function. 

\medskip
\noindent\emph{First step and generated regressor.}
The first step $g_0$ solves \eqref{orth1} for some generalized error
$\epsilon(W,g_0)$ and a linear, closed subspace $\Delta_1\subseteq L_2(Z)$. This encompasses, for example:
\begin{itemize}
    \item \emph{Hd-PS:}
    $\epsilon(W,g_0)=D-\Lambda(g_0(Z))$, with $\Delta_1$ the mean-square limit
    of sparse linear combinations $\sum_{k=1}^K \gamma_{0k} c_k(Z)$, and
    $V = \Lambda(g_0(Z))$ (for treatment effects, sample selection, etc.).
    \item \emph{Dimension reduction / learned confounders (e.g., autoencoders):}
    $\epsilon(W,g_0)=Z-d_0(e_0(Z))$, $g_0=(d_0,e_0)$, with $\Delta_1$ defined in Section~\ref{sec:ex_PLMauto}, and 
    $V = e_0(Z)$ equal to the encoder.
    \item \emph{Control-function residuals:}
    $\epsilon(W,g_0)=D-g_0(Z)$, with $\Delta_1=L_2(Z)$, and $V = D-g_0(Z)$ equal to the residual of the first-stage regression of $D$ on $Z$.
\end{itemize}

\medskip
\noindent\emph{Second and third steps.}
In the partially linear model, we set $S=(Y,D)$ and
$X=\emptyset$ in \eqref{orth2}, and define
\begin{equation} \label{eq:PLM_2step}
h_{0Y}(v)\equiv \E[Y\mid V=v],
\qquad
h_{0D}(v)\equiv \E[D\mid V=v],
\end{equation}
so that $h_0\equiv (h_{0Y},h_{0D})$, 
$\Delta_2(g_0)=L_2(V)^{1+\dim(D)}$. Following the downstream local robustness principle, the identifying moment is \citep[cf.][]{robinson1988root}
\begin{equation}
\label{eq:PLM_general_moment}
m(W,g_{0},h_{0},\theta _{0})
=
\big( Y-h_{0Y}(V)-\theta
_{0}' ( D-h_{0D}(V))\big) \cdot (D - h_{0D}(V)).
\end{equation}
When $h_{0Y}$ and $h_{0D}$ are estimated by cross-fitted ML, the estimator based on \eqref{eq:PLM_general_moment} is the DML estimator of \cite{chernozhukov2018double}.

We now indicate two leading special cases.

\medskip
\noindent\textbf{(a) Hd-PS regression adjustment.} 

Let $D$ be a binary treatment and $Y$ satisfy the potential-outcome model
$Y=Y_0 + D(Y_1-Y_0)$, where $Y_0$ and $Y_1$ are the potential outcomes under
control and treatment, respectively. Under strong ignorability,
$(Y_0, Y_1)\perp D\mid Z$, the propensity score $\E[D\mid Z]$ is a
balancing score \citep{rosenbaum1983central}, and classical matching and reweighting methods can be based on it
\citep{heckman1998matching,hirano2003efficient}.

With high-dimensional $Z$, we estimate the propensity score by Logit--Lasso (Hd-PS), where $\Delta_1$ is the mean-square limit of sparse linear combinations of a dictionary $(c_k)_{k=1}^\infty$ in $L_2(Z)$, and the generated regressor is $V=\varphi(D,Z,g_0)=\Lambda(g_0(Z))$. For expositional clarity, we consider that the generated regressor recovers the propensity score: $V=\Lambda(g_0(Z))=\E[D|Z]$. This simplifies the second step (now $h_0=h_{0Y}$, since $h_{0D}(V)=V$ is known), while it accommodates a rich set of controls through the high-dimensional first step. The estimand $\theta_0$ has a transparent causal (weighted-variance) interpretation because $V$ is a balancing score. In the general case of $V\neq\E[D|Z]$, our three-step inference procedure applies, though interpreting the target as causal requires additional assumptions.  

\begin{comment}
   
\emph{Identification and overlap-weighted interpretation.}
Let $\tau(v)\equiv \E[Y_1-Y_0\mid \pi_0(Z)=v]$ denote the conditional average
treatment effect given the propensity score. The partially linear coefficient
$\theta_0$ can be written as an overlap-weighted average of $\tau(V)$:
\begin{equation*}
  \theta_0
  =\frac{\E\!\left[Var(D\mid V)\,\tau(V)\right]}{\E\!\left[Var(D\mid V)\right]}
  =\frac{\E\!\left[V(1-V)\,\tau(V)\right]}{\E\!\left[V(1-V)\right]}.
\end{equation*}
The weighting kernel $V(1-V)$ emphasizes regions with good overlap and
automatically downweights propensity-score extremes, providing a stability
rationale for regression adjustment relative to estimators that rely on
$1/V$ or $1/(1-V)$ weighting.
\end{comment}
This regression-adjustment formulation is closely related to the estimating
equations in \citet{robins1992estimating}, which exploit the ``exposure residual''
$D-V$. In particular,
orthogonalized moment conditions can be based on products of
$(Y-\theta D)$ and $(D-V)$, yielding robustness to first-step estimation error.
Our three-step locally robust GMM framework constructs such orthogonal moments
for $\theta_0$ when $V$ is estimated by ML and $\kappa_0$ is
flexible, delivering valid inference in this widely used workflow. The debiased GMM estimator can be easily implemented as a (cross-fitted) Ordinary Least Squares (OLS) estimator, see equation~\eqref{DATE} and below.

A nonparametric version of the Hd-PS regression adjustment example is provided in Section~\ref{ATEPSM} of Appendix~\ref{sec:AdditionalExamples}. This example generalizes \cite{heckman1998matching}, \cite{hahn2013asymptotic}, and \cite{mammen2016semiparametric} to a machine-learning propensity score estimator and locally robust estimation and inference.

\medskip
\noindent\textbf{(b) Learned confounders via autoencoders and other embeddings.}
\smallskip

Researchers controlling for high-dimensional unstructured data (images, text, audio, or video) often employ a low-dimensional learned representation. Let
$e_0(Z)$ be such an embedding (e.g., the encoder from an autoencoder), let $g_0=(d_0,e_0)$, and define
\[
V = \varphi(D,Z,g_0) = e_0(Z).
\]
The second and third steps remain as in
\eqref{eq:PLM_2step}--\eqref{eq:PLM_general_moment}. We construct locally
robust estimators that account for the estimation of these embeddings in Section~\ref{sec:ex_PLMauto}.

There is a growing literature using deep latent-variable models and learned representations for causal inference. \citet{louizos2017causal} use variational autoencoders to learn latent confounders from proxy variables, while \citet{klaassen2024doublemldeep} and \citet{schulte2025adjustment} study treatment-effect estimation with multimodal or non-tabular data based on pre-trained or jointly trained neural-network representations within a DML framework. Surveys such as \citet{scholkopf2021toward} review causal representation learning more broadly. Unlike this literature, we treat the learned embedding $e_0(Z)$ as an ML-generated regressor and derive three-step locally robust GMM estimators that explicitly account for its estimation in downstream inference. Further details about this example, including the construction of locally robust estimators, are provided in Section~\ref{sec:ex_PLMauto}. 
\end{ex}

\begin{comment}
\begin{ex}[Nonparametric ATE]
\label{sec:ATE_non}
Among existing causal methods that use machine learning to estimate the propensity score, twang (gradient boosting) is widely used in empirical work, particularly in combination with doubly robust augmented inverse probability weighted (AIPW) estimation; see \cite{leite2024machine}. However, the high sensitivity of AIPW to the estimation of the propensity score has been well documented in the literature; see, in particular, the recent analysis by \cite{yadlowsky2022explaining} based on high-dimensional asymptotics. Here, we propose an alternative nonparametric estimator of the ATE based on nonparametric regression on a gradient-boosting propensity score estimator. That is, the first step learns $V=g_0(Z)$ from a gradient-boosting nonparametric estimation of $D$ onto $Z$, corresponding to $\epsilon(W,g)=D-g(Z)$. The second step fits $h_0(d,v) = \E[Y\mid D=d, V=v]$, so $\Delta_2(g_0) = L_2(D,V)$. 
We also investigate the case where the second step is misspecified, which does not fall under the influence function calculations of \cite{hahn2013asymptotic} and is quite relevant for applied work. Our proposal here relates to the $L_2$-boosting treatment effect estimation results of \cite{kueck2023estimation} and to the literature on balancing weights, see, e.g., \cite{athey2018approximate}, with the key difference that, in our setting, the estimated propensity score is an ML-generated regressor.
\end{ex}
\end{comment}

\begin{ex}[CASF with a control-function approach in a non-separable model]
\label{ex:CF}

We observe $W=(Y,D,Z)$ satisfying the model $Y=H(X,U)$, for an
unknown function $H$ and unobserved error term $U$. The main feature of this model is that $D$, a
component of $X$, may be an endogenous regressor. We assume that the
endogenous regressor satisfies $D=g_{0}(Z)+V$, with $U$ and $V$ being
unobserved correlated error terms. The function $g_{0}$ can be identified
by a conditional mean restriction, as in equation~\eqref{orth1} with $\epsilon(W, g_0) = D - g_0(Z)$. We assume a
control-function approach: $U\mid D,Z\sim U\mid X,V\sim U\mid V$, where $\sim$ denotes
equality in distribution. Thus, the generated regressor is a first-step residual:
\begin{equation*}
V\equiv \varphi (X,Z,g_{0}) = D-g_{0}(Z).
\end{equation*}

As in \citet{blundell2003endogeneity}, the control-function
assumption implies
\begin{align*}
\mathbb{E}[Y\mid X=x,V=v]& =\mathbb{E}[H(X,U)\mid X=x,V=v] \\
& =\mathbb{E}[H(x,U)\mid V=v]\equiv h_{0}(x,v).
\end{align*}%
This defines the second step, which satisfies \eqref{orth2} with $\Delta
_{2}(g_{0})=L_{2}(X,V)$.

The control-function assumption allows us to identify the Average Structural
Function (ASF) at a point $x\in \mathbb{R}^{\operatorname{dim}(X)}$:
\begin{equation*}
\operatorname{ASF}_{0}(x)\equiv \mathbb{E}[H(x,U)]=\mathbb{E}[\mathbb{E}[H(x,U)\mid V]]=%
\mathbb{E}[h_{0}(x,V)].
\end{equation*}%
Some well-known conditions on the support of the random vectors are needed
for the above equation to hold
\citep[see][]{blundell2004endogeneity,imbens2009identification}.

In this setup, a parameter of interest is the CASF, given by
\begin{equation*}
\theta _{0}=\int \operatorname{ASF}_{0}(x^{\ast })dF^{\ast }(x^{\ast }),
\end{equation*}%
for a counterfactual distribution $F^{\ast }$. When $F^{\ast }$ is implied
by a certain policy, the CASF may be used to measure the effect of the
policy
\citep[see][]{stock1989nonparametric,stock1991nonparametric,blundell2004endogeneity}.
By Fubini's Theorem, the CASF can be written as
\begin{equation*}
\theta _{0}=\int \mathbb{E}[h_{0}(x^{\ast },D-g_{0}(Z))]dF^{\ast }(x^{\ast
})=\mathbb{E}\left[ \int h_{0}(x^{\ast },D-g_{0}(Z))dF^{\ast }(x^{\ast })%
\right] .
\end{equation*}%
Hence, the moment function that identifies the CASF is:
\begin{equation}
\label{eq:CASF_moment}
m(w,g_0,h_0,\theta_0 )=\int h_0(x^{\ast },d-g_0(z))dF^{\ast }(x^{\ast })-\theta_0 .
\end{equation}%

We propose in (\ref{eq:CASF_D}) a debiased estimator for the CASF, allowing for and accounting for ML first and second steps. The CASF estimator here generalizes the automatic debiased estimator of trained regression averages under covariate shifts, as in \citet{chernozhukov2023automatic}, to the empirically important case in which the covariate shift arises through an endogenous regressor. 

A remarkable feature of the CASF example is that, to evaluate the moment condition at a point $w=(y,d,z)$, one needs the entire second-step nuisance function $h_0$. Thus, even in the low-dimensional case, it is not encompassed by the setup of \citet{hahn2013asymptotic,hahn2019three} (nor are Examples~\ref{ex:PLM_ML}(a)--(b)).
\end{ex}
%%%%%%%%%%%%%%%%%%%%%%%%%%%%%%%%%%%%%%%%%%%%%%%%%%%%%%%%%
%%%%%%%%%%%%%%%%%%%%%%%%%%%%%%%%%%%%%%%%%%%%%%%%%%%%%%%%%
\section{Debiased estimation with ML-generated regressors}
\label{sec:debiased_est}

A fundamental property that allows us to develop debiased estimators is Neyman orthogonality, also referred to as local robustness \citep[see][]{neyman1959optimal,chernozhukov2018double,chernozhukov2022locally}. In our three-step setting, Neyman-orthogonal moments are obtained by augmenting the original identifying moments with influence-function (IF) corrections associated with the first ($g_0$) and second ($h_0$) steps. The second-step IF accounts for the effect of estimating the second-step nuisance $h_0$ and corresponds to the classical correction in \citet{newey1994asymptotic}.

A key difference relative to standard two-step problems is that, here, the first-step nuisance $g_0$ enters the moment condition in two ways: directly through $m(W,g_0,h_0,\theta)$ and indirectly through the fact that estimation of $h_0$ depends on the generated regressor $V=\varphi(D,Z,g_0)$. Thus, estimation error in $g_0$ affects the target parameter through a \emph{direct} (or \emph{evaluation}) effect (in the toy example, $\beta_0 v$ evaluated at $v=V(g)$) and an \emph{indirect} (or \emph{conditioning}) effect that operates through $h_0$ (in the toy example, $g\mapsto h(g)(v)=\beta_g v$); see Figure~\ref{fig:chainrule} and Section~\ref{sec:first_second_IF}. This indirect effect is absent in standard two-step locally robust problems but is unavoidable whenever the conditioning variable in a regression is itself ML-generated.

We show that the debiased moment function takes the generic form
\begin{equation}\label{orth}
\begin{aligned}
    \psi (w,g_{0},h_{0},\alpha _{0},\theta ) &= m(w,g_{0},h_{0},\theta ) \\
    &\quad+ \underbrace{\alpha _{01}(z)\cdot
\epsilon (w,g_{0})}_{=\phi _{1}(w,g_{0},\alpha _{01})} \\
 &\quad+ \underbrace{%
\alpha _{02}(x,\varphi (d,z,g_{0}))^{\prime }[s-h_{0}(x,\varphi (d,z,g_{0}))]%
}_{=\phi _{2}(w,g_0,h_{0},\alpha _{02})},
\end{aligned}
\end{equation}%
where $\alpha_0 \equiv (\alpha_{01}, \alpha_{02})$ are the Riesz representers associated with the first and second steps, respectively. In the case of multiple moment conditions ($q>1$), each component of $m$ is debiased separately.

The function $\phi_{1}$ in \eqref{orth} is the first-step IF and captures the effect of the generated regressors on the identifying moments. It is generally nonzero, so inference that ignores generated regressors is typically invalid.\footnote{One instance where $\phi _{1}=0$ and inference that does not account for generated regressors is valid is when the sample size used to construct the generated regressors is asymptotically larger than the sample size used to estimate the main parameter (see Remark~\ref{Parfirststep} for a formal statement).} The explicit analytic expression for $\alpha_{01}$ is typically complicated (see equation~\eqref{eq:alpha1_defi} in Appendix~\ref{sec:Proofs}), but we construct automatic estimators that do not require this expression. The second-step IF $\phi _{2}$ is of the usual form \citep{newey1994asymptotic}, but automatic estimation of the corresponding Riesz representer $\alpha_{02}$ must be generalized to allow for generated regressors as inputs.

A central insight of this paper is that the \emph{indirect} contribution of $g_0$ operates entirely through the second-step. By the functional chain rule, this implies that, when the moment is orthogonal with respect to $h$ (so $\alpha_{02}=0$), the indirect effect of the generated regressor is zero and the first-step IF $\phi_1$ simplifies. We refer to this property as \emph{downstream local robustness}. It generalizes an observation in \citet{hahn2013asymptotic} to a general three-step ML framework and to a much broader class of problems beyond generated regressors (cf.\ Proposition~\ref{prop:zero_indirect}).

%%%%%%%%%%%%%%%%%%%%%%%%%%%%%%%%%%%%%%%%%%%%%%%%%%%%%%%%%%%%%%%%%%%%%%%%%%%%%%%%%%%%%%%%%%%%%%%%%%%%%%
\subsection{The debiased estimator}

Automatic debiased estimation with generated regressors is based on the moment condition in equation~\eqref{orth}, where the Riesz representers $\alpha_{01}$ and $\alpha_{02}$ are estimated automatically (see Section~\ref{sec:est_auto}). We construct sample analogues using cross-fitting, as in \citet{chernozhukov2018double}: the sample is split into $L$ folds $I_\ell$, and for each fold we evaluate $\psi(W_i,g_0,h_0,\alpha_0,\theta)$ only on observations $i\in I_\ell$ that were not used to estimate $(g_0,h_0,\alpha_0)$. Formally, we partition $(W_i)_{i=1}^n$ into $L$ groups $I_\ell$, for $\ell=1,\dots,L$. For each group, we have estimators $\hat{g}_\ell$, $\hat{h}_\ell$, and $\hat{\alpha}_\ell=(\hat{\alpha}_{1\ell},\hat{\alpha}_{2\ell})$ based only on observations outside $I_\ell$.

The debiased sample moment function is
\begin{equation*}
\hat{\psi}(\theta )\equiv \frac{1}{n}\sum_{\ell =1}^{L}\sum_{i\in I_{\ell }}%
\hat{\psi}_{i\ell }(\theta ),
\end{equation*}%
with
\begin{equation}
\hat{\psi}_{i\ell }(\theta )\equiv m(W_{i},\hat{g}_{\ell },\hat{h}_{\ell
},\theta )+\hat{\alpha}_{1\ell }(Z_{i})\cdot \epsilon(W_{i},\hat{g}_{\ell })+%
\hat{\alpha}_{2\ell }(X_{i},\hat{V}_{i\ell })^{\prime } (S_{i}-\hat{h}%
_{\ell }(X_{i},\hat{V}_{i\ell })),  \label{eq:dmon_i}
\end{equation}%
for $\hat{V}_{i\ell }\equiv \varphi (D_{i},Z_{i},\hat{g}_{\ell })$. When there is more than one moment condition, each component of $m$ is debiased by its own Riesz representers, so as many $\hat\alpha_{\ell}$'s must be estimated as there are moment conditions.

The three-step debiased GMM estimator is then defined as
\begin{equation}  \label{dgmm}
\hat{\theta}=\operatornamewithlimits{argmin}_{\theta\in\Theta}\hat{\psi}%
(\theta)^{\prime}\hat{\Upsilon}\hat{\psi}(\theta),
\end{equation}
where $\hat{\Upsilon}$ is a positive semi-definite weighting matrix of
dimension $q\times q$. Under regularity conditions (see
Section~\ref{sec:asymptotic}), $\hat{\theta}$ is asymptotically normal with the usual GMM asymptotic variance.

\begin{ex}[continues=ex:PLM_ML]
\label{ex:ATE_PLM1}
We illustrate the construction of a three-step debiased and cross-fitted ML estimator for the partially linear model in Example~\ref{ex:PLM_ML}. For expositional clarity, we consider $\operatorname{dim}(D)=1$. Suppose we have a generated regressor $\hat{V}_{i\ell} = \varphi(D_i, Z_i, \hat{g}_\ell)$. Within each fold, the second step estimates $h_{0Y}(v)=\E[Y\mid V=v]$ and $h_{0D}(v)=\E[D\mid V=v]$ by regressing $Y_i$ and $D_i$ on $\hat V_{i\ell}$ using a dictionary $\mathbf{b}_{J}(v)=(b_{1}(v),\ldots,b_{J}(v))^{\prime }$; for instance, one may take $b_j(v)=v^{j-1}$ and conduct an $\ell_1$-penalized least squares. This yields $\hat{h}_{\ell,Y}$ and $\hat{h}_{\ell,D}$ and defines $\hat h_\ell=(\hat h_{\ell,Y},\hat h_{\ell,D})$.

The partially linear moment in \eqref{eq:PLM_general_moment} is orthogonal with respect to $h$, implying $\alpha_{02}=0$. Therefore, a natural application of the DML estimator of \citet{chernozhukov2018double} yields the closed-form expression
\begin{equation}
\label{eq:ATE_PLM_DML}
\hat{\theta}_{DML} = \frac{\sum_{\ell =1}^{L}\sum_{i\in I_{\ell }}%
\left( Y_i - \hat{h}_{\ell,Y}(\hat{V}_{i\ell})\right)\left( D_{i}-\hat{h}_{\ell,D}(\hat{V}_{i\ell})\right)}{\sum_{\ell
=1}^{L}\sum_{i\in I_{\ell }}\left( D_{i}-\hat{h}_{\ell,D}(\hat{V}_{i\ell})\right) ^{2}}.
\end{equation}
This estimator is locally robust with respect to the second step but does not account for the effect of learning the generated regressor $\hat V_{i\ell}$.

Using the general debiased moment \eqref{orth}, the three-step debiased estimator takes the form
\begin{equation}
\label{DATE}
\hat{\theta}=\hat{\theta}_{DML} + \frac{\sum_{\ell =1}^{L}\sum_{i\in I_{\ell }}%
\hat{\alpha}_{1\ell }(Z_{i})\cdot \epsilon(W_{i},\hat{g}_{\ell})}{\sum_{\ell
=1}^{L}\sum_{i\in I_{\ell }}\left( D_{i}-\hat{h}_{\ell,D}(\hat{V}_{i\ell})\right) ^{2}}.
\end{equation}

Since the partially linear moment is orthogonal to $h$, downstream local robustness implies that the indirect contribution of the generated regressor through the second step vanishes. Hence, the first-step Riesz representer $\alpha_{01}$ also simplifies considerably. We construct an automatic cross-fitted estimator $\hat\alpha_{1\ell}$ in Section~\ref{sec:auto_1step}, with a special case provided below in equation~\eqref{eq:alpha1ell_riesz_lasso}.

\medskip
\noindent\textbf{Hd-PS regression adjustment.}

Let $\mathbf{c}_K(z) \equiv (c_1(z), \dots, c_K(z))'$ be a dictionary with $K$ atoms (in the high-dimensional case, this can simply collect the regressors $z_j$). The first step estimates the propensity score $\E[D\mid Z]$ via Lasso--Logit: $\hat{g}_\ell(z) =\mathbf{c}_K(z)'\widehat{\boldsymbol{\gamma}}_{K\ell}$, where
\begin{equation*}
    \widehat{\boldsymbol{\gamma}}_{K\ell}
    =
    \operatornamewithlimits{argmin}_{\boldsymbol{\gamma }_{K}\in \mathbb{R}^{K}}\left\{ -\sum_{i\notin I_{\ell }} \left[ D_{i}\log\Lambda(\mathbf{c}_{K}(Z_{i})^{\prime }\boldsymbol{\gamma }_{K})+(1-D_{i})\log(1-\Lambda(\mathbf{c}_{K}(Z_{i})^{\prime }\boldsymbol{\gamma }_{K})) \right]+\lambda \lVert \boldsymbol{\gamma }_{K}\rVert
_{1}\right\},
\end{equation*}
$\lVert \cdot\rVert _{1}$ is the $\ell_1$ norm, and $\lambda$ is a penalization parameter. The score of this problem leads to the orthogonality condition in equation~\eqref{orth1} with $\epsilon(W,g_0)=D-\Lambda(g_0(Z))$ and $\Delta_1$ the mean-square limit of sparse linear combinations of $(c_k)_{k=1}^\infty$. The generated regressor is the estimated propensity score
\[
\hat{V}_{i\ell}=\Lambda(\hat{g}_{\ell }(Z_{i})).
\]

In the Hd-PS regression adjustment, estimation of the second step simplifies to $\hat{h}_{\ell,D}(\hat{V}_{i\ell})=\hat{V}_{i\ell}$, since $\E[D\mid V]=V$. We provide a simple weighted Lasso estimator $\hat{\alpha}_{1\ell }$ as follows:
\begin{equation}\label{eq:alpha1ell_riesz_lasso}
\hat{\alpha}_{1\ell}(z)\equiv\mathbf{c}_K(z)'\widehat{\boldsymbol{\beta}}_{K\ell},\quad
\widehat{\boldsymbol{\beta}}_{K\ell}\in\operatornamewithlimits{argmin}_{\boldsymbol{\beta}_K\in\mathbb{R}^K}\left\{\sum_{\ell'\neq\ell}\sum_{i\in I_{\ell'}}\omega_{i\ell\ell'}\bigl(\mathcal{E}_{i\ell\ell'}-\mathbf{c}_K(Z_i)'\boldsymbol{\beta}_K\bigr)^2+\lambda\|\boldsymbol{\beta}_K\|_1\right\}.
\end{equation}
where the ``dependent'' variable is $\mathcal{E}_{i\ell\ell'}=-[Y_{i}-\hat{h}_{\ell \ell
^{\prime }}(\hat{V}_{i\ell \ell ^{\prime }})]$ and the weights are $\omega_{i\ell\ell'} \equiv \hat{V}_{i\ell \ell^{\prime }} (1-\hat{V}_{i\ell \ell^{\prime }})$. Here, $\hat{h}_{\ell \ell^{\prime }}$ and $\hat{V}_{i\ell \ell^{\prime }}$ are estimators that use only observations not in $I_\ell\cup I_{\ell^{\prime }}$. The arguments leading to this construction are detailed in Section~\ref{sec:ex_ATE_PLM}. In general, $\alpha_{01}\neq 0$, and hence, without our correction, inference is generally invalid. The DML estimator $\hat{\theta}_{DML}$ must be debiased to obtain an estimator that is locally robust to the generated propensity score. This estimator $\hat{\theta}$ can be easily implemented as a (cross-fitted) OLS of $Y_i - \hat{h}_{\ell,Y}(\hat{V}_{i\ell})+\hat{\alpha}_{1\ell }(Z_{i})$ on $D_{i}-\hat{V}_{i\ell}$.
\end{ex}

\begin{ex}[continues=ex:CF]
The moment condition defining the CASF is not orthogonal to the second step, so a debiasing term for each step is needed. We illustrate how to build a three-step debiased estimator of the CASF, starting from a plug-in estimator and comparing it to the natural extension of the DML estimator.

The first step recovers the control function $\hat{V}_{i\ell} = D_i - \hat{g}_\ell(Z_i)$, with $\hat{g}_\ell(z) =\mathbf{c}_K(z)'\widehat{\boldsymbol{\gamma}}_{K\ell}$ being a Lasso fit of $D_i$ on a dictionary $\mathbf{c}_K(Z_i)$. The second step estimates $h_0(x,v) = \E[Y\mid X=x, V=v]$ via Lasso. For a dictionary $\mathbf{b}_J(x, v)$ with $J$ atoms, we get $\hat{h}_\ell(x, v) = \mathbf{b}_J(x, v)'\widehat{\boldsymbol{\eta}}_J$ by $L_1$-penalized least squares of $Y_i$ on the dictionary evaluated at $X_i$ and the generated control function $\hat{V}_{i\ell}$.

To estimate the CASF according to equation~\eqref{eq:CASF_moment}, we compute the integral by Monte Carlo integration, since the counterfactual distribution $F^*$ is fixed by the researcher. Let $(X_s^*)_{s=1}^S$ be a sample from $F^*$, independent of the original sample ($S\gg n$). The cross-fitted plug-in estimator for the CASF is:
\begin{equation} \label{eq:CASF_plugin}
\hat{\theta}_{PI}=\frac{1}{nS}\sum_{\ell =1}^{L}\sum_{i\in I_{\ell
}}\sum_{s=1}^{S}\hat{h}_{\ell }(X_{s}^{\ast },\hat{V}_{i\ell }).
\end{equation}%
A DML estimator that accounts for estimation of $h_0$ in the second step, but not for the generated control function, it is given by
\begin{equation} \label{eq:CASF_DML}
    \hat{\theta}_{DML}= \hat{\theta}_{PI} + \frac1n \sum_{\ell =1}^{L}\sum_{i\in I_{\ell}} \hat{\alpha}_{2\ell}(X_i,\hat{V}_{i\ell}) \cdot (Y_i - \hat{h}_\ell(X_i, \hat{V}_{i\ell})).
\end{equation}
The three-step debiased estimator accounts for the generated regressor by extending the moment condition:
\begin{equation} \label{eq:CASF_D}
    \hat{\theta}= \hat{\theta}_{DML} + \frac1n \sum_{\ell =1}^{L}\sum_{i\in I_{\ell}} \hat{\alpha}_{1\ell}(Z_i) \cdot (D_i - \hat{g}_\ell(Z_i)).
\end{equation}
Automatic estimation of the Riesz representers $\alpha_{01}$ and $\alpha_{02}$ is detailed in Section~\ref{sec:CASF_details}.
\end{ex}

\subsection{Monte Carlo simulations}
\label{sec:MC}

We give an overview of two Monte Carlo studies: estimation of Hd-PS regression adjustment in the partially linear model and estimation of the CASF with a control-function approach. We evaluate the finite-sample performance of several estimation procedures. First, the plug-in estimator that uses the original moment condition. For inference based on the plug-in estimator, we consider both accounting and not accounting for estimation effects in the asymptotic variance. Second, the natural application of the DML procedure of \citet{chernozhukov2018double,chernozhukov2022locally}, which corrects only for the second step in parameter and asymptotic-variance estimation. Third, our proposed three-step debiased (3SD) estimator with an asymptotic-variance estimator (see Section~\ref{sec:asymptotic}). A detailed description of the setups, estimation procedures, and results is provided in Appendix~\ref{sec:MC_details}.

\subsubsection{Hd-PS regression adjustment in the partially linear model}

The available data are $(Y, D, Z)$, with $Z \equiv (Z_j)_{j=1}^{10}$. The outcome and treatment equations are:
\begin{align*}
    Y &= D + Z_1 + Z_2 + \varepsilon, \\
    D &= \mathbf{1}\left( C \nu \leq  Z_1 + Z_2 + Z_3 + Z_4 + Z_5 + Z_6 \right).
\end{align*}
The error terms $\varepsilon$ and $\nu$ are independent, with $\varepsilon \sim N(0, 1)$. The distribution of $\nu$ varies with the specification: it can be logistic or standard normal. The regressors $Z$ are independent of each other and are uniformly distributed on $[-1,1]$. The regressors are also independent of $\nu$. On the other hand, the first two regressors $(Z_1, Z_2)$ and $\varepsilon$ are correlated, rendering the treatment $D$ endogenous. The constant $C$ is chosen so that the propensity score is supported on $[0.01, 0.99]$. Here, $\theta_0 = 1$.

\begin{figure}[h!]
\centering
\begin{subfigure}{.33\textwidth}
    \centering
  \includegraphics[width=\textwidth]{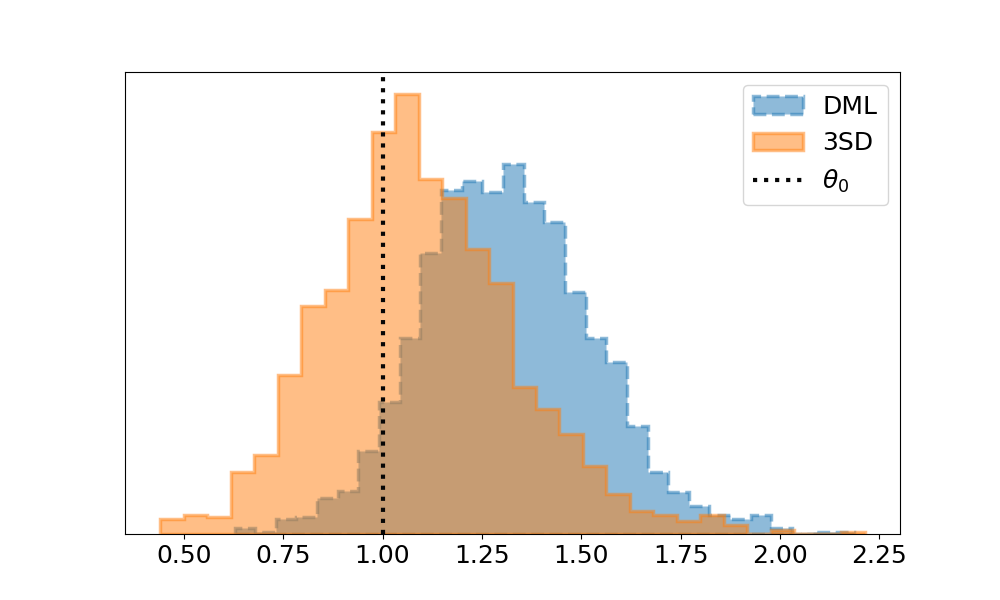}
  \caption{$n=100$.}
  
\end{subfigure}\hfill
\begin{subfigure}{.33\textwidth}
\centering
  \includegraphics[width=\textwidth]{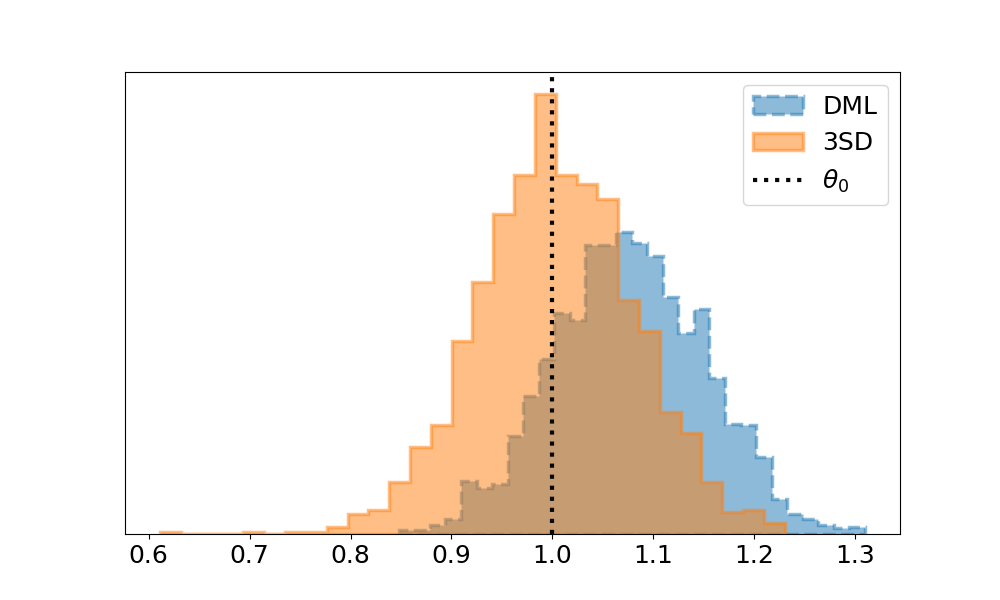}
  \caption{$n=500$.}
  
\end{subfigure}\hfill
\begin{subfigure}{.33\textwidth}
  \includegraphics[width=\textwidth]{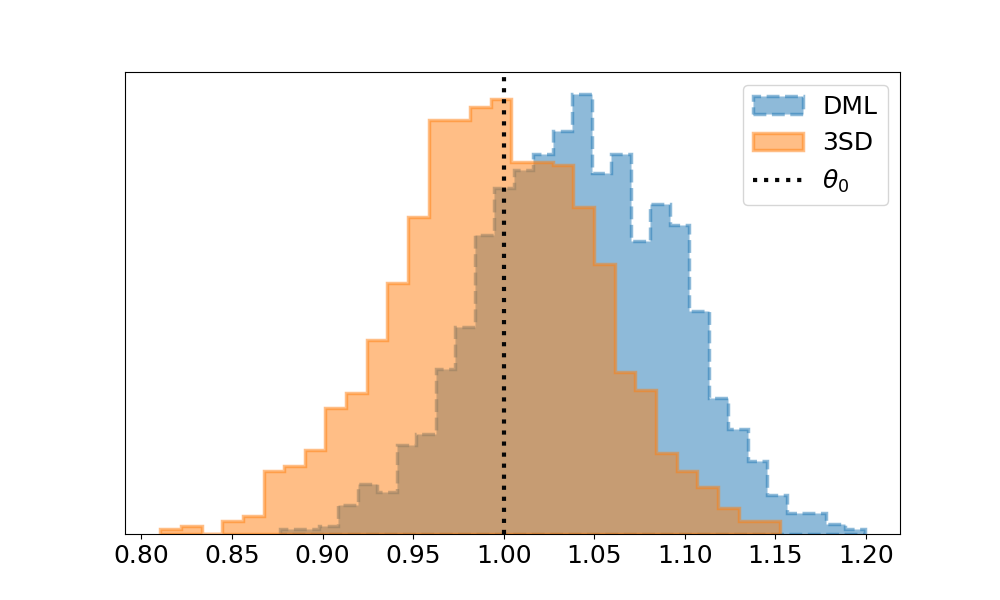}
  \caption{$n=1000$.}
  
\end{subfigure}

\caption{Histograms of the estimators for $\theta_0$ in a partially linear model framework (logistic $\nu$) with Hd-PS first-step. Number of replications is $2000$. \textit{DML} = Double/Debiased Machine Learning estimator, \textit{3SD} = Three-Step Debiased estimator. Note that the DML estimator equals the plug-in estimator.}
\label{fig:histogram_ATE}
\end{figure}

Results regarding the mean bias are similar across specifications. For a small sample ($n=100$), the DML estimator (the same as the plug-in estimator here) is heavily biased, while the three-step debiased estimator performs well (see Table~\ref{tab:MC_ATE} in Appendix~\ref{sec:MC_details}). When $n=100$, debiasing removes around 85\% of the bias present in the DML estimator. As the sample size increases, the bias of the DML estimator becomes smaller. Nevertheless, the bias of the DML estimator remains orders of magnitude larger than that of the three-step debiased estimator. Figure~\ref{fig:histogram_ATE} displays histograms of both estimators for the logistic-$\nu$ specification. We see that, when $n=1000$, the DML estimator is still biased. The distribution of the three-step debiased estimator is centered around $\theta_0 = 1$.

Table~\ref{tab:MC_ATE} in Appendix~\ref{sec:MC_details} shows that the coverage of the three-step debiased estimator is close to the nominal 95\%, even for $n=100$. On the other hand, the DML asymptotic-variance estimator tends to overestimate the true asymptotic variance. This leads to coverage rates that exceed the nominal level, except for the $n=100$ case, where the bias dominates. In addition, the plug-in estimator for $\theta_0$ (which equals the DML estimator) shows poor performance even when using the correct asymptotic variance for inference. Its coverage is below 90\% even when $n=1000$.

\subsubsection{CASF with a control-function approach}

The available data are $(Y, D, Z)$, with $Z \equiv (Z_j)_{j=1}^6$. The variables $D$ and $Y$ are generated by:
\begin{align*}
	Y = \sum_{k=1}^5 Z_k + 2D + U \text{ and } D = \sum_{k=1}^6 Z_k + V.
\end{align*}
The error terms $U$ and $V$ are correlated, with $U, V \sim N(0,1)$, so $D$ is endogenous. The regressors $Z$ are standard normal and are independent of each other and of the errors $(U,V)$. In this case, $X = (Z_1, \dots, Z_5, D)$. We estimate the CASF for the following counterfactual distribution $F^*$: (i) the distribution of $(Z_1,\dots, Z_5)$ remains unchanged and (ii) $D$ is normal with mean $1$ (instead of $0$) and the same variance as in the DGP. Therefore, the true parameter is $\theta_0=2$.

The plug-in estimator is severely biased across all sample sizes (see Table~\ref{tab:MC_CASF} in Appendix~\ref{sec:MC_details} and Figure~\ref{fig:histogram_CASF} below). However, the comparison between the DML and the three-step debiased estimator differs from that in the previous example. For small samples ($n=100$), both estimators have similar bias. As the sample size increases, the bias of the three-step debiased estimator decreases, while the bias of the DML estimator remains sizable. This confirms the presence of an asymptotic bias in the DML estimator.

\begin{figure}[h!]
\centering
\begin{subfigure}{.33\textwidth}
    \centering
  \includegraphics[width=\textwidth]{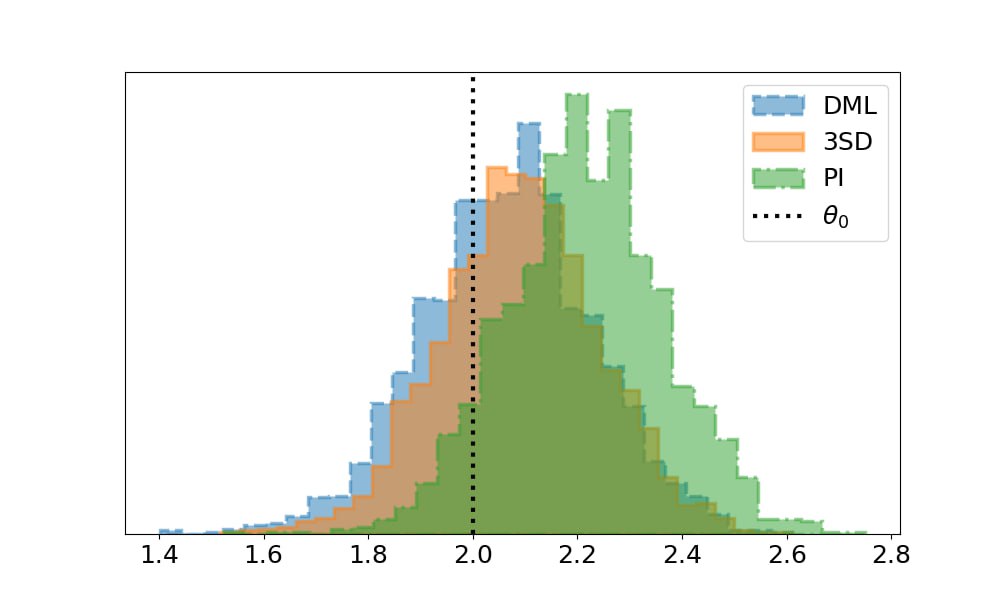}
  \caption{$n=100$.}
  
\end{subfigure}\hfill
\begin{subfigure}{.33\textwidth}
\centering
  \includegraphics[width=\textwidth]{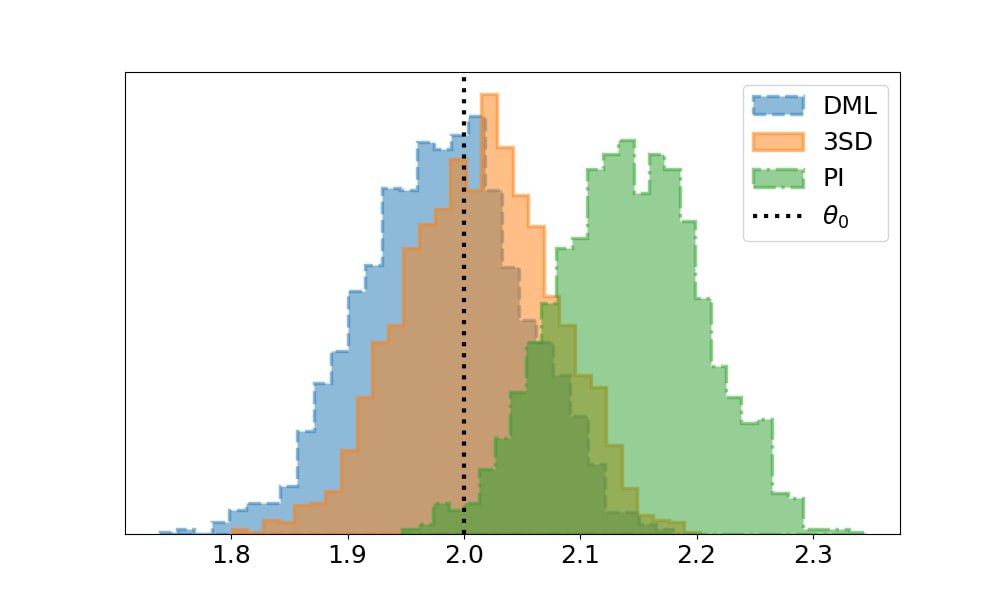}
  \caption{$n=500$.}
  
\end{subfigure}\hfill
\begin{subfigure}{.33\textwidth}
  \includegraphics[width=\textwidth]{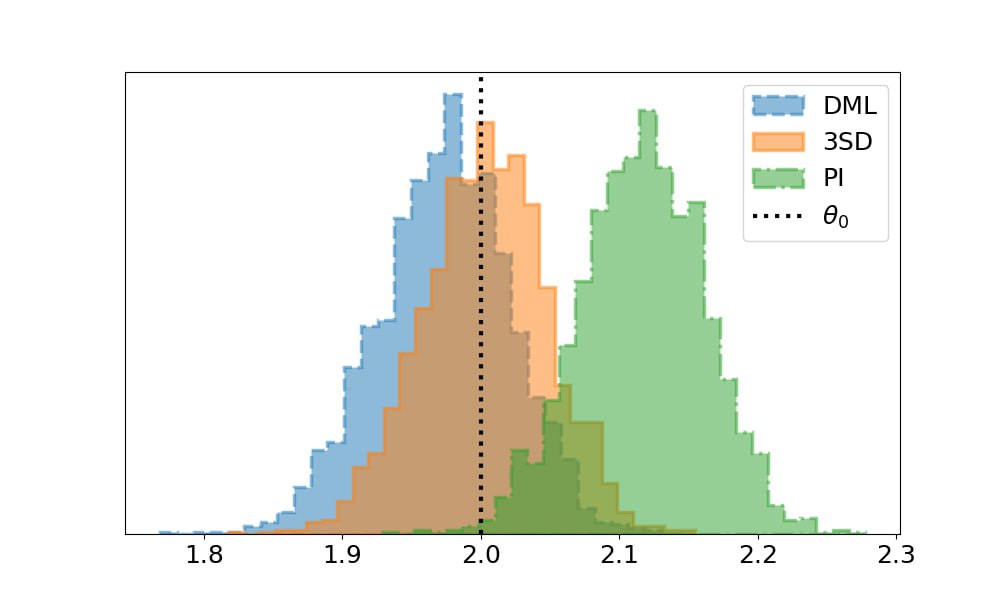}
  \caption{$n=1000$.}
  
\end{subfigure}

\caption{Histograms of the CASF estimators with a Control-Function Approach. Number of replications is $2000$. \textit{PI} = Plug-in estimator, \textit{DML} = Double/Debiased Machine Learning estimator, and \textit{3SD} = Three-Step Debiased estimator.}
\label{fig:histogram_CASF}
\end{figure}

Results regarding coverage also differ from those of the previous example (see Table~\ref{tab:MC_CASF} in Appendix~\ref{sec:MC_details}). The three-step debiased estimator shows good coverage, close to the nominal 95\% level when $n=500$ or $n=1000$. In this case, the DML asymptotic-variance estimator underestimates the true asymptotic variance. Thus, its coverage is well below the nominal 95\% level across all sample sizes. In estimating the CASF, the plug-in estimator performs poorly due to the large asymptotic bias. Even when using the correct asymptotic variance for inference, its coverage ranges from 59.4\% when $n=100$ to 24.5\% when $n=1000$.

Summarizing, not accounting for the generated regressor leads to large biases in finite samples. In contrast, our three-step debiased procedure substantially reduces bias and delivers robust and valid inference. The following sections show how the Riesz representers needed to build the debiased moment function are identified and estimated. These sections are more technical than the previous ones; thus, an applied reader may wish to jump directly to Section~\ref{sec:Examples}, where additional details about the examples are gathered.

%%%%%%%%%%%%%%%%%%%%%%%%%%%%%%%%%%%%%%%%%%%%%%%%%%%%%%%%%%%%%%%%%%%%%%%%%%%%%%%%%%%%%%%%%%%%%%%%%%%%%%
\subsection{First- and second-step influence functions}
\label{sec:first_second_IF}

This section provides a detailed construction of orthogonal moment functions in our three-step setting with generated regressors. We begin by introducing additional concepts and notation. Let $F$
denote a possible cdf for a data observation $W$. We denote by $g(F)$ the
probability limit of an estimator $\hat{g}_\ell$ of the first step when the true
distribution of $W$ is $F$, i.e., under general misspecification
\citep[see][]{newey1994asymptotic}. That is, $F$ is unrestricted except for
regularity conditions such as existence of $g(F)$ and finiteness of the expectation of
certain functions of the data. For example, if $\hat{g}_\ell(z)$ is a
nonparametric estimator of $\mathbb{E}[D\mid Z=z]$, then $g(F)(z)=\mathbb{E}%
_{F}[D\mid Z=z]$ is the conditional expectation function when $F$ is the true
distribution of $W$. We denote expectation under $F$ by $\mathbb{E}_{F}$, which is well defined
under the regularity condition that $\mathbb{E}_{F}[|D|]$ is finite. We assume that $g(F)$ is identified as the solution in $g$ to
\begin{equation*}
\mathbb{E}_{F}[\delta _{1}(Z)\epsilon (W,g)]=0\text{ for all }\delta _{1}\in
\Delta _{1}.
\end{equation*}%
Our notation is consistent with $g(F_{0})=g_{0}$ being the probability limit of $\hat{g}$ when $F_{0}$ is the cdf of $W$.

To study the effect of the second step, suppose again that $W$ is distributed
according to $F$, but the first-step nuisance is independently fixed
to $g\in \Delta _{1}$. Let $h(F,g)$ be the solution in $h\in \Delta _{2}(g)$
to
\begin{equation}
\mathbb{E}_{F}\left[ \delta _{2}(X,V(g))\{S-h(X,V(g))\}\right] =0\text{ for
all }\delta _{2}\in \Delta _{2}(g),  \label{eq:ortho_genFg}
\end{equation}%
where $V(g)\equiv \varphi (D,Z,g)$ and $\Delta _{2}(g)$ is a linear and closed subspace of $L_{2}(X,V(g))$ for each $g\in \Delta _{1}$. The
solution of the above equation is a function of $(x,v)$, written as $h(F,g)(x,v)$. In the toy example, $h(F,g)(x,v)=\beta_g(F)v$, where $\beta_g(F)=\E_{F}[YV(g)]/\E_{F}[V(g)^2]$. We use the short notation $h_{0}(x,v)\equiv h(F_{0},g_{0})(x,v)$. Thus, henceforth, a subscript $0$ in $h$ means that the conditioning variable is the true generated regressor $V\equiv V(g_{0})$; for example, $h_{0}(x,v)=\mathbb{E}[Y\mid X=x,V=v]$ when $\Delta_{2}(g)=L_{2}(X,V(g))$. We may think of the mapping $h(F,g)$ as the probability
limit of an estimator of $h_{0}$ under the following conditions: (i) the
true distribution of $W$ is $F$ and (ii) the estimator is constructed with the
first-step nuisance fixed at $g\in \Delta_{1}$. A feasible estimator $\hat{h}_\ell$ of $h_{0}$
will, however, rely on the estimator $\hat{g}_\ell$ with probability limit $g(F)$.
Therefore, we assume that the probability limit of $\hat{h}_\ell$ under general
misspecification is $h(F,g(F))$.

\begin{ex}[continues=ex:PLM_ML]
\label{ex:PLM_ortho}
In the partially linear model with generated regressors, $h$ has two components,
$h=(h_{Y},h_{D})$. For $S$ equal to $Y$ or $D$, denote $h_{S}(F,g)(v)=\mathbb{E}%
_{F}[S\mid V(g)=v]$. The first step $g$ enters each second step $h_{S}$ in
two ways: (i) indirectly, through the conditioning variable $V(g)$, and (ii)
directly, when we evaluate $v$ at $V(g)$. Following our notation,
$h_{S}(F_{0},g_{0})(V(g_{0}))$ simplifies to $h_{0S}(V)$, and
$h_{0}=(h_{0Y},h_{0D})$.
\end{ex}

Let $H$ be some alternative distribution
that is unrestricted except for regularity conditions, and define $F_{\tau }\equiv
(1-\tau )F_{0}+\tau H$ for $\tau \in [0,1]$. We assume that $H$ is
chosen so that $g(F_{\tau })$ and $h(F_{\tau },g(F_{\tau }))$ exist for sufficiently small $\tau$, and that other regularity conditions are satisfied. The effect of \emph{both first- and second-step estimation} on the moment condition is measured by the derivative with respect to $\tau$ at $\tau=0$ of $\bar{m}(g(F_\tau), h(F_\tau, g(F_\tau)))$, with
\begin{equation*}
    \bar{m}(g,h) \equiv \mathbb{E}[m(W,g,h,\theta_0)].
\end{equation*}
We study these effects separately. By the chain rule,
\begin{align}
\frac{d}{d\tau }\bar{m}(g(F_{\tau }),h(F_{\tau },g(F_{\tau })))& =%
\frac{d}{d\tau }\bar{m}(g(F_{\tau }),h(F_{0},g(F_{\tau })))
\label{eq:FSD} \\
& \quad +\frac{d}{d\tau }\bar{m}(g_{0},h(F_{\tau },g_{0})),  \label{eq:SSD}
\end{align}
where, henceforth, $d/d\tau$ denotes the right derivative with respect to $\tau$, evaluated at $\tau=0$. In the display above, the first derivative on the right-hand side (RHS)
accounts for the first step. As in \citet{hahn2013asymptotic}, the first
step affects the moment condition in two ways (see Figure~\ref{fig:chainrule}). We have a \textit{direct impact} on $\bar{m}$, quantified by the derivative of $\bar{m}(g(F_{\tau }),h_{0})$. This direct impact includes the \textit{effect of evaluating $h$} at the generated regressor. We also have an
\textit{indirect effect} on the moment that arises because $g$ affects
\emph{estimation} of $h_{0}$ in the second step (through conditioning), quantified by the derivative of $\bar{m}(g_{0},h(F_{0},g_{\tau }))$. Both effects (direct and
indirect) are considered in \eqref{eq:FSD}. The derivative in
\eqref{eq:SSD} accounts for the effect of the second step. This effect is
independent of the first step and therefore treats $g_{0}$ as known.

\begin{figure}[h!]
\centering
\begin{tikzpicture}[->,>=stealth',auto,node distance=2.6cm,
		thick]
		
		\node (par) {$\tau$};
		\node (dist) [align=center, right of=par] {$F_\tau$};
		\node (g) [align=center,right of=dist] {$g(F_\tau)$};
		\node (h) [align=center, right of=g] {$h(F_\tau, g(F_\tau))$};
		\node (exp) [align=center, below of=h] {$\bar{m}(g(F_\tau), h(F_\tau, g(F_\tau)))$};
		
		\draw [->] (par) -- (dist);
		\draw [->] (dist) -- (g);
		\draw [->] (g) -- (h) node [midway, fill=white, above=0.25em] {(I)};
		\draw [->] (h) -- (exp) node [midway, fill=white, right = 0.25em] {(2S)};
		\draw [->] (g) -- (exp) node [midway, fill=white, above right =0.01em and 0.01em] {(D)};
		\draw [->] (dist) to [out=90,in=90] (h);
	\end{tikzpicture}
\caption{The effect of a deviation $F_\protect\tau$ on the moment condition.
(2S) represents the second-step effect. (D) represents the direct effect of
the first step. The path (I)-(2S) represents the indirect estimation effect
of the first step.}
\label{fig:chainrule}
\end{figure}

To debias the moment conditions, we compute separate IFs for each estimation step. That is, we seek functions $\phi _{1}(w,g,\alpha
_{1})$ and $\phi _{2}(w,g,h,\alpha _{2})$ such that, all $H$ defining a regular path $F_{\tau }\equiv
(1-\tau )F_{0}+\tau H$,
\begin{align}
\frac{d}{d\tau }\bar{m}(g(F_{\tau }),h(F_{0},g(F_{\tau })))& =\int
\phi _{1}(w,g_{0},\alpha _{01})\,dH(w) \text{ and}  \label{eq:IF1} \\
\frac{d}{d\tau }\bar{m}(g_{0},h(F_{\tau },g_{0}))& =\int \phi
_{2}(w,g_0, h_{0},\alpha _{02})\,dH(w).  \label{eq:IF2}
\end{align}%
Additionally, we require the IFs to have zero mean and finite variance. Note that $(\alpha_{01}, \alpha_{02})$ are the Riesz representers of the above derivatives, which are evaluated at $(g_0, h_0, \theta_0)$, and thus may depend on $(g_0, h_0, \theta_0)$. For functions $\phi_1$ and $\phi_2$ satisfying the above conditions, the moments
\begin{equation*}
\psi (w,g,h,\alpha ,\theta )\equiv m(w,g,h,\theta )+\phi _{1}(w,g,\alpha
_{1}) + \phi _{2}(w,g,h,\alpha _{2})
\end{equation*}
are orthogonal/locally robust/debiased. The next section shows that, under some conditions, the IFs have the form displayed in equation~\eqref{orth}. It also illustrates the separate automatic estimation of each Riesz representer $\alpha_{01}$ and $\alpha_{02}$.

It is worth highlighting that when the second-step effect is zero, the first-step indirect effect is zero by the chain rule. This is a special case of a more general result that applies beyond generated regressors.
\begin{prop}[Downstream local robustness]
\label{prop:zero_indirect}
Assume that $\bar m$ is Hadamard differentiable in the second step $h$ at $(g_0,h_0)$, with derivative $D_{02}$, and that $h(F, g)$ is Hadamard differentiable in $g$ at $(F_0, g_0)$. If the moment is locally robust with respect to the second step, i.e., $D_{02} = 0$,
then, for any regular path $\tau \mapsto F_\tau$ through $F_0$,
\begin{equation*}
\frac{d}{d\tau}\bar m\big(g(F_0),h(F_0,g(F_\tau))\big)
= 0.
\end{equation*}
\end{prop}

\begin{rem}[Comparison with two-step approaches]
\label{twostep}
The dependence of the second step on the first step makes the results in \cite{chernozhukov2022locally} not applicable in our setting. In particular, we require that for each $g\in \Delta _{1}$, $h(X,V(g))\in \Delta _{2}(g)$, so the parameter space of the second step depends on the first step. In a two-step setting with multiple parameters, one would need $(g,h)$ to live in a linear product space \citep[see Theorem~3 in][]{chernozhukov2022locally}. The dependence of $\Delta_2(g)$ on $g$ breaks this structure. Our results rely on alternative assumptions (see Assumption~\ref{ass:inclusion}) to handle this parameter space.
\end{rem}
%%%%%%%%%%%%%%%%%%%%%%%%%%%%%%%%%%%%%%%%%%%%%%%%%%%%%%%%%%%%%%%%%%%%%%%%%%%%%%%%%%%%%%%%%%%%%%%%%%%%%%%%%%%%%%%%%%%%%%%%%%%%%%%%%%%%%%%%%%%%%%%%%%%%%%%%%%%%%%%%%%%%%%%%%%%%%%%%%%%%%%%%%%%%%%%%%%%%%%%%%%%%%%%%%%%%%%%%%%%%%%%%%%%%%%%%%%%%%%%%%%%%%%%%%%%%%%%%%%%%%%%
\section{Automatic estimation of the Riesz representers}
\label{sec:Automatic}

The orthogonal moments require a consistent estimator $\hat{\alpha%
}_\ell$ of the Riesz representers $\alpha _{0}\equiv (\alpha _{01},\alpha _{02})$%
. When the shape of $\alpha _{0}$ is known, one can plug-in nonparametric
estimators of the unknown components of $\alpha _{0}$ to form $\hat{\alpha}_\ell$%
. In the
generated regressors setup, however, the nuisance parameters (especially $%
\alpha _{01}$) have a complex analytical shape (see the result in equation~%
\eqref{eq:alpha1_defi} in Appendix \ref{sec:Proofs}). Therefore, the plug-in estimators may be cumbersome to compute in practice.

To ease exposition and without loss of generality, in this section, we consider that there is a single moment condition ($p=q=1$). Recall that in the multi-dimensional case one must estimate Riesz representers $\alpha _{0}$ for each moment condition.

%%%%%%%%%%%%%%%%%%%%%%%%%%%%%%%%%%%%%%%%%%%%%%%%%%%%%%%%%%%%%%%%%%%%%%%%%%%%%%%%%%%%%%%%%%%%%%%%%%%%%%
\subsection{Separate identification of Riesz representers}
\label{sec:ident_alphas}

We provide separate orthogonality conditions that will serve as a basis
for the identification and automatic estimation of the Riesz representers $\alpha _{01}$ and $%
\alpha _{02}$. Define the following moment functions: $\psi _{1}(w,g,\alpha
_{1},\theta )\equiv m(w,g,h(F_{0},g),\theta )+\phi _{1}(w,g,\alpha
_{1})$ for the first step, and $\psi _{2}(w,h,\alpha _{2},\theta
)\equiv m(w,g_{0},h(F,g_0),\theta )+\phi _{2}(w,g_0,h(F,g_0),\alpha _{2})$ for the
second step. 
Since, individually, the spaces $\Delta _{1}$ and $\Delta _{2}(g_{0})$ are linear, an application of
Theorem~3 in \cite{chernozhukov2022locally} to each step leads to
\begin{align}
\frac{d}{d\tau }\mathbb{E}[\psi _{1}(W,g_{0}+\tau \delta _{1},\alpha
_{01},\theta_0 )]& =0\text{ for all }\delta _{1}\in \Delta _{1}\text{ and}
\label{eq:auto_momentF} \\
\frac{d}{d\tau }\mathbb{E}[\psi _{2}(W,h_{0}+\tau \delta _{2},\alpha
_{02},\theta_0 )]& =0\text{ for all }\delta _{2}\in \Delta _{2}(g_{0}),
\label{eq:auto_momentS}
\end{align}%
where $\delta_{1}$ represents a possible
direction of deviation of $g(F)$ from $g_{0}$ and $\delta _{2}$ represents a
possible deviation of $h(F,g_{0})$ from $h_{0}$. The innovation relative to \cite{chernozhukov2022locally} is that we can
compute the IFs $\phi_1$ and $\phi_2$ by separately studying $\psi _{1}$ and $\psi _{2}$, respectively. This means we can
separately identify $\alpha _{01}$ and $\alpha _{02}$ from (\ref{eq:auto_momentF}) and (\ref{eq:auto_momentS}), even though $\psi _{1}$ and $\psi _{2}$ are not LR moment functions ($\psi _{1}$ and $\psi _{2}$ are not LR to $h_0$ and $g_0$, respectively). 

Likewise, rather than joint
identification from the analytical derivatives of the original identifying
moments as in \cite{chernozhukov2022locally}, which are not be available with generated regressors, we propose an approach that uses the linearization and orthogonality of $\psi _{1}
$ and $\psi _{2}$ with respect to $g$ and $h$, respectively, to construct
separate estimators of $\alpha _{01}$ and $\alpha _{02}$. This approach does not
require knowing the shape of $\alpha _{0}$. It is \textquotedblleft
automatic" in only requiring the orthogonal moment functions and data for
the construction of $\hat{\alpha}_\ell$. Moreover, an automatic estimator can be
constructed separately for each step. 

The key ingredients for our approach are (i) the shape of the IFs and (ii) a consistent estimator of the linearization of the moment condition with respect to each
parameter ---$g$ for the first step and $h$ for the second. Section~\ref%
{sec:Linearization} provides the formal development. For a detailed construction of the automatic estimators, we refer to Section~%
\ref{sec:est_auto}.

%%%%%%%%%%%%%%%%%%%%%%%%%%%%%%%%%%%%%%%%%%%%%%%%%%%%%%%%%%%%%%%%%%%%%%%%%%%%%%%%%%%%%%%%%%%%%%%%%%%%%%
\subsection{First- and second-step linearization}
\label{sec:Linearization}

We start with the linearization of the second-step effect because this will
show up in the first-step linearization. The linearization of
the second step with a known first step is a well-established result in the
literature \citep[see,
e.g.,][Equation~4.1]{newey1994asymptotic}, and it will follow immediately if
$\bar{m}(g_{0},h)$ is linear in $h$.

Before introducing the result, we note that throughout this section, for $F_\tau\equiv (1-\tau)F_0+\tau H$, we consider that $%
\tau\mapsto h_\tau\equiv h(F_\tau,g_0)$ and $\tau \mapsto g_\tau \equiv g(F_\tau)$ denote differentiable paths in $L_2(X,V)$ and $L_2(Z)$, respectively; i.e., $0\mapsto h_0$ and $%
dh_\tau/d\tau$ exists (equivalently for $g_\tau$). When an assumption is stated for $h_0$ or $h_\tau$, it is understood that it applies to each of its components.

We assume that $\bar{m}$ can be linearized with respect to the second step
parameter:

\begin{ass}
\label{ass:Diff_m_h} There exists a function $D_{02}(w,h)$ such that
\begin{equation*}
\frac{d}{d\tau}\bar{m}(g_0,h_\tau)=\frac{d}{d\tau}\mathbb{E}[D_{02}(W,h_\tau)].
\end{equation*}
 Moreover, $h\mapsto \mathbb{E}[D_{02}(W,h)]$
is linear and continuous in $L_2(X,V)$.
\end{ass}

The same assumption has been considered in \cite{newey1994asymptotic}. A necessary and sufficient condition for the linearity and continuity part is the existence of $r_{02} \in L_2(X,V)$ such that $\mathbb{E}[D_{02}(W,h)]=\mathbb{E}[r_{02}(X,V)h(X,V)]$ for all $h\in L_2(X,V)$. We can then get the shape of the second step IF:

\begin{prop}
{\label{prop:second_step}} Under Assumption~\ref{ass:Diff_m_h}, there exists an $\alpha _{02}\in \Delta
_{2}(g_{0})$, given by the orthogonal projection of $r_{02}$ onto $\Delta_{2}(g_{0})$,  such that the function
\begin{equation*}
\phi _{2}(w,g_0,h_{0},\alpha _{02})=\alpha _{02}(x,\varphi
(d,z,g_{0}))^{\prime }\{s-h_{0}(x,\varphi (d,z,g_{0}))\},
\end{equation*}%
satisfies equation~\eqref{eq:IF1} and is thus the second-step IF.
\end{prop}

An important observation is that if $r_{02}$ is zero, then $\alpha _{02}$ (and hence $\phi_{2}$) is also zero. We also note that $\bar{m}$ is linearized at $(g_{0},h_{0},\theta_0)$, so $D_{02}$, $r_{02}$, and $\alpha _{02}$ may also depend on $(g_{0},h_{0},\theta_0)$. This is omitted for notational simplicity, but it is of course accounted for in the theory of this paper, and it
will become relevant to construct feasible automatic estimators (see Section~%
\ref{sec:est_auto}).

We now move to the more complicated linearization of the first-step effect. Note that if the chain rule can be applied:
\begin{equation}\label{eq:1st_step_decomposition}
\begin{aligned}
    \frac{d}{d\tau} \bar{m}(g(F_\tau),h(F_0, g(F_\tau))) &= \frac{d}{d\tau%
} \bar{m}(g(F_\tau),h_0) \\
&+\frac{d}{d\tau} \bar{m}(g_0,h(F_0, g(F_\tau))).
\end{aligned}
\end{equation}
The first derivative in the RHS can be easily analyzed if we linearize $\bar{%
m}(g,h_{0})$ in $g$:
\begin{ass}
\label{ass:Diff_m_g} There exists a function $D_{dir}(w,g)$ such that
\begin{equation*}
\frac{d\bar{m}(g_\tau,h_0)}{d\tau}=\frac{d\mathbb{E}%
[D_{dir}(W,g_\tau)]}{d\tau}.
\end{equation*}
Moreover, $g\mapsto \mathbb{E}%
[D_{dir}(W,g)]$ is linear and continuous in $L_{2}(Z)$.
\end{ass}
The term $D_{dir}$ is responsible for the direct effect of the first step (the evaluation effect). Again, for simplicity of notation, we drop the dependence of $D_{dir}$ on $(g_0,h_0, \theta_0)$, though our theory accounts for this dependence. 

To study the indirect effect, $d\bar{m}(g_0,h(F_0, g(F_\tau)))/d\tau$, we generalize the
key Lemma~1 in \cite{hahn2013asymptotic} to allow for ML second
steps as in equation~\eqref{orth2}. The lemma is stated for one-dimensional $S$. For higher dimensions, it must be applied component-wise.
\begin{lma}
{\label{lma:generalized_HR}} Assume that the chain rule can be applied along
the path $\tau \mapsto g_{\tau }$. Then, for every $\delta _{2}\in L_{2}(X,V)
$ satisfying that there exists an $\varepsilon >0$ such that $\delta _{2}\in
\cap _{\tau <\varepsilon }\Delta _{2}(g_{\tau })$:%
\begin{equation*}
\frac{d}{d\tau }\mathbb{E}[\delta _{2}(X,V)\cdot h(F_{0},g_{\tau })(X,V)]= \frac{d}{d\tau }\mathbb{E}[\delta_{2}(X,V(g_{\tau})) \cdot (S-h_{0}(X,V(g_{\tau})))]
\end{equation*}
\end{lma}

The condition that the function $\delta_2$ belongs to every set $\Delta(g)$
for $g$ close to $g_0$ is related to ``regularity" of $\Delta_2(g)$. If the
functions in the sets $\Delta_2(g)$, with $g \in \Delta_{1}$,
 have the same shape, one would expect
that many $\delta_2$'s satisfy the condition in the above lemma. The
condition allows to take derivatives in equation~\eqref{eq:ortho_genFg}
along the path $(F_0, g_\tau)$.

To linearize the first step, we ask $\alpha_{02}$ to satisfy the condition for $\delta_{2}$ in Lemma~\ref{lma:generalized_HR}. We
 also impose some additional assumptions on the paths $\tau \mapsto
h(F_0, g_\tau)$. This allows us to express $d\bar{m}(g_0,h(F_0,
g(F_\tau)))/d\tau$ as an inner product.

\begin{ass}
\label{ass:inclusion} For every path $\tau \mapsto g_\tau$ there exits an $%
\varepsilon>0$ such that \mbox{} \\[-20pt]

\begin{enumerate}[label=\textbf{\alph*.},ref=\ref{ass:inclusion}.\alph*]
		\item \label{ass:inclusion:alpha} $\alpha_{02}\in \cap_{\tau<\varepsilon}\Delta_2(g_\tau)$, and
            \item \label{ass:inclusion:h} $h(F_0, g_\tau)\in \Delta_2(g_0)$ for all $0\leq\tau<\varepsilon$.
\end{enumerate}
\end{ass}

As we have emphasized, this assumption is related to \textquotedblleft
regularity" in the shape of the functions in $\Delta _{2}(g)$. It is needed to deal with a non-linear parameter space for $(g, h)$. In both the
nonparametric case $\Delta _{2}(g)=L_{2}(X,V(g))$ and the partially linear
case $\Delta _{2}(g)=\{\beta ^{\prime }x+\kappa (v)\colon \beta \in \mathbb{R%
}^{p},\kappa \in L_{2}(V(g))\}$ the assumption translates into
square-integrability conditions (see Appendix~\ref{sec:app_inclusion} for a detailed discussion). What
Assumption~\ref{ass:inclusion} rules out, for example, it is to specify a partially linear model for some $g$ and a nonparametric regression for others.

Once we can apply Lemma~\ref{lma:generalized_HR}, the remaining step is to
linearize the terms $h_0(X,\varphi(D,Z,g(F_\tau)))$ and $\alpha_{02}(X,%
\varphi(D,Z,g(F_\tau)))$. To achieve this, we require $h_0$, $\alpha_0$, and
$\varphi$ to be differentiable in an appropriate sense:

\begin{ass}
\label{ass:diff_h_alpha_phi} $h_0(x,v)$ and $\alpha_{02}(x,v)$ are almost surely differentiable w.r.t. $v$ with square-integrable derivatives. Moreover, the mapping $g\mapsto \varphi(d,z,g)$, from $L_2(Z)$ to $L_2(D,Z)$%
, is Hadamard differentiable at $g_0$, with derivative $D_\varphi$.
\end{ass}

The Hadamard derivative of $\varphi $ is a linear and continuous map $%
D_{\varphi }\colon L_{2}(Z)\rightarrow L_{2}(D,Z)$ such that
\begin{equation*}
\frac{d}{d\tau }\varphi (d,z,g_{\tau })=\frac{d}{d\tau }D_{\varphi }g_{\tau
}.
\end{equation*}%
To illustrate, if $\varphi (d,z,g)=g(z)$ (first step prediction)
or $\varphi (d,z,g)=d-g(z)$ (first step residual), then 
$D_{\varphi }g=g$ or $D_{\varphi }g=-g$, respectively.

To identify the first-step Riesz representer $\alpha_{01}$ while allowing for general residuals $\epsilon(W,g)$, we need the following assumption \citep[cf.][Ass.~2 and the discussion below]{ichimura2022influence}.
\begin{ass}
\label{rho}
The mapping $g \mapsto \E[\epsilon(W, g)]$ is Hadamard differentiable at $g_0$. The Riesz representer of the derivative ($r_e$) satisfies $r_{e}(z)<0$ and is bounded and bounded away from zero.
\end{ass}
The usual first-step error $\epsilon(W,g)=D-g(Z)$ has $r_e(Z)=-1$, satisfying the above assumption. The Logit-Lasso error $\epsilon(W, g)=D-\Lambda(g_0(Z))$ has $r_e(Z) = -\Lambda(g_0(Z))(1-\Lambda(g_0(Z))$ and satisfies the above assumption if the propensity score is bounded away from zero and one. 

The next theorem gives the shape of the first-step IF:
\begin{thm}
{\label{thm:first_step}} Under Assumptions \ref{ass:Diff_m_h}-\ref{rho}:

\begin{itemize}
\item[\textsc{\textbf{(Lin)}}] The function
\begin{equation}
{\label{eq:D1_defi}}D_{01}(w,g)\equiv D_{dir}(w,g)+\frac{%
\partial }{\partial v}\left[ \alpha_{02}(x,v)' (s-h_{0}(x,v))%
\right] \cdot D_{\varphi }g,
\end{equation}%
where the derivative is evaluated at $v=\varphi (d,z,g_{0})$, satisfies
\begin{equation*}
\frac{d}{d\tau }\bar{m}(g(F_{\tau }),h(F_{0},g(F_{\tau })))=\frac{d}{%
d\tau }\mathbb{E}[D_{01}(W,g(F_{\tau }))].
\end{equation*}

\item[\textsc{\textbf{(IF)}}] There exists an $\alpha _{01}\in \Delta _{1}$, given by  equation~\eqref{eq:alpha1_defi} in Appendix \ref{sec:Proofs}, such that the function
\begin{equation*}
\phi _{1}(w,g_{0},\alpha _{01})=\alpha _{01}(z)\cdot \epsilon
(w,g_{0}),
\end{equation*}%
satisfies equation~\eqref{eq:IF2} and is thus the first-step IF.
\end{itemize}
\end{thm}

The shape of the first step Riesz representer $\alpha _{01}$ has a rather complex form. Indeed, the linearization with respect to the first step
effect is also complex (c.f. equation~\eqref{eq:D1_defi}).
The first term corresponds to the linearization of the \textit{direct}
effect of $g$. It is given by $D_{dir}$, the linearization of $d\bar{m}%
(g_{\tau },h_{0})/\tau $. The second term corresponds to the \textit{%
indirect} effect. Consistent estimation of the second term generally
requires estimators for (i) $g_{0}$, (ii) $h_{0}$, (iii) $\partial
h_{0}/\partial v$, (iv) $\alpha _{02}$, and (v) $\partial \alpha
_{02}/\partial v$. Section~\ref{sec:auto_1step} provides the details on how
to estimate $\mathbb{E}[D_{01}(W,g)]$. We also note that some simplifications
and variations on the expression for $\mathbb{E}[D_{01}(W,g)]$ and for $\phi_{1}$ occur under different 
scenarios.

\begin{rem}[Relation to Hahn and Ridder (2013)]\label{rem:HR}
Theorem~5 in \cite{hahn2013asymptotic} studies a three-step generated-regressor problem with
$V=g_0(Z)$ (i.e.\ $\varphi(d,z,g)=g(z)$), $\Delta_1=L_2(Z)$, $\Delta_2=L_2(X,V)$, and third-step moment
\[
m(w,g,h,\theta)=\eta\!\left(w,\,h(x,g(z))\right)-\theta,
\]
so that $(g,h)$ enters through the scalar $h(x,g(z))$.

In our notation, their second-step derivative corresponds to $D_{02}$. If $\eta$ is differentiable in its second argument,
\[
D_{02}(w,\delta)
=
\frac{\partial \eta}{\partial y}\!\left(w,h_0(x,v)\right)\delta(x,v),
\qquad v=g_0(z),
\]
with Riesz representer
\[
\alpha_{02}(x,v)
=
\E\!\left[\left.\frac{\partial \eta}{\partial y}\!\left(W,h_0(X,V)\right)\right|X=x,V=v\right].
\]
Let $r_{02}(w)=\partial\eta/\partial y\bigl(w,h_0(x,v)\bigr)$, so that
$\alpha_{02}(x,v)=\E[r_{02}(W)\mid X=x,V=v]$.

In this evaluation-functional setting the first-step derivative is separable,
\[
D_{01}(w,g)=r_{01}(w)\,g(z),
\]
where
\[
r_{01}(w)
=
\bigl(y-h_0(x,v)\bigr)\,\partial_v\alpha_{02}(x,v)
+
\bigl(r_{02}(w)-\alpha_{02}(x,v)\bigr)\,\partial_v h_0(x,v).
\]
Since $\Delta_1=L_2(Z)$, the first-step Riesz representer is
\[
\alpha_{01}(z)=\E[r_{01}(W)\mid Z=z].
\]
As noted by \citet[Remark~3]{hahn2013asymptotic}, if $\alpha_{02}=0$ then
$r_{01}(w)=r_{02}(w)\,\partial_v h_0(x,v)$, illustrating a special case of downstream local robustness.

Relative to \cite{hahn2013asymptotic}, our results:  
(i) allow general generated regressors $V=\varphi(D,Z,g_0)$ for Hadamard differentiable $\varphi$;  
(ii) allow general Hadamard differentiable functionals $\bar m(g,h)=\E[m(W,g,h,\theta_0)]$, not only evaluation functionals; and  
(iii) allow more general spaces $\Delta_1$ and $\Delta_2(g)$ (e.g.\ sparse or sieve structures).  
Beyond influence functions, we provide an automatic Riesz-representer implementation suitable for high-dimensional/non-Donsker ML estimators and establish asymptotic normality of the resulting debiased estimators.
\end{rem}
\begin{rem}[Other First Steps]
\label{OtherFS} 
There are examples of first steps that are not included in (\ref{orth1}), such as some parametric estimators, functions
identified by orthogonality conditions with instruments, where $g$ depends on
other variables different from $Z$, or the control function approach of %
\citet{imbens2009identification}, among others. Nevertheless, much of our analysis is still useful for these other cases. In particular, the expression for $D_{01}(w,g)$ remains the same, and our results can be readily extended to other first steps by characterizing the corresponding first step IF $\phi_{1}$ such that 
\begin{equation*}
\frac{d}{d\tau }\mathbb{E}[D_{01}(W,g_{\tau })]=\int \phi _{1}(w,g_{0},\alpha _{01})dH(w).
\end{equation*}
We illustrate the application of this equation with parametric first steps in the next remark.
\end{rem}

\begin{rem}[General Parametric First Steps]
\label{Parfirststep} Suppose we replace our definition of $g_{0}$ and $\Delta_{1}$ in (\ref{orth1}) by a generic parametric fit $g_{0}(z)=G(z,\zeta_{0})$, where $G$ has a known functional form and $\zeta_{0}$ is an unknown finite-dimensional parameter in a parameter space $B\subseteq \mathbb{R}^{\func{dim}(\zeta)}$. We allow for $\zeta_{0}$ to be identified by parametric or semiparametric restrictions. For example, this setting includes semiparametric estimators (e.g., single-index models with $g_{0}(z)=z'\zeta_{0}$). Let $\hat{\zeta}$ be a regular estimator for $\zeta_{0} $ satisfying
\begin{equation*}
\sqrt{n}(\hat{\zeta}-\zeta_{0})=\frac{1}{\sqrt{n}}\sum_{i=1}^{n}\psi_{\zeta}
(D_{i},Z_{i},\xi_0)+o_{P}(1),
\end{equation*}%
where the IF $\psi_\zeta$ has zero mean and finite variance, and $\xi_0$ contains $\zeta_{0}$ and may contain additional nuisance parameters. We assume that the pathwise derivative of $\mathbb{E}[\psi_\zeta
(D_{i},Z_{i},\xi)]$ with respect to these additional nuisance parameters at $\xi_{0}$ is zero. Then, by standard arguments in regular estimation, see \cite{newey1994asymptotic}, and the previous remark, all our results apply in the parametric first step case with the adjustment term $\phi
_{1}=-\alpha_{01}'\psi_\zeta$, where 
\begin{equation*}
\alpha_{01}=\mathbb{E}[D_{01}(W,\dot{G})],
\end{equation*}%
and $\dot{G}(Z,\zeta_{0})=\partial G/\partial \zeta (Z,\zeta_{0})$. In particular, if $\psi_\zeta=0$, i.e., if $\sqrt{n}(\hat{\zeta}-\zeta_{0})=o_{P}(1),$ and other mild conditions are satisfied (to apply a Delta method), then there is no estimation effect from the generated regressors. This is typically the case when $\hat{\zeta}$ is constructed from a large sample (with a large sample size relative to $n$).    
\end{rem}

\begin{rem}[Simplifications]
\label{DRcase} If the original identifying moments are such that $\alpha
_{02}= 0,$ then the first step linearization simplifies to:
\begin{equation*}
D_{01}(w,g) = D_{dir}(w,g).
\end{equation*}%
Another simplification occurs under the index restriction $\mathbb{E}[S|D,Z]=\mathbb{E}%
[S|X,V]$, which implies:
\begin{equation*}
D_{01}(w,g) = D_{dir}(w,g)-\frac{\partial h_{0}}{\partial v%
}(x,v)\alpha _{02}(x,v)D_{\varphi }g.
\end{equation*}%
In both cases, the corresponding $\alpha _{01}\in \Delta _{1}$, given by  (\ref{eq:alpha1_defi}) in Appendix \ref{sec:Proofs}, simplifies accordingly.
\end{rem}
%%%%%%%%%%%%%%%%%%%%%%%%%%%%%%%%%%%%%%%%%%%%%%%%%%%%%%%%%%%%%%%%%%%%%%%%%%%%%%%%%%%%%%%%%%%%%%%%%%%%%%
\subsection{Cross-fitted automatic estimators}
\label{sec:est_auto}

The debiased sample moment functions are estimated using cross-fitting, where we partition the sample $(W_i)_{i=1}^n$ into $L$ groups $I_\ell$, for $\ell = 1, \dots, L$. Estimation of the debiased moment function $\psi$ for an observation $i \in I_\ell$ requires estimators of the Riesz representers $(\hat{\alpha}_{1\ell},\hat{\alpha}_{2\ell})$ based only on observations not in $I_\ell$. This
section is devoted to the construction of automatic estimators satisfying
this property. Through the section, we consider that the researcher has
at her disposal first and second step estimators, $\hat{g}_{\ell\ell^{\prime
}}$ and $\hat{h}_{\ell\ell^{\prime }}$, and a preliminary estimator $\tilde{%
\theta}_{\ell\ell^{\prime }}$, that use only observations not in $I_\ell
\cup I_{\ell^{\prime }}$; and estimators $(\hat{g}_{\ell\ell^{\prime
}\ell^{\prime \prime }}$, $\hat{h}_{\ell\ell^{\prime }\ell^{\prime \prime
}}, \tilde{\theta}_{\ell\ell^{\prime }\ell^{\prime \prime }})$ that use only
observations not in $I_\ell\cup I_{\ell^{\prime }} \cup I_{\ell^{\prime
\prime }}$. Depending on the application, some of the preliminary estimators may not be needed (see, e.g., the debiased estimator in the partially linear model).

Our approach to automatically estimate the Riesz representers relies on the orthogonality conditions discussed in Section~\ref{sec:ident_alphas}. We can combine the orthogonality conditions with the linearization results in Section~\ref{sec:Linearization} to obtain sets of moment conditions for the estimation of the Riesz representers. In particular, a combination of equation~\eqref{eq:auto_momentF} and Theorem~\ref{thm:first_step} gives
\begin{equation}
\mathbb{E}[D_{01}(W,\delta _{1})]=\mathbb{E}[-\alpha
_{01}(Z)r_{e}(Z)\delta _{1}(Z)],\text{ for each }\delta _{1}\in \Delta _{1},
\label{eq:Lin_projection1}
\end{equation}
where $D_{01}$ is the linearization of the identifying moment function $\bar{m}$ with respect to the first step and $r_e$ gives the linearization of the generalized error function $\epsilon(w,g)$ (cf. Assumption \ref{rho}). Varying $\delta_1$, the above equation provides a set of moment conditions that identify $\alpha_{01}$. Likewise, identification of the second-step Riesz representer $\alpha _{02}$ follows from equation~\eqref{eq:auto_momentS} and Proposition~\ref{prop:second_step}:
\begin{equation}
\mathbb{E}[D_{02}(W,\delta _{2})]=\mathbb{E}[\alpha
_{02}(X,V)\delta _{2}(D,Z)]=0,\text{ for each }\delta _{2}\in \Delta
_{2}(g_{0}),  \label{eq:Lin_projection2}
\end{equation}%
where $D_{02}$ is the linearization of the identifying moment function $\bar{m}$ with respect to the second step. The shape of the linearizations $D_{01}$, $D_{02}$, and $r_{e}$ may vary with the problem (see Section~\ref{sec:Examples} for some examples).

Equations \eqref{eq:Lin_projection1} and \eqref{eq:Lin_projection2} form the basis for automatic estimation of the Riesz representers. These require finding consistent estimators of the linearizations of the identifying moment functions and the generalized error. In
this section, we will write $D_{02}(w,h|g_{0},h_{0},\theta_0)$ to make
explicit that the linearization with respect to $h$ may depend on $(h_{0},g_{0},\theta_0)$. For the linearization of the effect of first-step estimation,
we will write $D_{01}(w,g|g_{0},h_{0},\alpha _{02},\theta_0 )$, to emphasize
that it may also depend on the second-step Riesz representer. $D_{01}$
generally also depends on the derivatives $\partial h_{0}/\partial v$ and $%
\partial \alpha _{02}/\partial v$. We do not make this explicit, but we will
address the issue in this section. We also write $r_e(z|g_0)$ to express that the generalized error is linearized at $g_0$.

%%%%%%%%%%%%%%%%%%%%%%%%%%%%%%%%%%%%%%%%%%%%%%%%%%%%%%%%%%%%%%%%%%%%%%%%%%%%%%%%%%%%%%%%%%%%%%%%%%%%%%
\subsubsection{Automatic estimation of the second-step Riesz Representer}

We start with the automatic estimator for $\alpha_{02}$. We assume that there is a dictionary $(b_{j})_{j=1}^{\infty
} $ whose closed linear span is $\Delta _{2}(g_{0})$. That is, any function
in $\Delta _{2}(g_{0})$ can be approximated, in the $L_{2}$ sense, by a
linear combination of the atoms. Thus, $\alpha _{02}$ can be approximated by
$\mathbf{b}_{J}^{\prime }\boldsymbol{\rho }_{0J}$, where $\mathbf{b}%
_{J}=(b_{1},...,b_{J})^{\prime }$ and $\boldsymbol{\rho }_{0J}=(\rho
_{01},...,\rho _{0J})^{\prime }$. We can now plug in $\mathbf{b}_{J}^{\prime
}\boldsymbol{\rho }_{0J}$ into equation~\eqref{eq:Lin_projection2} for $%
\delta _{2}=b_{j}$, $j=1,...,J$. This gives the following $J$ moment
conditions:
\begin{equation*}
\mathbb{E}[D_{02}(W,\mathbf{b}_{J})]=\mathbb{E}[\mathbf{b}_{J}(X,V)\mathbf{b}%
_{J}(X,V)^{\prime }]\boldsymbol{\rho }_{0J},
\end{equation*}%
where $D_{02}(w,\mathbf{b}_{J})\equiv
(D_{02}(w,b_{1}),...,D_{02}(w,b_{J}))^{\prime }$.

The above moment conditions can be used to construct an OLS-like estimator
of $\boldsymbol{\rho }_{0J}$. Note, however, that in high-dimensional
settings $\mathbb{E}[\mathbf{b}_{J}(X,V)\mathbf{b}_{J}(X,V)^{\prime }]$ may
be near singular. Therefore, we use the regularized estimator solving 
\begin{equation}
\min_{\boldsymbol{\rho }_{J}\in \mathbb{R}^{J}}\left\{ -2\mathbb{E}[D_{02}(W,%
\mathbf{b}_{J})^{\prime }]\boldsymbol{\rho }_{J}+\boldsymbol{\rho }%
_{J}^{\prime }\mathbb{E}[\mathbf{b}_{J}(X,V)\mathbf{b}_{J}(X,V)^{\prime }]%
\boldsymbol{\rho }_{J}+\lambda \lVert \boldsymbol{\rho }_{J}\rVert
_{q}^{q}\right\},  \label{eq:Auto2_objective}
\end{equation}%
where $\lVert \boldsymbol{\rho }_{J}\rVert _{q}\equiv
(\sum_{j=1}^{J}|\rho _{j}|^{q})^{1/q}$ for $q\geq 1$ and $\lambda \geq 0$ is a tuning
parameter. For $q=1$, the above is the Lasso objective function, while $q=2$
corresponds to Ridge Regression. Additionally, we could consider elastic-net-type penalties, where $\lambda (\xi \lVert \boldsymbol{\rho }_{J}\rVert
_{2}^{2}+(1-\xi )\lVert \boldsymbol{\rho }_{J}\rVert _{1})$, for $\xi \in
\lbrack 0,1]$, replaces the $L_q$ penalization.

For a given $\ell\in{1,...,L}$, the automatic estimator $\hat{\alpha}_{2\ell}$ is based on the sample version of the objective function in equation~\eqref{eq:Auto2_objective}. We estimate $\mathbb{E}[D_{02}(W,\mathbf{b}_{J})]$ by
\begin{equation*}
\hat{D}_{2\ell }\equiv \frac{1}{n-n_{\ell }}\sum_{\ell ^{\prime }\neq \ell
}\sum_{i\in I_{\ell ^{\prime }}}D_{02}(W_{i},\mathbf{b}_{J}|\hat{g}_{\ell
\ell ^{\prime }},\hat{h}_{\ell \ell ^{\prime }},\tilde{\theta}_{\ell \ell
^{\prime }}),
\end{equation*}%
where $n_{\ell }$ is the number of observations in $I_{\ell }.$ In turn, $%
\mathbb{E}[\mathbf{b}_{J}(X,V)\mathbf{b}_{J}(X,V)^{\prime }]$ is estimated
by
\begin{equation*}
\hat{B}_{\ell }\equiv \frac{1}{n-n_{\ell }}\sum_{\ell ^{\prime }\neq \ell
}\sum_{i\in I_{\ell ^{\prime }}}\mathbf{b}_{J}(X_{i},\varphi (D_{i},Z_{i},%
\hat{g}_{\ell \ell ^{\prime }}))\mathbf{b}_{J}(X_{i},\varphi (D_{i},Z_{i},%
\hat{g}_{\ell \ell ^{\prime }}))^{\prime }.
\end{equation*}

With this, we can build an automatic estimator of the second-step Riesz representer that only uses observations not in $I_{\ell }$. It is given by 
\begin{equation}\label{eq:min_rho}
\hat{\alpha}_{2\ell}=\mathbf{b}_J'\widehat{\boldsymbol{\rho}}_{J\ell},\quad 
\widehat{\boldsymbol{\rho}}_{J\ell}=\operatornamewithlimits{argmin}_{\boldsymbol{\rho}_J\in\mathbb{R}^J}
\left\{-2\hat{D}_{2\ell}'\boldsymbol{\rho}_J+\boldsymbol{\rho}_J'\hat{B}_\ell\boldsymbol{\rho}_J
+\lambda\|\boldsymbol{\rho}_J\|_q^q\right\}.
\end{equation}

The tuning parameter $\lambda $ can be chosen by cross-validation.

\subsubsection{Automatic estimation of the first-step Riesz representer}
\label{sec:auto_1step}

We also assume that there is a dictionary $(c_{k})_{k=1}^{\infty }$ that spans $%
\Delta _{1}$. This means that $%
\alpha _{01}$ can be approximated by $\mathbf{c}_{K}^{\prime }\boldsymbol{%
\beta }_{0K}$, where $\mathbf{c}_{K}=(c_{1},...,c_{K})^{\prime }$ and $%
\boldsymbol{\beta }_{0K}=(\beta _{01},...,\beta _{0K})^{\prime }$. We can
now plug in $\mathbf{c}_{K}^{\prime }\boldsymbol{\beta }_{0K}$ into equation~%
\eqref{eq:Lin_projection1} for $\delta _{1}=c_{k}$, $k=1,...,K$. This gives
the following $K$ moment conditions:
\begin{equation*}
\mathbb{E}[D_{01}(W,\mathbf{c}_{K})]=-\mathbb{E}[r_e(Z)%
\mathbf{c}_{K}(Z)\mathbf{c}_{K}(Z)^{\prime }]\boldsymbol{\beta }_{0K},
\end{equation*}%
where $D_{01}(w,\mathbf{c}_{K})\equiv
(D_{01}(w,c_{1}),...,D_{01}(w,c_{K}))^{\prime }$. Recall that $r_e$ gives the derivative of the generalized error $\epsilon$ (see Assumption~\ref{rho}). 

We use these conditions as a basis to construct
the objective function to estimate $\boldsymbol{\beta }_{0K}$:
\begin{equation*}
\min_{\boldsymbol{\beta }_{K}\in \mathbb{R}^{K}}\left\{ -2\mathbb{E}[D_{01}(W,%
\mathbf{c}_{K})^{\prime }]\boldsymbol{\beta }_{K} + \boldsymbol{\beta }%
_{K}^{\prime }\mathbb{E}[-r_e(Z)\mathbf{c}_{K}(Z)\mathbf{c}%
_{K}(Z)^{\prime }]\boldsymbol{\beta }_{K}+\lambda \lVert \boldsymbol{\beta }%
_{K}\rVert _{q}^{q}\right\} , 
\end{equation*}%
where the tuning parameter $\lambda $ may be different from that of the
second step. The automatic estimator for the first-step Riesz representer is built with the sample version of the above equation.  The estimator is given by
\begin{equation}\label{eq:min_beta}
\hat{\alpha}_{1\ell}=\mathbf{c}_K'\widehat{\boldsymbol{\beta}}_{K\ell},\quad
\widehat{\boldsymbol{\beta}}_{K\ell}=\operatornamewithlimits{argmin}_{\boldsymbol{\beta}_K\in\mathbb{R}^K}
\left\{-2\hat{D}_{1\ell}'\boldsymbol{\beta}_K+\boldsymbol{\beta}_K'\hat{C}_\ell\boldsymbol{\beta}_K
+\lambda\|\boldsymbol{\beta}_K\|_q^q\right\}.
\end{equation}
with
\begin{equation*}
\hat{C}_{\ell }\equiv \frac{-1}{n-n_{\ell }}\sum_{\ell ^{\prime }\neq \ell
}\sum_{i\in I_{\ell ^{\prime }}}r_e(Z_{i}|\hat{g}_{\ell \ell ^{\prime
}})\mathbf{c}_{K}(Z_{i})\mathbf{c}_{K}(Z_{i})^{\prime },
\end{equation*}%
and
\begin{equation*}
\hat{D}_{1\ell }\equiv \frac{1}{n-n_{\ell }}\sum_{\ell ^{\prime }\neq \ell
}\sum_{i\in I_{\ell ^{\prime }}}D_{01}(W_{i},\mathbf{c}_{K}|\hat{g}_{\ell
\ell ^{\prime }},\hat{h}_{\ell \ell ^{\prime }},\hat{\alpha}_{2\ell \ell
^{\prime }},\tilde{\theta}_{\ell \ell ^{\prime }}).  
\end{equation*}%

Note that the estimator for the first-step linearization $\hat{D}_{1\ell}$ may require estimators of the second-step Riesz representer $\hat{\alpha}_{2\ell\ell'}$ that do not include observations in $I_\ell \cup I_{\ell'}$. These estimators can be obtained using the
methodology of the previous section. To construct $\hat{\alpha}_{2\ell \ell
^{\prime }}=\mathbf{b}_{J}^{\prime }\widehat{\boldsymbol{\rho }}_{J\ell \ell
^{\prime }}$, we let  $\widehat{%
\boldsymbol{\rho }}_{J\ell \ell ^{\prime }}$ solve the optimization problem in equation~\eqref{eq:min_rho}, with $\hat{D}_{2\ell}$ and $\hat{%
B}_{\ell}$ replaced by
\begin{align*}
\hat{D}_{2 \ell\ell^{\prime }} &\equiv \frac{1}{n-n_{\ell}-n_{\ell^{\prime }}%
} \sum_{\ell^{\prime\prime}\notin \{\ell,\ell^{\prime }\}}\sum_{i\in
I_{\ell^{\prime\prime}}} D_{02}(W_i, \mathbf{b}_J | \hat{g}_{\ell\ell^{\prime
}\ell^{\prime \prime }}, \hat{h}_{\ell\ell^{\prime }\ell^{\prime \prime }},
\tilde{\theta}_{\ell\ell^{\prime }\ell^{\prime \prime }}) \text{ and } \\
\hat{B}_{\ell\ell^{\prime }} &\equiv \frac{1}{n-n_{\ell}-n_{\ell^{\prime }}}
\sum_{\ell^{\prime\prime}\notin \{\ell,\ell^{\prime }\}}\sum_{i\in
I_{\ell^{\prime\prime}}} \mathbf{b}_J(X_i,\varphi(D_i,Z_i,\hat{g}%
_{\ell\ell^{\prime }\ell^{\prime \prime }}))\mathbf{b}_J(X_i,\varphi(D_i,Z_i,%
\hat{g}_{\ell\ell^{\prime }\ell^{\prime \prime }}))^{\prime },
\end{align*}
respectively.

Furthermore, $D_{01}$ may depend on the derivatives $\partial h_{0}/\partial v$ and $%
\partial \alpha _{02}/\partial v$ (see equation~\eqref{eq:D1_defi}). We thus need to provide consistent estimators of these derivatives to build $\hat{D}_{1\ell }$. It is straightforward to construct
an estimator $\partial\hat{\alpha}_{2\ell\ell^{\prime }}/\partial v$ of the
derivative of $\alpha_{02}$ based on the cross-fitted Lasso estimator. Since we have already estimated $\hat{\alpha}%
_{2\ell\ell}= \mathbf{b}_J^{\prime }\widehat{\boldsymbol{\rho}}%
_{J\ell\ell^{\prime }}$, if each $b_j$ is differentiable w.r.t. $v$, we have
that $\partial\hat{\alpha}_{2\ell\ell^{\prime }}/\partial v\equiv (\partial%
\mathbf{b}_J/\partial v)^{\prime }\widehat{\boldsymbol{\rho}}_{J\ell\ell^{\prime
}}$.

Estimation of $\partial h_{0}/\partial v$ may be trickier. It will depend
on the shape of the estimator $\hat{h}_{\ell \ell }$. Note that, since $%
h_{0}\in \Delta _{2}(g_{0})$, we may use the dictionary $(b_{j})_{j=1}^{%
\infty }$ to approximate the parameter. In this case, $\hat{h}_{\ell \ell }$
will be a Lasso or Ridge Regression estimator and we can estimate the
derivative of $h_{0}$ as we have estimated the derivative of $\alpha _{02}$.
Moreover, if estimating $h_{0}$ involves a nonparametric regression problem with a low-dimensional covariate, we can often take $\hat{h}_{\ell \ell }$ as a Kernel or a Local
Linear Regression estimator, as in \cite{heckman1998matching}. Then, the derivatives of $h_{0}$ can be
estimated by finding the analytical expression of the derivatives of the
kernel function.

For a general ML estimator $\hat{h}_{\ell\ell^{\prime }}$ (e.g., Random
Forest), we propose a numerical derivative approach to estimate $\partial
h_0/\partial v$. Let $t_n$ be a tuning parameter depending on the sample
size with $t_n \downarrow 0 $. We propose to estimate $\partial
h_0(x,v)/\partial v$ by
\begin{equation}
\frac{\partial \hat{h}_{\ell\ell^{\prime }}}{\partial v}(x,v)\equiv \frac{%
\hat{h}_{\ell\ell^{\prime }}(x,v+t_n)-\hat{h}_{\ell\ell^{\prime }}(x,v)}{t_n}%
.  \label{eq:deriv_h_est}
\end{equation}
This approach has been used and justified theoretically in \cite{bravo2020two} in a two-step setting. Note that, usually, we need to compute the derivative evaluated at $(X_i,
\varphi(D_i,Z_i, \hat{g}_{\ell\ell^{\prime }}))$. Alternative ML estimators that achieve optimal rates for partial derivatives are discussed in \cite{dai2016optimal}.

%%%%%%%%%%%%%%%%%%%%%%%%%%%%%%%%%%%%%%%%%%%%%%%%%%%%%%%%%%%%%%%%%%%%%%%%%%%%%%%%%%%%%%%%%%%%%%%%%%%%%%
\section{Examples: Three-step debiased estimators}
\label{sec:Examples}

\subsection{Hd-PS regression adjustment}
\label{sec:ex_ATE_PLM}
The debiased three-step estimator is
\begin{equation*}
    \hat{\theta} = \frac{\sum_{\ell =1}^{L}\sum_{i\in I_{\ell }} \left[
\left( Y_i - \hat{h}_\ell(\hat{V}_{i\ell})\right)\left( D_{i}-\hat{V}_{i\ell}\right) + \hat{\alpha}_{1\ell }(Z_{i})\cdot \left(D_{i}-\hat{V}_{i\ell}\right) \right]}{\sum_{\ell
=1}^{L}\sum_{i\in I_{\ell }}\left( D_{i}-\hat{V}_{i\ell}\right) ^{2}}, 
\end{equation*}
where $\hat{V}_{i\ell} = \Lambda(\hat{g}_\ell(Z_i))$ and $\Lambda$ is the Logistic function. Cross-fitted estimators $(\hat{g}_\ell, \hat{h}_\ell)$ are discussed in Example~\ref{ex:ATE_PLM1} (p.~\pageref{ex:ATE_PLM1}). In this section, we detail the estimation of the Riesz representer $\alpha_{01}$. As discussed in Section~\ref{sec:est_auto}, we propose to estimate the Riesz representer by $\hat{\alpha}_{1\ell }(z)=\mathbf{c}_{K}(z)^{\prime }\widehat{\boldsymbol{\beta }%
}_{K\ell }$ with $\widehat{\boldsymbol{\beta }}_{K\ell }$ solving \eqref{eq:min_beta}. We show how to construct $\hat{C}_\ell$ and $\hat{D}_{1\ell}$.

The term $\hat{C}_\ell$ depends on the linearization of the first-step generalized error $\epsilon(w, g) = d - \Lambda(g(z))$. Since
\begin{equation*}
    \difTau \E[\epsilon(W, g_\tau)] = \E\left[-\Lambda(g_0(Z))[1-\Lambda(g_0(Z))] \difTau g_\tau\right] \text{ and } r_e(Z)= -\Lambda(g_0(Z))[1-\Lambda(g_0(Z))].
\end{equation*}
Therefore,
\begin{equation*}
    \hat{C}_\ell =  \frac{1}{n-n_{\ell }}\sum_{\ell ^{\prime }\neq \ell
}\sum_{i\in I_{\ell ^{\prime }}}\hat{V}_{i\ell\ell'}(1-\hat{V}_{i\ell\ell'})\mathbf{c}_{K}(Z_{i})\mathbf{c}_{K}(Z_{i})^{\prime }, \text{ where } \hat{V}_{i\ell\ell'} \equiv \Lambda(\hat{g}_{\ell\ell'}(Z_i)).
\end{equation*}

To find $\hat{D}_{1\ell}$, note that, since $\alpha_{02}=0$, we have that $D_{01} = D_{dir}$ and
\begin{align*}
    \frac{\partial}{\partial\tau}\E[m(W, g_\tau, h_0, \theta_0)]=\frac{\partial}{\partial\tau}\E\left[ \{Y - h_0(\Lambda(g_\tau(Z))) -\theta_0 (D - \Lambda(g_\tau(Z)))\} \cdot (D - \Lambda(g_\tau(Z)))\right] \\
    = \E \left[ \left\{ -\left(Y - h_0(V) -\theta_0(D-V)\right) + \left( -\dot{h}_0(V) + \theta_0 \right) \cdot (D - V) \right\} \dot{\Lambda}(g_0(Z)) \frac{\partial g_\tau}{\partial\tau} \right]
    \\ = \E \left[  -\left(Y - h_0(V) \right) \dot{\Lambda}(g_0(Z)) \frac{\partial g_\tau}{\partial\tau} \right],
\end{align*}
where $\dot{h}_0 \equiv dh_0/dv$ and $\dot{\Lambda} \equiv d\Lambda/du = \Lambda \cdot (1 - \Lambda)$. From this representation and $\varepsilon=Y - h_0(V) -\theta_0(D-V)$, it follows that the effect of the first step is zero if $\E[\varepsilon|D,Z] = 0$, which we do not assume as it imposes strong restrictions on heterogeneity. To estimate $\alpha_{01}$, the linearization $D_{dir}$ is projected onto $\Delta_1 \subseteq L_2(Z)$. Therefore, as $\E[D|Z] = V$, the expression for $D_{dir}$ simplifies, and we consider the following estimator for the linearization of the first step:
\begin{equation*}
D_{01}(W_{i},\mathbf{c}_{K}|\hat{g}_{\ell \ell ^{\prime }},\hat{h}_{\ell \ell
^{\prime }})=-\left[Y_{i}-\hat{h}_{\ell \ell
^{\prime }}(\hat{V}_{i\ell \ell ^{\prime }}) \right] \hat{V}_{i\ell \ell
^{\prime }} \left( 1- \hat{V}_{i\ell \ell
^{\prime }} \right) \mathbf{c}_{K}(Z_i).
\end{equation*}
Note that, in this case, the linearization $D_{01}$ does not depend on $\theta_0$ and $\alpha_{02}$. Hence, no additional estimators $\tilde{\theta}_{\ell\ell'}$ and $\hat{\alpha}_{2\ell\ell'}$ are needed.

For the Hd-PS regression adjustment estimator, the estimation of the first-step Riesz representer can be reframed as a weighted Lasso regression, which is defined in equation~\eqref{eq:alpha1ell_riesz_lasso}. 

\subsection{Partially linear model with generated regressors: Autoencoders}
\label{sec:ex_PLMauto}

The partially linear model is a workhorse for debiased machine learning methods, see \cite{ahrens2025introduction} and references therein. Here we propose a debiased estimator for the partially linear model that is robust to ML-generated regressors. We first consider a general $V=\varphi(D,Z,g_0)$, where $g_0$ is identified by \eqref{orth1}. Then, we illustrate the framework with learned confounders via autoencoders.

Suppose $\dim(D)=p$. For the partially linear model with generated regressors, introduce the second-step nuisances $h_{0S}(v)=\E[S\mid V=v]$ for $S=Y$ or $S=D_j$, for $j=1,\dots,p$. Let $h_{0D} \equiv (h_{0D_1}, \dots, h_{0D_p})'$. The DML-type identifying moment is
\begin{equation*}
m(W,g_{0},h_{0},\theta _{0})=( Y-h_{0Y}(V)-\theta
_{0}' ( D-h_{0D}(V))) \cdot (D - h_{0D}(V)),
\end{equation*}%
where $h_0\equiv(h_{0Y},h_{0D})$. We assume $\kappa_0\in\Delta_2(g_0)$ so this moment identifies $\theta_0$ for the relevant second-step space. In this example, $\alpha_{02}=0$, but $\alpha_{01}$ is generally nonzero (cf.\ Section~\ref{sec:ex_ATE_PLM}), so standard DML inference that ignores generated regressors is not generally valid.

We therefore use a debiased three-step estimator that is robust to the first step.  The estimator solves
\begin{equation*}
\hat{\theta}=\operatornamewithlimits{argmin}_{\theta\in\Theta}\hat{\psi}%
(\theta)^{\prime}\hat{\Upsilon}\hat{\psi}(\theta),
\end{equation*}
with $\hat{\psi}(\theta) = n^{-1} \sum_{\ell=1}^L\sum_{i\in I_\ell} \hat{\psi}_{i\ell}(\theta)$, weighting matrix $\hat\Upsilon$, and debiased moments
\begin{equation}\label{eq:PLM_moments}
\hat{\psi}_{i\ell}(\theta)=
\left(Y_i-\hat h_{\ell, Y}(\hat V_{i\ell})-\theta'(D_i-\hat h_{\ell, D}(\hat V_{i\ell}))\right)
(D_i-\hat h_{\ell, D}(\hat V_{i\ell}))
+\widehat{\boldsymbol{\alpha}}_{1\ell}(Z_i)\,\epsilon(W_i,\hat g_\ell),
\end{equation}
where $\widehat{\boldsymbol{\alpha}}_{1\ell}(z)$ is the $p$-vector of automatic first-step Riesz-representer estimators (one per moment component). Each component is constructed as in Section~\ref{sec:auto_1step}:
$\hat\alpha_{1j\ell}(z)=\mathbf c_K(z)'\widehat{\boldsymbol\beta}_{Kj\ell}$, with
$\widehat{\boldsymbol\beta}_{Kj\ell}$ solving \eqref{eq:min_beta}. The construction of $\hat C_\ell$ depends on $\epsilon(w,g)$. For example, in a control-function setup with $V=D-g_0(Z)$ one has $\epsilon(w,g)=d-g(z)$ and $r_e(z)=-1$.

Construction of $\hat D_{1j\ell}$, for each $j=1,\dots, p$, parallels Section~\ref{sec:ex_ATE_PLM}. Since $\alpha_{02}=0$, the indirect effect is zero and only the direct effect remains. The linearization of the $j$-th moment condition is
\begin{equation} \label{eq:PLM_linearization}
\begin{aligned}
        D_{01j}(W_i, g) = r_{dir}(W_i)D_\varphi g, \text{ with } r_{dir}(W_i) = -\dot h_{0D_j}(V_i) \cdot\varepsilon+\dot\varepsilon \cdot [D_{ji}-h_{0D_j}(V_i)], \\
\varepsilon=Y_i-h_{0Y}(V_i)-\theta_0'(D_i-h_{0D}(V_i)) \text{ and } \dot\varepsilon \equiv \frac{\partial\varepsilon}{\partial v} = \dot{h}_{0Y}(V_i) + \theta_0'\dot{h}_{0D}(V_i),
\end{aligned}
\end{equation}
where $\dot{h}_{0S} \equiv dh_{0S}/dv$. To build $D_{01j}(W_i, \mathbf{c}_K|\hat{g}_{\ell\ell'}, \hat{h}_{\ell\ell'}, \tilde{\theta}_{\ell\ell'})$, we replace these terms in equation~\eqref{eq:PLM_linearization}: (i) $\varepsilon$ by $\hat\varepsilon_{i\ell\ell'}=Y_i-\hat h_{\ell\ell',Y}(\hat V_{i\ell\ell'})-\tilde\theta_{\ell\ell'}'(D_i-\hat h_{\ell\ell',D}(\hat V_{i\ell\ell'}))$, (ii) $\dot\varepsilon$ by $\dot\varepsilon_{i\ell\ell'}=-\dot h_{\ell\ell',Y}(\hat V_{i\ell\ell'})+\tilde\theta_{\ell\ell'}'\dot h_{\ell\ell',D}(\hat V_{i\ell\ell'})$, (iii) $\dot{h}_{0S}$ by the corresponding cross-fitted estimators $\dot h_{\ell\ell',S}$, and (iv) $D_\varphi\mathbf{c}_K$ by a cross-fitted estimator $\hat D_{\varphi i\ell\ell'}\mathbf{c}_K$ of the linearization of $\varphi$ w.r.t. $g$; e.g., for $\varphi(d,z,g)=\Lambda(g(z))$, $\hat D_{\varphi i\ell\ell'}\mathbf{c}_K=\Lambda(\hat g_{\ell\ell'}(Z_i))(1-\Lambda(\hat g_{\ell\ell'}(Z_i)))\mathbf{c}_K(Z_i)$, while for $\varphi(d,z,g)=d-g(z)$, $\hat D_{\varphi i\ell\ell'}=-\mathbf{c}_K(Z_i)$.

In general, $\alpha_{01}$ is nonzero. It is the orthogonal projection onto $\Delta_1$ of
$D_\varphi^*r_{dir}$, where $D_\varphi^*$ is the adjoint of $D_\varphi$. Even if $\E[\varepsilon\mid D,Z]=0$, $\alpha_{01}$ typically remains nonzero, so inference that ignores generated regressors is invalid. For $V=g_0(Z)$, $\Delta_1=L_2(Z)$ and $\Delta_2=L_2(V)$, these influence function calculations are covered by \cite{hahn2013asymptotic}.

\medskip
\noindent\textbf{Learned confounders via autoencoders.}

\begin{figure}[htbp]
  \centering
  \includegraphics[width=0.65\linewidth]{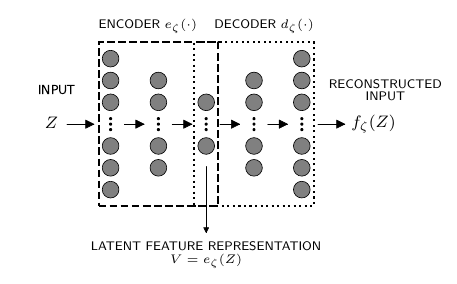}
  \caption{Autoencoder architecture.}
  \label{fig:autoencoders}
\end{figure}

An autoencoder consists of an encoder $e_0(\cdot)$, a low-dimensional representation $V=e_0(Z)$ (our generated regressor), and a decoder $d_0(\cdot)$, identified by
\begin{equation*}
(e_{0},d_{0})=\operatornamewithlimits{argmin}_{e\in \mathcal{E},d\in \mathcal{D}}\mathbb{E}[\mathcal{L}(Z,d(e(Z)))],
\end{equation*}%
where $\mathcal{L}$ is a loss and $\mathcal E$ and $\mathcal D$ are function classes (see Figure~\ref{fig:autoencoders}). For concreteness, we take $\mathcal{L}(Z,f)=|Z-f|^2$ and feed-forward neural networks indexed by $\zeta\in\mathbb{R}^{\dim(\zeta)}$, so that $e_0(Z)=e_{\zeta_0}(Z)$ and $d_0(V)=d_{\zeta_0}(V)$ for some minimizer $\zeta_0$. A key feature of autoencoders is the bottleneck $\dim(V)\ll\dim(Z)$, which yields nonlinear dimension reduction \citep{bengio2013representation}.

To write this example in our setting, define $g_0=(d_0,e_0)$ and $V=\varphi(D,Z,g_0)=e_0(Z)$, and assume w.l.o.g.\ $\dim(V)=1$. The generalized error is $\epsilon(W,g)=Z-d(e(Z))$, $g=(d,e)$, and the first-step identifying condition is
\begin{equation*}
\E[\delta_1(Z)\epsilon(W,g_0)]=0\quad\text{for all }\delta_1\in\Delta_1,
\end{equation*}%
where $\Delta_1$ is the linear span generated by the columns of the Jacobian
\begin{equation*}
\dot f_0(Z)=\left.\frac{\partial f_\zeta}{\partial\zeta}(Z)\right|_{\zeta=\zeta_0},
\text{ with } f_\zeta(Z)=d_\zeta(e_\zeta(Z)).
\end{equation*}%
Define also $\mathbf e_K \equiv \left.\partial e_\zeta/\partial\zeta\right|_{\zeta=\zeta_0}$, which shows up in the linearization of the moment condition w.r.t. $g$ (the direct effect).
The implementation follows the generic construction with $\mathbf c_K(Z)=\dot f_0(Z)$, $K=\dim(\zeta)$, and objective
\begin{equation}\label{AutoObj}
\min_{\boldsymbol{\beta }_{K}\in \mathbb{R}^{K}}\left\{ -2\E[D_{01j}(w,\mathbf{e}_{K})']\boldsymbol{\beta }_{K}+\boldsymbol{\beta }%
_{K}^{\prime }\E[\mathbf{c}_{K}(Z)\mathbf{c}_{K}^{\prime }(Z)]%
\boldsymbol{\beta }_{K}+\lambda \lVert \boldsymbol{\beta }_{K}\rVert
_{q}^{q}\right\},
\end{equation}%
with $D_{01j}$ given by equation~\eqref{eq:PLM_linearization} with $D_\varphi\mathbf{e}_K = \mathbf{e}_K$.

Since $\zeta_0$ is unknown, we use $\hat\zeta_{\ell\ell'}$, which is estimated without observations in $I_\ell\cup I_{\ell'}$, and compute Jacobians by backpropagation:
\begin{equation*}
\hat{\mathbf c}_{K,\ell\ell'}=\left.\frac{\partial f_\zeta}{\partial\zeta}\right|_{\zeta=\hat\zeta_{\ell\ell'}}' 
 \text{ and }
\hat{\mathbf e}_{K,\ell\ell'}=\left.\frac{\partial e_\zeta}{\partial\zeta}\right|_{\zeta=\hat\zeta_{\ell\ell'}}.
\end{equation*}%
Then,
\begin{align*}
    \hat{C}_{\ell }&=\frac{1}{n-n_{\ell }}\sum_{\ell ^{\prime }\neq \ell
}\sum_{i\in I_{\ell ^{\prime }}}\hat{\mathbf c}_{K,\ell\ell'}(Z_i)\hat{\mathbf c}_{K,\ell\ell'}(Z_i)^{\prime} \text{ and} \\
\hat{D}_{1j\ell }&=\frac{1}{n-n_{\ell }}\sum_{\ell ^{\prime }\neq \ell
}\sum_{i\in I_{\ell ^{\prime }}}D_{01j}(W_{i},\hat{\mathbf e}_{K,\ell\ell'}\mid\hat g_{\ell\ell'},\hat h_{\ell\ell'},\tilde\theta_{\ell\ell'}),
\end{align*}
with $\hat V_{i\ell\ell'}=e_{\hat\zeta_{\ell\ell'}}(Z_i)$. Solving the sample analog of \eqref{AutoObj} yields $\widehat{\boldsymbol\beta}_K$ and $\widehat{\boldsymbol\alpha}_{1\ell}(Z)=\hat{\mathbf c}_{K,\ell\ell'}(Z)'\widehat{\boldsymbol\beta}_K$. The three-step debiased estimator uses the moment function in \eqref{eq:PLM_moments}, with $\epsilon(W_i, \hat{g}_\ell) = Z_i - f_{\hat{\zeta}_\ell}(Z_i)$ and $\hat{\zeta}_\ell$ estimated without observations in $I_\ell$.

%%%%%%%%%%%%%%%%%%%%%%%%%%%%%%%%%%%%%%%%%%%%%%%%%%%%%%%%%%%%%%%%%%%%%%%%%%%%%%%%%%%%%%%%%%%%%%%%%%%%%%
\subsection{CASF in a non-separable model}
\label{sec:CASF_details}

The three-step debiased estimator of the CASF is given in equation~\eqref{eq:CASF_D}. We provide the ingredients to build the estimators $\hat{\alpha}_{1\ell}$ and $\hat{\alpha}_{2\ell}$. Recall that the moment function defining the CASF is
\begin{equation*}
m(w,g,h,\theta )=\int h(x^{\ast },\varphi (d,z,g))dF^{\ast }(x^{\ast
})-\theta .
\end{equation*}%

This moment is already linear in $h$ and hence
\begin{equation*}
D_{02}(w,b_{j})=\int b_{j}(x^{\ast }, \varphi(d,z,g_{0}))dF^{\ast }(x^{\ast }),
\end{equation*}%
for each atom $b_{j}$ in the dictionary. We follow the same strategy as before and approximate $D_{02}$ by Monte Carlo
integration. Let $(X_{s}^{\ast })_{s=1}^{S}$ be a sample drawn from $F^{\ast
}$. To construct the objective function to estimate $\widehat{\boldsymbol{\rho }}%
_{J\ell }$, for an observation $i\in I_{\ell ^{\prime }}$, we set
\begin{equation*}
D_{02}(W_{i},b_{j}|\hat{g}_{\ell \ell ^{\prime }})=\frac{1}{S}%
\sum_{s=1}^{S}b_{j}(X_{s}^{\ast },D_{i}-\hat{g}_{\ell \ell ^{\prime
}}(Z_{i})),
\end{equation*}%
for each $j=1,\dots ,J$. Here we emphasize that the linearization does not depend on $h_0$ and $\theta_0$, it only depends on $g_0$. With this we construct $\hat{\alpha}_{2\ell }=%
\mathbf{b}_{J}^{\prime }\widehat{\boldsymbol{\rho }}_{J\ell }$ following (\ref%
{eq:min_rho}).

It is straightforward to show that the linearization of the moment condition w.r.t. $g$ is $D_{dir}(w,g)=r_{dir}(w)g(z)$, with
\begin{equation*}
r_{dir}(w)= -\int \frac{\partial h_{0}}{\partial v}(x^{\ast
},d-g_{0}(z))dF^{\ast }(x^{\ast }).
\end{equation*}
 We can now plug in the expression for $D_{dir}$ into equation~\eqref{eq:D1_defi}, where the linearization of the first step effect is defined. Recall that $D_{\varphi }g=-g$. Then, for the CASF, equation~\eqref{eq:D1_defi} becomes
\begin{equation*}
D_{01}(w,g)\equiv \left\{ r_{dir}(w)+\frac{\partial h_{0}}{\partial v}%
(x,v)\alpha _{02}(x,v)\right\} g(z).
\end{equation*}
The linearization depends on $h_{0}$ and $\alpha _{02}$,
and the derivative of $h_{0}$ w.r.t. $v$. It also depends on $g_{0}$, as $v\equiv d-g_{0}(z)$. However, it does not depend on $\theta_0$. Note that $\E[\partial\alpha_{02}/\partial v \cdot (Y - h_0)]=0$ by the control-function assumption.

We approximate $r_{dir}(W_i)$, with $i\in I_{\ell ^{\prime }}$, by
\begin{equation*}
-\frac{1}{S}\sum_{s=1}^{S}\frac{\partial \hat{h}_{\ell \ell ^{\prime }}}{%
\partial v}(X_{s}^{\ast },D_{i}-\hat{g}_{\ell \ell ^{\prime }}(Z_{i})),
\end{equation*}%
where, ${\partial \hat{h}_{\ell \ell ^{\prime }}}/{\partial v}%
=(\partial \mathbf{b}_{J}/\partial v)^{\prime }\widehat{\boldsymbol{\eta}}_{\ell
\ell ^{\prime }}$. The parameters $\boldsymbol{\hat{\eta}}_{\ell \ell
^{\prime }}$ are Lasso cross-fitted slope estimates for the second step $%
h_{0}$. To estimate $D_{1\ell }$, it remains to show how to estimate $\alpha_{02} \cdot \partial h_0/\partial v$ for an observation $i\in I_{\ell ^{\prime }}$. Being $%
V_{i\ell \ell ^{\prime }}\equiv D_{i}-\hat{g}_{\ell \ell ^{\prime }}(Z_{i})$, we can estimate it by
\begin{equation*}
\mathbf{b}_{J}(X_{i},\hat{V}_{i\ell \ell ^{\prime }})^{\prime }\widehat{%
\boldsymbol{\rho }}_{J\ell \ell ^{\prime }}\cdot \frac{\partial \hat{h}%
_{\ell \ell ^{\prime }}}{\partial v}(X_{i},\hat{V}_{i\ell \ell ^{\prime }}).
\end{equation*}%
Therefore, we have that, for $i\in I_{\ell ^{\prime }}$,
\begin{align*}
D_{01}(W_{i},c_{k}|\hat{g}_{\ell \ell ^{\prime }},\hat{h}_{\ell \ell ^{\prime
}},\hat{\alpha}_{2\ell \ell ^{\prime }})=c_{k}(Z_{i})\cdot & \left\{ -\frac{1}{S}\sum_{s=1}^{S}\frac{\partial
\hat{h}_{\ell \ell ^{\prime }}}{\partial v}(X_{s}^{\ast },\hat{V}_{i\ell \ell
^{\prime }}) +\mathbf{b}_{J}(X_{i},\hat{V}_{i\ell \ell ^{\prime }})^{\prime }%
\widehat{\boldsymbol{\rho }}_{J\ell \ell ^{\prime }}\cdot \frac{\partial \hat{h}%
_{\ell \ell ^{\prime }}}{\partial v}(X_{i},\hat{V}_{i\ell \ell ^{\prime
}})\right\} ,
\end{align*}%
for each $k=1,...,K$. Finally, note that $\epsilon(w,g)=d-g(z)$ and, hence, $r_e(z) = -1$. These results can then be used to construct the objective
function to estimate $\widehat{\boldsymbol{\beta }}_{K\ell }$ in (\ref%
{eq:min_beta}) and then $\hat{\alpha}_{1\ell }=\mathbf{c}_{K}^{\prime }\widehat{\boldsymbol{\beta }}_{K\ell }$.

\section{Asymptotic theory}
\label{sec:asymptotic}

%%%%%%%%%%%%%%%%%%%%%%%%%%%%%%%%%%%%%%%%%%%%%%%%%%%%%%%%%%%%%%%%%%%%%%%%%%%%%%%%%%%%%%%%%%%%%%%%%%%%%%
\subsection{General results}

This section gives general conditions for asymptotic normality of the
automatic debiased GMM and conditions for consistent estimation of its
asymptotic variance. The conditions are based on the mean-square
consistency, small interaction of estimation biases, and locally robust
conditions. These asymptotic results generalize \cite{chernozhukov2022locally} to our three-step setting with generated regressors. Estimation rates for the
Riesz representers $(\alpha_{01}, \alpha_{02})$ require (i) that the
dictionaries approximate well the Riesz representers and (ii) being able to
estimate the linear approximations of $\bar{m}(g,h)$ given by $%
\E[D_{01}(W,g)]$ and $\E[D_{02}(W,h)]$ at a certain rate %
\citep[see][]{chernozhukov2022automatic}.

In the presence of generated regressors, the theory needs to account for
the fact that the estimator of the correction term (and probably that of the
moment condition) evaluates the estimators $\hat{h}_\ell$ and $\hat{\alpha}%
_{2\ell}$ in the generated regressor $\hat{V}_{i\ell} \equiv \varphi(D_i,
Z_i, \hat{g}_\ell)$ (c.f., equation~\eqref{eq:dmon_i}). We modify the
expansion of $\hat{\psi}_{i\ell}(\theta_0)-\psi(W_i,
g_0,h_0,\alpha_{0},\theta_0)$ given by \cite{chernozhukov2022locally} to deal with this fact. After a first order expansion, which forms the basis of the local robustness property, remainders implying the generated regressor are of a particularly complex form. In the case of downstream local robustness, when $\alpha_{02} = 0$, the remainder simplifies. In any other cases, we rely on smoothness conditions on the dictionaries and $g \mapsto \varphi(D,Z, g)$ to bound the remainder (c.f. Assumption~\ref{ass:gen_reg}).

We begin with assumptions on the dictionaries. The first assumption formally
states that the dictionaries $(b_j)_{j=1}^\infty$ and $(c_k)_{k=1}^\infty$
span $\Delta_{2}(g_0)$ and $\Delta_1$, respectively.\footnote{%
In this section, for a measurable function $f$, $\lVert f \rVert_2 \equiv
\sqrt{\E[f(W)^2]}$ denotes its $L_2$-norm. Also, for a $m\times n$
matrix $A=(A_{i,j})_{i=1,j=1}^{m,n}$, $\lVert A\rVert_\infty \equiv
\max_{i,j} |A_{ij}|$.}
\begin{ass} \label{ass:span} \mbox{} \\[-20pt]
	\begin{enumerate}[label=\textbf{\alph*.},ref=\ref{ass:span}.\alph*]
		\item \label{ass:span:b} For every $j$, $b_j\in \Delta_{2}(g_0)$. Also, $\forall \delta_{2}\in \Delta_{2}(g_0)$ and for every $\varepsilon>0$, there exist $J$ and $\boldsymbol{\rho}_J$ such that $\lVert \delta_2 - \mathbf{b}_J'\boldsymbol{\rho}_J \rVert_2 < \varepsilon$.
		\item \label{ass:span:c} For every $k$, $c_k\in \Delta_{1}$. Also, $\forall \delta_{1}\in \Delta_{1}$ and for every $\varepsilon>0$, there exist $K$ and $\boldsymbol{\beta}_K$ such that $\lVert \delta_1 - \mathbf{c}_K'\boldsymbol{\beta}_K \rVert_2 < \varepsilon$.
	\end{enumerate}
\end{ass}

We also assume bounded dictionaries
\citep[see, for
instance,][]{newey1997convergence}:
\begin{ass} \label{ass:bounded_dicts}
	$\sup_{j\in\mathbb{N}} |b_j(X,V)|<\infty$ and $\sup_{k\in\mathbb{N}} |c_k(Z)|<\infty$.
\end{ass}
The assumption translates into consistency of $\hat{B}_\ell$ and $\hat{C}%
_\ell$. Also, on top of the following assumption, it will guarantee that the Riesz representers are bounded:
\begin{ass}
	\label{ass:abs_sumability} For the real-valued sequences $(\rho_{0j})_{j=1}^\infty$ and $(\beta_{0k})_{k=1}^\infty$  such that  $\alpha_{02}(x,v)=\sum_{j=1}^{\infty} \rho_{0j} b_j(x,v)$ and $\alpha_{01}(z)=\sum_{k=1}^{\infty} \beta_{0k} c_k(z)$:
	\begin{enumerate}[label=\textbf{\alph*.},ref=\ref{ass:abs_sumability}.\alph*]
		\item \label{ass:abs_sumability:main} $\sum_{j=1}^{\infty} |\rho_{0j}|<\infty$ and $\sum_{k=1}^{\infty} |\beta_{0k}|<\infty$.
		
		\item \label{ass:abs_sumability:largest_terms} For a $C>0$, the atoms $b_j$ and $c_k$ corresponding to the largest $C\sqrt{n}$ values of $\rho_{0j}$ and $\beta_{0k}$ are included in $\mathbf{b}_J$ and $\mathbf{c}_K$.
	\end{enumerate}
\end{ass}
This assumption keeps the $L_1$-norm of the coefficient of the Lasso
penalized regression under control. The result is relevant to estimate the
asymptotic variance \citep[see][]{chernozhukov2022automatic}. We also note that the absolute
summability of the coefficients imposes a sparsity condition on the relevant
terms to approximate $\alpha_{01}$ and $\alpha_{02}$ %
\citep[see][p.~985]{chernozhukov2022automatic}.

We require the following estimation rates:
\begin{ass} \label{ass:est_convergence} There is $1/3 < r < 1/2$ such that
\begin{enumerate}[label=\textbf{\alph*.},ref=\ref{ass:est_convergence}.\alph*]
	\item \label{ass:est_convergence:estimators} $\lVert \hat{g}_\ell-g_0 \rVert_2 = O_p(n^{-r})$ and $\lVert \hat{h}_\ell-h_0 \rVert_2 = O_p(n^{-r})$.
	
	\item \label{ass:est_convergence:linearizations}  $\norm{\hat{D}_{1\ell}-\mathbb{E}[D_{01}(W, \mathbf{c}_K)]}_\infty=O_p(n^{-r})$ and $\norm{\hat{D}_{2\ell}-\mathbb{E}[D_{02}(W, \mathbf{b}_J)]}_\infty=O_p(n^{-r})$.
\end{enumerate}
\end{ass}
This assumption imposes standard rate conditions on the estimators of the
nuisance parameters and on the linearization of the moment condition. For general results on rates with generated regressors see \cite{mammen2012nonparametric}; for Lasso rates, see \cite{bickel2009simultaneous,bunea2007sparsity,zhang2008sparsity}, and references
therein; for $L_2$-rates with boosting with high-dimensional regressors see \cite{kueck2023estimation}; for deep neural networks with a ReLU activation function,  see \cite{farrell2021deep}.  Under some regularity conditions on the linearizations %
\citep[see][Ass.~12]{chernozhukov2022automatic}, Assumption~\ref{ass:est_convergence:linearizations} can be derived
from the rate conditions on the estimators of the nuisance parameters.

%Moreover, the above assumption also holds when $(\hat{g}_\ell, \hat{h}_\ell, \tilde{\theta}_\ell)$ are replaced by $(\hat{g}_{\ell\ell'}, \hat{h}_{\ell\ell'}, \tilde{\theta}_{\ell\ell'})$ and $(\hat{g}_{\ell\ell'\ell''}, \hat{h}_{\ell\ell'\ell''}, \tilde{\theta}_{\ell\ell'\ell''})$.

We also ask for the following rates for the Lasso penalty and the number of
terms in the dictionaries:
\begin{ass}
	\label{ass:rates} \mbox{} \\[-20pt]
	\begin{enumerate}[label=\textbf{\alph*.},ref=\ref{ass:rates}.\alph*]
		\item \label{ass:rates:penalty} The Lasso penalty term $\lambda=\lambda(n)$ for estimation of $(\alpha_{01},\alpha_{02})$ satisfies: $n^{-r}=o(\lambda)$ and $\lambda=o(n^{c-r})$ for every $c>0$.
		
		\item \label{ass:rates:numterms} The number of terms in the dictionaries satisfy $J, K = O(n^\kappa)$ for a constant $\kappa>0$.
	\end{enumerate}
\end{ass}
This assumption asks for the Lasso penalty to go to zero slightly slower
than $n^{-r}$, where $r$ is the rate from Assumption \ref{ass:est_convergence}. For instance, a rate of $\log(n)/n^{r}$ is allowed. Moreover,
it requires polynomial rates in the growth of the number of terms in the
dictionaries.

The above are general conditions imposed on the dictionaries and the tuning
parameters for the Lasso penalized regression. The specific problem at hand
only appears in two instances. First, Assumption~\ref{ass:span} requires
that the dictionaries approximate well the correction-term Riesz representers (living in $\Delta_1$ and $\Delta_2(g_0)$, respectively). Second,
Assumption~\ref{ass:est_convergence} requires (i) mean-square rates for the
estimators of $g_0$ and $h_0$ and (ii) to be able to estimate the
linearizations at the same rate. As discussed before, these conditions
provide rates of estimators of the Riesz representers $\alpha_{01}$ and $\alpha_{02}$ \citep[see][]{chernozhukov2022automatic}. For
instance, the convergence rate of $\hat{\alpha}_{1\ell}$ will be fast enough
to guarantee that the interaction term satisfies $\lVert \hat{\alpha}%
_{1\ell}-\alpha_{01}\rVert_2 \cdot \lVert \hat{g}_\ell-g_0\rVert_2 =
o_p(n^{-1/2})$ \citep[c.f. Assumption~2 in][]{chernozhukov2022locally}.

We now provide assumptions on the moment condition. The first is a
mean-square consistency condition similar to Assumption~1 in \cite{chernozhukov2022locally}:
\begin{ass} \label{ass:sq_convergence} \mbox{} \\[-20pt]
	\begin{enumerate}[label=\textbf{\alph*.},ref=\ref{ass:sq_convergence}.\alph*]
		\item \label{ass:sq_convergence:m_bound} $\mathbb{E}[|m(W, g_0,h_0,\theta_0)|^2]<\infty$.
		
		\item \label{ass:sq_convergence:m_conv} $\int |m(w,\hat{g}_\ell,\hat{h}_\ell,\theta_0)-m(w,g_0,h_0,\theta_0)|^2 dF_0(w)\xrightarrow{P} 0$.
		
		\item \label{ass:sq_convergence:m_conv_theta} $\int |m(w,\hat{g}_\ell,\hat{h}_\ell,\tilde{\theta}_\ell)-m(w,\hat{g}_\ell,\hat{h}_\ell,\theta_0)|^2 dF_0(w)\xrightarrow{P} 0$.
		
		\item \label{ass:sq_convergence:var_bound} $\mathbb{E}[(S-h_0(X,V))^2|D,Z]$ and $\mathbb{E}[\epsilon(W, g_0)^2|Z]$ are bounded almost surely.
	\end{enumerate}
\end{ass}
Assumption~\ref{ass:sq_convergence:m_bound} is necessary for regular
estimation of $\theta_0$. Assumptions~\ref{ass:sq_convergence:m_conv} and %
\ref{ass:sq_convergence:m_conv_theta} are mean-square consistency conditions
for the moment condition. Boundedness of the conditional errors
(Assumption~\ref{ass:sq_convergence:var_bound}) translates into
mean-square consistency conditions for the debiasing term $\phi$. We repeat
here that the boundedness of $\alpha_{01} $ and $\alpha_{02}$ is implied by Assumptions~\ref%
{ass:bounded_dicts} and \ref{ass:abs_sumability:main}.

We also need to strengthen Assumption~\ref{rho} to control the remainder for linealizing the generalized error:
\begin{ass}
\label{ass:general_erro}
    The mapping $g \mapsto \E[\epsilon(W, g)]$ is Frechet differentiable at $g_0$, with derivative $D_e$. Moreover, the Riesz representer of the derivative ($r_{e}$) satisfies $r_{e}(z)<0$ and is bounded and bounded away from zero. The remainder from the linearization is quadratic: there exists $\varepsilon > 0$ and $C> 0$ such that, for $\norm{g-g_0}_2 < \varepsilon$,
    \begin{equation*}
        \lvert \E[\epsilon(W, g) - \epsilon(W, g_0) - D_e(g-g_0)] \rvert \leq C \norm{g-g_0}_2^2 
    \end{equation*}
\end{ass}

The following assumption is standard in the literature, see, e.g.,
\cite{newey1994asymptotic}. It imposes a quadratic remainder bound for the
first-order linearization of $\bar m(g,h)$ and therefore strengthens
Assumptions~\ref{ass:Diff_m_h} and \ref{ass:Diff_m_g}. Consider the linearization
\[
\bar{\psi}(g,h)
\equiv
\mathbb{E}\!\left[
m(W,g,h,\theta_0)-m(W,g_0,h_0,\theta_0)
-D_{dir}(W,g-g_0)
-D_2(W,h-h_0)
\right].
\]
Note that the linearization treats both $g$ and $h$ as ``independent" nuisance, i.e., it does not account for the fact that $g$ affects estimation of $h$. We assume the following:

\begin{ass}\label{ass:linear_approx}
For each $\ell=1,\dots,L$, one of the following conditions holds:
\begin{enumerate}[label=(\roman*)]
    \item For a $C>0$, with probability tending to one, $|\bar{\psi}(\hat g_\ell,\hat h_\ell)|
\le
C(
\|\hat g_\ell-g_0\|_2^2
+
\|\hat h_\ell-h_0\|_2^2)$ or
\item $\sqrt{n}\,\bar{\psi}(\hat g_\ell,\hat h_\ell) \xrightarrow{P} 0$.
\end{enumerate}
\end{ass}

To account for the generated regressors, we introduce the following assumption. Its goal is to guarantee that the remainder of the chain rule in our Lemma~\ref{lma:generalized_HR}, which accounts for the indirect effect, is quadratic. To state the assumption, we introduce the mapping $\nu(h, \alpha_2)\equiv[\partial\phi_2/\partial v](w, g_0, h, \alpha_2) = \partial/\partial v\{\alpha_2(x,v) \cdot (s - h(x,v)\}$ (c.f. Lemma~\ref{lma:generalized_HR}). Define $\hat{\nu}_\ell=\nu(\hat{h}_\ell, \hat{\alpha}_{2\ell})$ and $\nu_0=\nu(h_0, \alpha_{02})$.
\begin{ass}
    \label{ass:gen_reg} \mbox{}
    Either $\alpha_{02}(x, v) = 0$ or
    \begin{enumerate}[label=\textbf{\alph*.},ref=\ref{ass:gen_reg}.\alph*]

        \item \label{ass:gen_reg:phi_Fderiv} The mapping $g \mapsto \varphi(d, z, g)$, from $\Delta_1$ to $L_2(D,Z)$, is twice Frechet differentiable with continuous second derivative.
    
		\item \label{ass:gen_reg:twice_deriv} $h_0$ and $\alpha_{02}$ are almost surely twice continuously differentiable with respect to $v$.

        \item \label{ass:gen_reg:bounded_deriv}  $\partial h_{0}/\partial v$ and $\partial \alpha_{02}/\partial v$ are almost surely bounded.

        \item \label{ass:gen_reg:bounded_alpha2} $\hat{h}_\ell \in \Delta_2(g_0)$, almost surely. Moreover, $\hat{h}_\ell$ is almost surely twice continuously differentiable with respect to $v$.

        \item \label{ass:gen_reg:continuous_deriv} $\norm{\hat{\nu}_\ell-\nu_0}_2 \lVert \hat{g}_\ell-g_0 \rVert_2 = o_p(n^{-1/2}).$
	\end{enumerate}
\end{ass}
First, under downstream local robustness, the indirect effect is zero, and the above conditions are not needed. Regarding these conditions, Assumption~\ref{ass:gen_reg:phi_Fderiv} asks for a quadratic remainder in the linearization of the generated regressor. Assumptions~\ref{ass:gen_reg:phi_Fderiv}-\ref{ass:gen_reg:bounded_deriv} strengthen Assumption~\ref{ass:diff_h_alpha_phi}. Assumption~\ref{ass:gen_reg:bounded_alpha2} also strengthens Assumption~\ref{ass:inclusion} (see Appendix~\ref{sec:app_inclusion} for a general discussion of this assumption). Assumption~\ref{ass:gen_reg:continuous_deriv} is a product-rate condition to handle higher-order components from the generated regressors. When estimation of $h_0$ is conducted by (penalized) regression onto the dictionary $(b_j)_{j=1}^\infty$, this assumption may be understood as smoothness conditions on the dictionary. This is the case of the CASF example. We provide a detailed discussion of Assumption~\ref{ass:gen_reg:continuous_deriv} in Section~\ref{sec:CASF_asymp}, where we verify it for the CASF, and in Appendix~\ref{sec:lasso_derivatives}.

Finally, the GMM procedure requires consistent estimation of the Jacobian of the
moment condition. Being the following assumption specific to the GMM
procedure, it is stated for the case with an arbitrary number of parameters
and moment conditions.
\begin{ass} \label{ass:jacobian}
	There exists a neighborhood $\mathcal{N}$ of $\theta_0$ such that, for small $\lVert g - g_0 \rVert_2$ and $\lVert h - h_0 \rVert_2$:
	\begin{enumerate}[label=\textbf{\alph*.}, ref=\ref{ass:jacobian}.\alph*, series=JAC]
		\item \label{ass:jacobian:diff} $m(W, g, h, \theta)$ is almost surely differentiable in $\mathcal{N}$.
		\item \label{ass:jacobian:bound}There exists a $C>0$ and a function $d(W, g,h)$, with $\E[d(W,g,h)]<C$, such that for $\theta\in\mathcal{N}$
		\begin{equation*}
			\left\lVert  \frac{\partial m}{\partial\theta}(W, g,h, \theta) -  \frac{\partial m}{\partial\theta}(W, g,h, \theta_0)\right\lVert_\infty \leq d(W, g,h) \lVert \theta - \theta_0 \rVert_\infty^{1/C} \text{ almost surely}.
		\end{equation*}
	\end{enumerate}
	
	Moreover, we assume that:
	\begin{enumerate}[resume*=JAC]
		\item \label{ass:jacobian:existence} The expectation of the Jacobian, $\partial m/\partial\theta$, exists.
		\item \label{ass:jacobian:convergence} It holds that
		\begin{equation*}
			\int\left\lVert\frac{\partial m}{\partial\theta}(w, \hat{g}_\ell,\hat{h}_\ell, \theta_0) -  \frac{\partial m}{\partial\theta}(w, g_0,h_0, \theta_0)\right\rVert_\infty dF_0(w)\xrightarrow{P} 0.
		\end{equation*}
	\end{enumerate}
\end{ass}

Assumptions~\ref{ass:abs_sumability}, \ref{ass:est_convergence}, \ref%
{ass:sq_convergence}, and \ref{ass:linear_approx} are stated for a single
moment condition. In the presence of more than one condition, they must be understood to hold componentwise. Assumption~\ref{ass:jacobian}, since it refers to a GMM-specific situation, is already formulated in the general case. The remaining assumptions do not depend on the dimension of the moment condition (they depend, on the other hand, on the dimension of $Y$ and $D$).

Let $\Xi \equiv (M'\Upsilon M)^{-1}M'\Upsilon' \Psi \Upsilon M (M'\Upsilon M)^{-1}$ be the usual asymptotic variance of the GMM estimator based on the debiased moment functions, where
\begin{equation*}
    M \equiv \mathbb{E}\left[\frac{\partial m}{\partial\theta}%
(W,g_0,h_0,\theta_{0})\right] \text{ and }
\Psi \equiv \mathbb{E}[\psi(W,g_0,h_0,\alpha_{0},\theta_{0})\psi(W,g_0,h_0,%
\alpha_{0},\theta_{0})^{\prime }].
\end{equation*}
Define the plug-in estimator $\hat{\Xi}\equiv (\hat{M}^{\prime }\hat{%
\Upsilon}\hat{M})^{-1}\hat{M}^{\prime }\hat{\Upsilon}^{\prime }\hat{\Psi}%
\hat{\Upsilon}\hat{M}(\hat{M}^{\prime }\hat{\Upsilon}\hat{M})^{-1}$, where $\hat{M}$ and $\hat{\Psi}$ are given by the corresponding cross-fitted sample analogs
\begin{equation*}
\hat{M} \equiv \frac{1}{n}\sum_{\ell =1}^{L}\sum_{i\in I_{\ell }}\frac{%
\partial m}{\partial \theta }(W_{i},\hat{g}_{\ell },\hat{h}_{\ell },\tilde{%
\theta}_{\ell })\text{ and } 
\hat{\Psi} \equiv \frac{1}{n}\sum_{\ell =1}^{L}\sum_{i\in I_{\ell }}\hat{%
\psi}_{i\ell }(\tilde{\theta}_{\ell })\hat{\psi}_{i\ell }(\tilde{\theta}%
_{\ell })^{\prime }. 
\end{equation*}%
The following theorem ensures asymptotic normality of $\sqrt{n}(\hat{\theta}-\theta _{0})$:
\begin{thm} \label{thm:asymptotics}
	Consider that Assumptions \ref{ass:span}-\ref{ass:jacobian} are satisfied, $\hat{\Upsilon}\xrightarrow{P}\Upsilon$, and $M'\Upsilon M$ is non-singular. Then, the three-step debiased GMM estimator in equation~\eqref{dgmm} satisfies
	\begin{equation*}
		\sqrt{n}(\hat{\theta}-\theta_0) \xrightarrow{D} N(0, \Xi).
	\end{equation*}
	Moreover, the plug-in estimator for the asymptotic variance is consistent: $\hat{\Xi} \xrightarrow{P} \Xi$.
\end{thm}

%%%%%%%%%%%%%%%%%%%%%%%%%%%%%%%%%%%%%%%%%%%%%%%%%%%%%%%%%%%%%%%%%%%%%%%%%%%%%%%%%%%%%%%%%%%%%%%%%%%%%%
\subsection{Regularity conditions for some examples}

\subsubsection{Hd-PS regression adjustment}
\label{sec:asymp_hdps}

Here we verify Assumptions~\ref{ass:sq_convergence}, \ref{ass:general_erro}, \ref{ass:linear_approx}, \ref{ass:gen_reg}, and \ref{ass:jacobian}. These are the assumptions that explicitly depend on the identifying moment condition and the first and second estimation steps. We also provide sufficient conditions for Assumptions~\ref{ass:Diff_m_h} and \ref{ass:Diff_m_g} required for linearizing the moment condition. The moment condition that identifies $\theta_0$ in the partially linear model is:
\begin{equation*}
    m(w, g, h, \theta) = \left[ y - h(\Lambda(g(z))) - \theta (d - \Lambda(g(z))) \right] \cdot \left[ d - \Lambda(g(z)) \right],
\end{equation*}
were recall that $\Lambda$ stands for the logistic cdf and $V=\Lambda(g_0(z))$.

The following assumption gives the result:
\begin{ass} \label{ass:regularity_ATE} \mbox{} \\[-20pt]
    \begin{enumerate}[label=\textbf{\alph*.},ref=\ref{ass:regularity_ATE}.\alph*]

        \item \label{ass:regularity_ATE_compact} The propensity score $V=\Lambda(g_0(Z))$ is bounded away from $0$ and $1$.

        \item \label{ass:regularity_ATE_expectation} $\E[Y^2 |D, Z]$ is bounded almost surely.

        \item \label{ass:regularity_ATE_dictionary} The atoms in the dictionary $\mathbf{b}_J$ are continuously differentiable.
    \end{enumerate}
\end{ass}
Assumption~\ref{ass:regularity_ATE_compact} is the usual overlap assumption. Assumption~\ref{ass:regularity_ATE_expectation} bounds $\E[Y^2|D,Z]$ (note that $Y$ may still be supported on $\mathbb{R}$). If regressors $Z$ have compact support, continuity of $E[Y^2|D=d,Z=z]$ would be sufficient for Assumption~\ref{ass:regularity_ATE_expectation}. Assumption~\ref{ass:regularity_ATE_dictionary} imposes smoothness conditions on the atoms. Note that if $h_0(v) = \E[Y|V=v]$ is smooth enough, the econometrician can always choose a dictionary with smooth atoms to estimate it.

We show that the assumptions for Theorem~\ref{thm:asymptotics} holds in the Hd-PS setting:
\begin{prop} \label{prop:ATE_conditions}
    Suppose that the convergence conditions in Assumption~ \ref{ass:est_convergence:estimators} hold and assume the existence of a consistent preliminary estimator $\tilde{\theta}_\ell \xrightarrow{P} \theta_0$. Then, Assumption~\ref{ass:regularity_ATE} guarantees that Assumptions~\ref{ass:sq_convergence}, \ref{ass:general_erro}, \ref{ass:linear_approx}, \ref{ass:gen_reg}, and \ref{ass:jacobian} are satisfied for the moment condition identifying $\theta_0$ in the Hd-PS regression adjustment within the partially linear model.

\end{prop}

\subsubsection{CASF in a non-separable model}
\label{sec:CASF_asymp}

Here we verify Assumptions~\ref{ass:sq_convergence}, \ref{ass:general_erro}, \ref{ass:linear_approx}, \ref{ass:gen_reg}, and \ref{ass:jacobian} for the CASF. To achieve this, we require some regularity on the distribution of $(X,V)$, where $V = D - g_0(Z)$, on the second step $h_0(x,v) \equiv \E[Y|X=x, V=v]$, and on the dictionary that is used for second-step estimation.

\begin{ass} \label{ass:CASF_conditions} \mbox{} \\[-20pt]
    \begin{enumerate}[label=\textbf{\alph*.},ref=\ref{ass:CASF_conditions}.\alph*]

        \item \label{ass:CASF_conditions_alpha2} $(X, V)$ has joint density $f_{xv}$ w.r.t. an absolutely continuous measure $\mu$, with $\mu(x,v) = \mu_x(x) \times \mu_v(v)$. Also, $F^*$ has density $f^*$ w.r.t. $\mu_x$. With $f_v$ being the marginal density of $V$, it holds that $f^* f_{v} / f_{xv}$ is almost surely bounded and twice continuously differentiable with bounded first derivative w.r.t. $v$.
        
        \item \label{ass:CASF_conditions_expectation} $\E[Y^2 |D, Z]$ and $\E[D^2|Z]$ are bounded almost surely.

        \item \label{ass:CASF_conditions_h} $h_0$ is twice differentiable w.r.t. $v$, with $\partial h_0/\partial v$ and $\partial^2 h_0/\partial v^2$ bounded almost surely.

        \item \label{ass:CASF_conditions_dictionary} The atoms in the dictionary $\mathbf{b}_J$ are twice differentiable w.r.t. $v$ with $\partial\mathbf{b}_J/\partial v$ and $\partial^2\mathbf{b}_J/\partial v^2$ bounded almost surely.

        \item \label{ass:CASF_conditions_deriv} $\norm{\partial\hat{h}_\ell/\partial v - \partial h_0 / \partial v}_2 \norm{\hat{g}_\ell - g_0}_2=o_p(n^{-1/2})$ and $\norm{\partial\hat{\alpha}_{2\ell}/\partial v - \partial \alpha_{02} /\partial v}_2 \norm{\hat{g}_\ell - g_0}_2=o_p(n^{-1/2})$.
    \end{enumerate}
    \end{ass}
 Assumption~\ref{ass:CASF_conditions_alpha2} is guarantees regular
identification of the CASF. Note that, in the case of the CASF, the linearization of the second step can be written as $\E[D_2(w, h)] = \E[r_2(X,V)h(X, V)]$, with $r_2 = f^* f_{v} / f_{xv}$. This assumption ensures finite variance of $r_2$ and ask for additional smoothness conditions. Assumption~\ref{ass:CASF_conditions_expectation} is the usual bounded conditional variance assumption. Assumption~\ref{ass:CASF_conditions_h} also imposes smoothness conditions on $h_0$, while Assumption~\ref{ass:CASF_conditions_dictionary} requires the dictionary used to estimate $h_0$ to satisfy the same smoothness conditions. 

Assumption~\ref{ass:CASF_conditions_deriv} requires product-rate conditions
involving the estimation error of the derivatives of the second-step
nuisance functions. These conditions hold under standard sparse
high-dimensional assumptions when $\hat h_\ell$ and
$\hat\alpha_{2\ell}$ are estimated by Lasso on the dictionary
$\mathbf b_J(x,v)$ introduced in Section~\ref{sec:CASF_details}.
In particular, if $h_0$ and $\alpha_{02}$ admit sparse
approximations on $\mathbf b_J$, if the Gram matrices
$\E[\mathbf b_J(X,V)\mathbf b_J(X,V)']$ and
$\E[(\partial \mathbf b_J(X,V)/\partial v)
(\partial \mathbf b_J(X,V)/\partial v)']$
have eigenvalues bounded away from zero and infinity, and if the Lasso
estimators achieve the usual $L_1$ coefficient rates, then the
derivative estimation errors satisfy
$\|\partial\hat h_\ell/\partial v-\partial h_0/\partial v\|_2
=O_p(s_h\sqrt{\log J/n})$
and
$\|\partial\hat\alpha_{2\ell}/\partial v-\partial \alpha_{02}/\partial v\|_2
=O_p(s_\alpha\sqrt{\log J/n})$ \citep{bickel2009simultaneous}.
Hence, Assumption~\ref{ass:CASF_conditions_deriv} holds whenever these
rates multiplied by the first-step rate
$\|\hat g_\ell-g_0\|_2$ are $o_p(n^{-1/2})$.
A detailed verification is given in Appendix~\ref{sec:lasso_derivatives}.

We can then show that the assumptions for
Theorem~\ref{thm:asymptotics} hold for the moment condition defining the CASF.
\begin{prop} \label{prop:CASF_conditions}
 Suppose that the convergence conditions in Assumptions~\ref{ass:span}-\ref{ass:rates} hold and assume the existence of a consistent preliminary estimator $\tilde{\theta}_\ell \xrightarrow{P} \theta_0$. Then Assumption~\ref{ass:CASF_conditions} guarantees that Assumptions~\ref{ass:sq_convergence}, \ref{ass:general_erro}, \ref{ass:linear_approx}, \ref{ass:gen_reg}, and \ref{ass:jacobian} are satisfied for the moment condition identifying the CASF.

    \end{prop}

\section{Conclusion}

\label{sec:conclusion}

We propose Automatic Locally Robust estimators for structural parameters in
the presence of ML-generated regressors. We show that the debiasing correction
term can be decomposed into terms accounting for the first-step and the second-step
estimation. Each of the first- and second-step IFs depends on an additional
Riesz representers, which can be automatically estimated (i.e., estimated
without finding their analytic shape).

We apply our results to construct Automatic Locally Robust estimators for
causal treatment effects and the CASF under different modelling assumptions (partially linear and nonparametric models) and different generated regressors (Hd-PS, autoencoders, control function, etc). The analytic shape of the Riesz representers in these cases is
particularly complex. For the partially linear model, our automatic debiased estimator overcomes the large biases of the state-of-the-art method, the DML, which does not account for the generated regressors. For the CASF parameter, the moment condition depends on the whole shape of the second-step nuisance parameter (not only its pointwise value), making existing results on generated regressors not applicable even in low-dimensional scenarios. Therefore,
automatic estimation is particularly well suited for these problems. We have
shown that commonly used plug-in or DML methods lead to highly biased inferences with ML-generated regressors. Three-step debiased estimators correct the bias and deliver much more accurate inference in a complex setting with
ML-generated regressors.

\newpage \addcontentsline{toc}{section}{References} \makeatletter
\makeatother
\bibliographystyle{apalike}
\bibliography{references}

%%%%%%%%%%%%%%%%%%%%%%%%%%%%%%%%%%%%%%%%%%%%%%%%%%%%%%%%%%%%%%%%%%%%%%%%%%
%%%%%%%%%%%%%%%%%%%%%%%%%%%%%%%%%%%%%%%%%%%%%%%%%%%%%%%%%%%%%%%%%%%%%%%%%%
%%%%%%%%%%%%%%%%%%%%%%%%%%%%%%%%%%%%%%%%%%%%%%%%%%%%%%%%%%%%%%%%%%%%%%%%%%
\cleardoublepage
\begin{appendices}

% No numbering of title page
\thispagestyle{empty}

\begin{center}
    {\Large\bf Supplementary Material}\\[10pt]

    {\large\bf Supplementary Material to ``Automatic Locally Robust GMM with Machine-Learning-Generated Regressors''}\\[12pt]

    {\large Juan Carlos Escanciano \quad and \quad Telmo J.\ P\'erez-Izquierdo}\\[6pt]

    {\it Universidad Carlos III de Madrid \quad and \quad University of the Basque Country}\\[10pt]

    {\normalsize \today}
\end{center}

\section*{Overview of the supplementary material}

This Supplementary Material accompanies ``Automatic Locally Robust GMM with Machine-Learning-Generated Regressors'' and collects implementation details, additional examples, Monte Carlo designs, and proofs that are omitted from the main text for brevity. Notation, assumptions, and equation numbering follow the main paper unless explicitly stated otherwise.

Appendix~\ref{sec:algorithm} summarizes the construction of the cross-fitted debiased moment function and the resulting Automatic Debiased GMM estimator, including a diagram (Figure~\ref{fig:est_algo}) that details the automatic estimation of the Riesz representers entering the correction terms. Appendix~\ref{sec:AdditionalExamples} provides an additional worked example, the nonparametric ATE estimator based on a boosted propensity score. Appendix~\ref{sec:MC_details} reports the full Monte Carlo designs, tuning choices, and the exact estimators of asymptotic variances used in the simulations. Appendix~\ref{sec:app_inclusion} gives primitive sufficient conditions for the inclusion requirement (Assumption~\ref{ass:inclusion}). Appendix~\ref{sec:lasso_derivatives} discusses rates on the derivatives of Lasso estimators. Finally, Appendix~\ref{sec:Proofs} contains proofs of the results stated in the paper and in this supplement.

%\cleardoublepage

% Reset page counter
\setcounter{page}{1}

% Number figures and tables by appendix section
\renewcommand{\thefigure}{\thesection.\arabic{figure}}
\renewcommand{\thetable}{\thesection.\arabic{table}}

% Reset counters at the start of each appendix section
\counterwithin{figure}{section}
\counterwithin{table}{section}

%%%%%%%%%%%%%%%%%%%%%%%%%%%%%%%%%%%%%%%%%%%%%%%%%%%%%%%%%%%%%%%%%%%%%%%%%%%%%%%%%%%%%%%%%%%%%%%
\section{Estimation algorithm}

\label{sec:algorithm}

Here, we provide a summary of the algorithm to construct a cross-fitted debiased moment function. The inputs to the
algorithm are cross-fitted estimators of $g_{0}$ and $h_{0}$. A preliminary
estimator of $\theta _{0}$ must also be supplied. In the most general case, one must
provide a total of $L$ estimators $(\hat{g}_{\ell },\hat{h}_{\ell },\tilde{%
\theta}_{\ell })$ only using observations not in $I_{\ell }$, $L(L-1)/2$
estimators $(\hat{g}_{\ell \ell ^{\prime }},\hat{h}_{\ell \ell ^{\prime }},%
\tilde{\theta}_{\ell \ell ^{\prime }})$ only using observations not in $%
I_{\ell }\cup I_{\ell ^{\prime }}$, and $L(L-1)(L-2)/6$ estimators $(\hat{g}%
_{\ell \ell ^{\prime }\ell ^{\prime \prime }},\hat{h}_{\ell \ell ^{\prime
}\ell ^{\prime \prime }},\tilde{\theta}_{\ell \ell ^{\prime }\ell ^{\prime
\prime }})$ only using observations not in $I_{\ell }\cup I_{\ell ^{\prime
}}\cup I_{\ell ^{\prime \prime }}$.

Figure~\ref{fig:est_algo} provides a diagram showing how to compute $\hat{\psi}_{i\ell}(\theta)$ for an observation $i\in
I_\ell $. The debiased moment function is given in equation~%
\eqref{eq:dmon_i}. To this equation, the diagram below adds the discussion
about how to construct automatic estimators of the
Riesz representers ($\alpha_{01}$ and $\alpha_{02}$) in the correction
terms. The arrows in the diagram indicate how to estimate each term. Once the debiased moment condition $\hat{\psi}_{i\ell}(\theta)$ is built, Automatic
Loally Robust GMM estimation is conducted with the objective function in equation~%
\eqref{dgmm}.

\begin{sidewaysfigure}
	\centering
    \resizebox{\textwidth}{!}{
	\begin{tikzpicture}[->,>=stealth',auto,node distance=2.6cm, thick]
		
		%Main equation
		\node (psi) [lblock] {$\hat{\psi}_{i\ell}(\theta)\equiv$};
		\node (m) [cblock, right = of psi]{$m(W_{i},\hat{g}_{\ell},\hat{h}_\ell, \theta)+$};
		\node (alpha1) [cblock, right = of m] {$\hat{\alpha}_{1\ell}(Z_i)$};
		\node (err1) [cblock, right = of alpha1] {$\epsilon(W_i, \hat{g}_\ell)+$};
		\node (alpha2) [cblock, right = of err1] {$\hat{\alpha}_{2\ell}(X_i, \hat{V}_{i\ell})'$};
		\node (err2) [rblock, right = of alpha2] {$(S_i-\hat{h}_\ell(X_i, \hat{V}_{i\ell}))$};
		
		%Definition of alpha2
		\node (alpha2def) [unblock, above = 5em and 0em of alpha2] {$=\mathbf{b}_J(X_i, \hat{V}_{i\ell})'\widehat{\boldsymbol{\rho}}_{J\ell}$};
		
		%Computation of rho
		\node (rho2pen) [unblock,  left = 0em and 5em of alpha2def] {$+\lambda \lVert \boldsymbol{\rho}_J  \rVert_q^q\}$};
		\node (rho2B) [unblock, left = of rho2pen] {$\boldsymbol{\rho}_J'\hat{B}_\ell\boldsymbol{\rho}_J$};
		\node (rho2plus) [unblock, left = of rho2B] {$+$};
		\node (rho2D) [unblock, left = of rho2plus] {$-2\hat{D}_{2\ell}'\boldsymbol{\rho}_J$};
		\node (rho2min) [unblock, left = of rho2D] {$\argmin_{\boldsymbol{\rho}_J\in\mathbb{R}^J} \{ $};
		
		%Computation of D2 and B2
		\node (D2def) [unblock, align=center, above left = 3em and 1em of rho2D] { Average on $\ell'\neq\ell$, $s\in I_{\ell'}$, of \\ $D_{02}(W_s, \mathbf{b}_J | \hat{g}_{\ell\ell'}, \hat{h}_{\ell\ell'}, \tilde{\theta}_{\ell\ell'})$};
		\node (B2def) [unblock, align=center, above right = 3em and 1em of rho2B] { Average on $\ell'\neq\ell$, $s\in I_{\ell'}$, of \\ $\mathbf{b}_J(X_s,\hat{V}_{s\ell\ell'})\mathbf{b}_J(X_s,\hat{V}_{s\ell\ell'})'$};
		
		%Definition of alpha1
		\node (alpha1def) [unblock, below = 5em of alpha1] {$=\mathbf{c}_K(Z_i)'\widehat{\boldsymbol{\beta}}_{K\ell}$};
		
		%Computation of beta
		\node (beta1min) [unblock, right = 0em and 5em of alpha1def] {$\argmin_{\boldsymbol{\beta}_K\in\mathbb{R}^K} \{ $};
		\node (beta1D) [unblock, right = of beta1min] {$-2\hat{D}_{1\ell}'\boldsymbol{\beta}_K$};
		\node (beta1plus) [unblock, right = of beta1D] {$+$};
		\node (beta1C) [unblock, right = of beta1plus] {$\boldsymbol{\beta}_K'\hat{C}_\ell\boldsymbol{\beta}_K$};
		\node (beta1pen) [unblock,  right = of beta1C] {$+\lambda \lVert \boldsymbol{\beta}_K  \rVert_q^q\}$};
		
		%Computation of D1 and C1
		\node (D1def) [unblock, align=center, below left = 3em and 1.5em of beta1D] { Average on $\ell'\neq\ell$, $s\in I_{\ell'}$, of \\ $D_{01}(W_s, \mathbf{c}_K | \hat{g}_{\ell\ell'}, \hat{h}_{\ell\ell'}, \hat{\alpha}_{2\ell\ell'}, \tilde{\theta}_{\ell\ell'})$};
		\node (C1def) [unblock, align=center, below  = 3em and 0em of beta1C] { Average on $\ell'\neq\ell$, $s\in I_{\ell'}$, of \\ $-r_e(Z_s|\hat{g}_{\ell\ell'})\mathbf{c}_K(Z_s)\mathbf{c}_K(Z_s)'$};
		
		%Explanation of D1
		\node (D1phi) [unblock, align=center, left = 3em  of D1def] {$\cdot D_\varphi\mathbf{c}_K$};
		\node (D1derr) [unblock, align=center, left = of D1phi]{$(y-\hat{h}_{\ell\ell'}(X_s,v))]$};
		\node (D1alpha2) [unblock, align=center, left =  of D1derr]{$\hat{\alpha}_{2\ell\ell'}(X_s,v)$};
		\node (D1diff) [unblock, align=center, left =  of D1alpha2]{$\frac{\partial}{\partial v}[$};
		\node (D1D11) [unblock, align=center, left =  of D1diff]{$D_{dir}(W_s,\mathbf{c}_K)+$};
		
		%Evaluation point of the derivative
		\node (diffeval) [unblock, align=center, above = 1em of D1diff]{Evaluated at \\ $v=\hat{V}_{s\ell\ell'}$};
		
		%Definition of alpha2 for the first step
		\node (1Salpha2def) [unblock, below = 5em and 0em of D1alpha2] {$=\mathbf{b}_J(X_s, v)'\widehat{\boldsymbol{\rho}}_{J\ell\ell'}$};
		
		%Computation of rho for the first step
		\node (1Srho2min) [unblock, right = 0em and 5em of 1Salpha2def] {$\argmin_{\boldsymbol{\rho}_J\in\mathbb{R}^J} \{ $};
		\node (1Srho2D) [unblock, right = of 1Srho2min] {$-2\hat{D}_{2\ell\ell'}'\boldsymbol{\rho}_J$};
		\node (1Srho2plus) [unblock, right = of 1Srho2D] {$+$};
		\node (1Srho2B) [unblock, right = of 1Srho2plus] {$\boldsymbol{\rho}_J'\hat{B}_{\ell\ell}\boldsymbol{\rho}_J$};
		\node (1Srho2pen) [unblock,  right = of 1Srho2B] {$+\lambda \lVert \boldsymbol{\rho}_J  \rVert_q^q\}$};
		
		%Computation of D2 and B2 for the first step
		\node (1SD2def) [unblock, align=center, below left = 3em and 1.5em of 1Srho2D] { Average on $\ell''\notin \{\ell, \ell'\}$, $\iota\in I_{\ell''}$, of \\ $D_{02}(W_\iota, \mathbf{b}_J | \hat{g}_{\ell\ell'\ell''}, \hat{h}_{\ell\ell'\ell''}, \tilde{\theta}_{\ell\ell'\ell''})$};
		\node (1SB2def) [unblock, align=center, below right = 3em and 1em of 1Srho2B] { Average on $\ell''\notin \{\ell, \ell'\}$, $\iota\in I_{\ell''}$, of \\ $\mathbf{b}_J(X_\iota,\hat{V}_{\iota\ell\ell'\ell''})\mathbf{b}_J(X_\iota,\hat{V}_{\iota\ell\ell'\ell''})'$};
		
		%Alpha2 arrows	
		\draw [->] (alpha2) -- (alpha2def);
		\draw [->] (alpha2def) -- (rho2pen);
		\draw [->] (rho2D) to [out=90,in=-90] (D2def);
		\draw [->] (rho2B) to [out=90,in=-90] (B2def);
		
		%Alpha1 arrows
		\draw [->] (alpha1) -- (alpha1def);
		\draw [->] (alpha1def) -- (beta1min);
		\draw [->] (beta1D) to [out=-90,in=90] (D1def);
		\draw [->] (beta1C) to [out=-90,in=90] (C1def);
		\draw [->] (D1def) -- (D1phi);
		\draw [->] (D1diff) -- (diffeval);
		\draw [->] (D1alpha2) -- (1Salpha2def);
		\draw [->] (1Salpha2def) -- (1Srho2min);
		\draw [->] (1Srho2D) to [out=-90,in=90]  (1SD2def);
		\draw [->] (1Srho2B) to [out=-90,in=90]  (1SB2def);
		
	\end{tikzpicture}
    }
	\caption{Illustration of the algorithm to estimate the moment condition $\hat{\psi}_{i\ell}$ for an observation $i\in I_\ell$. In the diagram, $\hat{V}_{i\ell} \equiv \varphi(D_i, Z_i, \hat{g}_\ell)$, $\hat{V}_{s\ell\ell'} \equiv \varphi(D_s, Z_s, \hat{g}_{\ell\ell'})$, and $\hat{V}_{\iota\ell\ell'\ell''} \equiv \varphi(D_\iota, Z_\iota, \hat{g}_{\ell\ell'\ell''})$. See Section~\ref{sec:Automatic} for definitions of $D_{01}$, $D_{dir}$, $D_{02}$, $D_\varphi$, and $r_e$.}
	\label{fig:est_algo}	
\end{sidewaysfigure}

%%%%%%%%%%%%%%%%%%%%%%%%%%%%%%%%%%%%%%%%%%%%%%%%%%%%%%%%%%%%%%%%%%%%%%%%%%%%%%%%%%%%%%%%%%%%%%%
\section{Additional examples}
\label{sec:AdditionalExamples}
\subsection{Nonparametric ATE with boosted propensity score}
\label{ATEPSM}

The influence function for the nonparametric ATE for the case $\Delta_1 = L_2(Z)$ and $\Delta_2(g_0) = L_2(X,V)$, with propensity score $V = \varphi(D, Z, g_0) = g_0(Z) = \E[D|Z]$, was obtained by \cite{hahn2013asymptotic}. It is given by
\begin{equation*}
\psi(w,g,h,\alpha,\theta )=h(1,v)-h(0,v)-\theta +\alpha _{02}(d,v)\cdot(y-h(d,v)) +\alpha _{01}(z)\cdot (d-v),
\end{equation*}%
evaluated at $v=g(z).$ The expression for $\alpha _{02}$ is well
known from results in the Double Robustness literature, see %
\citet{robins1994estimation}, treating the generated regressors as given. It
corresponds to the Horvitz-Thompson weights. That is
\begin{equation*}
\alpha _{02}(d,v)=\frac{d-v}{v(1-v)}.
\end{equation*}%

We generalize these results to a general $\Delta_2(g_0)$, i.e., under potential misspecification of the outcome equation model (a practical case). In this more general case, the Riesz representer $\alpha _{02}$ is the orthogonal projection of the fully nonparametric one onto $\Delta_2(g_0)$. Relative to \cite{hahn2013asymptotic}, we also consider debiased automatic estimation and prove asymptotic normality with ML-generated regressors. Specifically, we illustrate with a boosted propensity score (twang) and we propose to estimate the second-step Riesz representer by $\hat{\alpha}_{2\ell }=%
\mathbf{b}_{J}^{\prime }\widehat{\boldsymbol{\rho }}_{J\ell }$, where $\widehat{%
\boldsymbol{\rho }}_{J\ell }$ solves the minimization problem in (\ref{eq:min_rho}) with $\hat{D}_{2\ell}$ computed using
\begin{equation*}
D_{02}(W_{i},b_{j}|\hat{g}_{\ell \ell ^{\prime }},\hat{h}_{\ell \ell ^{\prime
}},\tilde{\theta}_{\ell \ell ^{\prime }})=b_{j}(1,\hat{V}_{i\ell\ell'})-b_{j}(0,\hat{V}_{i\ell\ell'}) \text{ for } i \in I_{\ell'} \text{ and } j=1,\dots,J, 
\end{equation*}%
where $\hat{V}_{i\ell\ell'} = \hat{g}_{\ell \ell ^{\prime}}(Z_{i})$. 

The first-step Riesz representer is estimated by $\hat{\alpha}_{1\ell }=\mathbf{c}_{K}^{\prime }\widehat{\boldsymbol{\beta }}_{K\ell }$, where $\widehat{\boldsymbol{\beta }}_{K\ell }$ solves the minimization problem in \eqref{eq:min_beta}. To compute $\hat{C}_\ell$, since $\varepsilon(w, g) = d - g(z)$, we have that $r_e(z) = -1$.

Following the result in Theorem~\ref{thm:first_step}, estimation of the corresponding $\hat{D}_{1\ell}$ is based on 
\begin{align*}
    D_{01}(W_{i},c_{k}|\hat{g}_{\ell \ell ^{\prime }},\hat{h}_{\ell \ell ^{\prime}},\hat{\alpha}_{2\ell \ell ^{\prime }},\tilde{\theta}_{\ell \ell ^{\prime}}) =
    \left[ \dot{h}_{\ell\ell'}(1, \hat{V}_{i\ell\ell'}) - \dot{h}_{\ell\ell'}(0, \hat{V}_{i\ell\ell'}) \right. \\ 
    + \left. \dot{\alpha}_{2\ell\ell'}(D_i, \hat{V}_{i\ell\ell'}) (Y_i - \hat{h}_{\ell\ell'}(D_i, \hat{V}_{i\ell\ell'}))
    - \hat{\alpha}_{2\ell\ell'}(D_i, \hat{V}_{i\ell\ell'}) \dot{h}_{\ell\ell'}(D_i, \hat{V}_{i\ell\ell'}) \right]c_k(Z_i),
\end{align*}
for $i\in I_{\ell^{\prime }}$ and $k=1,\dots, K$. In the above equation, $\dot{h}_{\ell\ell'}$ and $\dot{\alpha}_{\ell\ell'}$ are cross-fitted estimators of the derivatives $\partial h_0/\partial v$ and $\partial \alpha_{02}/\partial v$, respectively. For a differentiable dictionary $\mathbf{b}_J$, we could estimate the later by $\dot{\alpha}_{2\ell\ell'}(d,v) = (\partial \mathbf{b}_J/\partial v(d,v))'\widehat{\boldsymbol{\rho}}_{J\ell\ell'}$. An estimator of $\partial h_0/\partial v$ could be constructed following equation~\eqref{eq:deriv_h_est} or using the results in \cite{fonseca2018boost} that are specific to boosting methods.

The debiased ATE estimator with generated regressors has the
expression%
\begin{equation*}
\hat{\theta}=\hat{\theta}_{DR}+\frac{1}{n}\sum_{\ell =1}^{L}\sum_{i\in
I_{\ell }}\hat{\alpha}_{1\ell }(Z_{i})\cdot (D_{i}-\hat{V}_{i\ell
} ),
\end{equation*}%
where $\hat{\theta}_{DR}$ is a cross-fitted Doubly Robust estimator that accounts for the estimation of the conditional means $h_{0}(1,v)$ and $h_{0}(0,v)$ but
does not account for the estimated propensity score, i.e.%
\begin{equation*}
\hat{\theta}_{DR}=\frac{1}{n}\sum_{\ell =1}^{L}\sum_{i\in I_{\ell }} \left[ \hat{h}%
_{\ell }(1,V_{i\ell })-\hat{h}_{\ell }(0,V_{i\ell })+\hat{\alpha}_{2\ell
}(X_{i},\hat{V}_{i\ell })\cdot (Y_{i}-\hat{h}_{\ell }(X_{i},\hat{V}_{i\ell
})) \right].
\end{equation*}

%%%%%%%%%%%%%%%%%%%%%%%%%%%%%%%%%%%%%%%%%%%%%%%%%%%%%%%%%%%%%%%%%%%%%%%%%%%%%%%%%%%%%%%%%%%%%%%%%%%%%%
\clearpage
\section{Details about the Monte Carlo simulation}
\label{sec:MC_details}

\subsection{Hd-PS regression adjustment in the partially linear model}

\subsubsection{Setup}

 The available data are $(Y, D, Z)$, with $Z \equiv (Z_j)_{j=1}^{10}$. Outcome and treatment equations are the following:
\begin{align*}
    Y &= D + Z_1 + Z_2 + \varepsilon, \\
    D &= \mathbf{1}\left( C \nu \leq  Z_1 + Z_2 + Z_3 + Z_4 + Z_5 + Z_6 \right).
\end{align*}
The error terms $\varepsilon$ and $\nu$ are independent, with $\varepsilon \sim N(0, 1)$. The distribution of $\nu$ varies with the specification: it can be logistic or standard normal. Regressors $Z$ are independent from each other and are uniformly distributed on $[-1, 1]$. Regressors are also independent of $\nu$. On the other hand, the first two regressors $(Z_1, Z_2)$ and $\varepsilon$ are linked by a Gaussian copula. That is, the distribution of $(\varepsilon, Z_1, Z_2)$ is
\begin{equation*}
\begin{aligned}
    F(\varepsilon, z_1, z_2) &= C(\Phi(\varepsilon), U(z_1), U(z_2)), \\
    C(u_1, u_2, u_3) &\equiv \Phi_\Sigma(\Phi^{-1}(u_1), \Phi^{-1}(u_2), \Phi^{-1}(u_3)), \\
    \Sigma &\equiv \begin{pmatrix} 1 & 1/2 & 1/2 \\
                                   1/2 & 1 & 0 \\
                                   1/2 & 0 & 1
    \end{pmatrix},
\end{aligned}
\end{equation*}
where $U$ is the cdf of a uniform on $[-1, 1]$, $\Phi$ is the standard normal cdf, and $\Phi_\Sigma$ is the cdf of a multivariate normal centered at zero and with covariance matrix $\Sigma$. Since all specifications include $(Z_1, Z_2)$ in the treatment equation, this makes treatment $D$ endogenous. The constant $C$ is chosen so that the propensity score is supported on $[0.01, 0.99]$. When $\nu$ is logistic, $C = 6/\Lambda^{-1}(0.99)$, with $\Lambda$ the logistic cdf. When $\nu$ is standard normal, $C = 6/\Phi^{-1}(0.99)$. In every specification $\theta_0 = 1$ and is identified by the moment condition in equation~\eqref{eq:PLM_general_moment}.

We evaluate the performance of three estimation procedures (recall that $\hat{V}_{i\ell} = \Lambda(\hat{g}_\ell(Z_i))$):
\begin{itemize}
    \item The DML estimator $\hat{\theta}_{DML}$, given in equation~\eqref{eq:ATE_PLM_DML} with $\hat{h}_{\ell,D}(\hat{V}_{i\ell})=\hat{V}_{i\ell}$, which equals the plug-in estimator for the partially linear model. The corresponding estimator of the asymptotic variance does not correct for the first step (see below). 
    \item The three-step debiased (3SD) estimator $\hat{\theta}$, given in equation~\eqref{DATE} with $\hat{h}_{\ell,D}(\hat{V}_{i\ell})=\hat{V}_{i\ell}$ and $\epsilon(W_i, \hat{g}_\ell) = \Lambda(\hat{g}_\ell(Z_i))$. The corresponding estimator of the asymptotic variance corrects for the first step (see below).
    \item A plug-in plus correct asymptotic variance (PI-CAV) procedure that uses the plug-in/DML estimator to recover the parameter, but accounts for the first step when estimating the asymptotic variance (see below).
\end{itemize}

For all estimation procedures, we set the number of cross-fitting partitions to $L = 10$. The propensity score is estimated by a Logit-Lasso of $D$ onto the regressors $Z$, with tuning parameter chosen by cross validation with 3 folds. That is, we choose the dictionary $\mathbf{c}_{10}(z) = (z_j)_{j=1}^{10}$. The second-step nuisance parameter $h_{0}(v)=\E[Y\mid V=v]$ is estimated by a Lasso of $Y$ onto a dictionary $\mathbf{b}_6(v) = (v^{j-1})_{j=1}^6$. The tuning parameter for the second step is also chosen by cross-validation with 3 folds.  Due to computational costs, we use $(\hat{g}_\ell, \hat{h}_\ell)$ to build the automatic estimator $\hat{\alpha}_{1\ell}$, instead of the doubly cross-fitted $(\hat{g}_{\ell\ell'}, \hat{h}_{\ell\ell'})$. Simulations with a smaller number of replications show that results were similar.

The asymptotic variance of each estimator is the usual GMM asymptotic variance. To be precise, let 
\begin{align*}
    \hat{M} &= \frac{1}{n}\sum_{\ell =1}^{L}\sum_{i\in I_{\ell }} \left(D_i - \hat{V}_{i\ell}) \right)^2, \\
    \hat{\Psi}_{DML} &= \frac{1}{n}\sum_{\ell =1}^{L}\sum_{i\in I_{\ell }} \left( Y_i - \hat{h}_{Y,\ell}(\hat{V}_{i\ell}) - \tilde{\theta}_\ell (D_i - \hat{V}_{i\ell}) \right)^2 \left(D_i - \hat{V}_{i\ell}\right)^2, \text{and} \\ 
    \hat{\Psi}&= \frac{1}{n}\sum_{\ell =1}^{L}\sum_{i\in I_{\ell }} \left( Y_i - \hat{h}_{Y,\ell}(\hat{V}_{i\ell}) - \tilde{\theta}_\ell (D_i - \hat{V}_{i\ell}) +\hat{\alpha}_{1\ell}(Z_i) \right)^2\left(D_i - \hat{V}_{i\ell}\right)^2,
\end{align*}
where  $\tilde{\theta}_\ell$ is the DML estimator that only uses observations not in $I_\ell$. The estimator of the asymptotic variance of the DML estimator is $\hat{\Psi}_{DML}/\hat{M}^2$. The estimator of the asymptotic variance of the three-step debiased estimator is $\hat{\Psi}/\hat{M}^2$. The estimator of the asymptotic variance in the PI-CAV procedure is $\hat{\Psi}/\hat{M}^2$.

\subsubsection{Results}

Table~\ref{tab:MC_ATE} reports the results for the two specifications (logistic and normal $\nu$) and a series of sample sizes ($n\in \{100, 500, 1000\}$). For each specification and sample size, we conducted $2000$ replications. We present the mean bias of each estimator (Mean Bias), the mean estimated asymptotic standard deviation (Asymp. SD), the standard error of the estimators (Std. Error), and the coverage (Coverage) of a 95\% confidence interval.

\begin{table}[!htb]
\centering
   \begin{tabular}{@{}lccccccccc@{}}
    \toprule
    \multicolumn{1}{c}{} & \multicolumn{2}{c}{Mean Bias}                          & \multicolumn{2}{c}{Asymp. SD}                     & \multicolumn{2}{c}{Std. Error}                            & \multicolumn{3}{c}{Coverage (\%)}                 \\ \cmidrule(lr{0.5em}){2-3}\cmidrule(lr{0.5em}){4-5}\cmidrule(lr{0.5em}){6-7}\cmidrule(lr{0.5em}){8-10}
    \multicolumn{1}{c}{} & \multicolumn{1}{c}{DML} & \multicolumn{1}{c}{3SD} & \multicolumn{1}{c}{DML} & \multicolumn{1}{c}{3SD} & \multicolumn{1}{c}{DML} & \multicolumn{1}{c}{3SD} & \multicolumn{1}{c}{DML} & \multicolumn{1}{c}{PI-CAV} & \multicolumn{1}{c}{3SD} \\ \midrule
    \multicolumn{9}{l}{Logistic $\nu$}                                                                                                                                                                                  \\
    $n=100$              & 0.323                   & 0.045                   & 0.322                   & 0.174                   & 0.215                   & 0.188                   & 88.9                    & 54.8                         & 92.0                    \\
$n=500$              & 0.076                   & 0.004                   & 0.144                   & 0.073                   & 0.074                   & 0.074                   & 99.5                    & 81.1                         & 93.9                    \\
$n=1000$             & 0.038                   & -0.002                  & 0.102                   & 0.051                   & 0.052                   & 0.051                   & 100                     & 88.4                         & 94.5                    \\ \midrule
    \multicolumn{9}{l}{Normal $\nu$}                                                                                                                                                                                    \\
    $n=100$              & 0.334                   & 0.043                   & 0.316                   & 0.170                   & 0.233                   & 0.180                   & 86.1                    & 51.9                         & 92.1                    \\
$n=500$              & 0.086                   & 0.004                   & 0.140                   & 0.071                   & 0.073                   & 0.071                   & 99.1                   & 75.8                         & 94.3                    \\
$n=1000$             & 0.045                   & -0.000                  & 0.099                   & 0.050                   & 0.051                   & 0.050                   & 99.9                   & 84.2                         & 95.3                   \\ \bottomrule
    \end{tabular}
\caption{Simulation results for the Hd-PS regression adjustment estimators in a partially linear model framework. Number of replications is $2000$. \textit{DML} = Double/Debiased machine learning estimator, \textit{3SD} = Three-Step debiased estimator, and \textit{PI-CAV} = Plug-in plus correct asymptotic variance procedure (Mean Bias and Std. Error equal to DML, Asymp. SD equal to 3SD).}
\label{tab:MC_ATE}
\end{table}

\subsection{CASF with a control-function approach}

\subsubsection{Setup}

The available data are $(Y, D, Z)$, with $Z \equiv (Z_j)_{j=1}^6$. Variables $D$ and $Y$ are generated by the following linear models:
\begin{align*}
	Y &= \sum_{k=1}^5 Z_k + 2D + U \text{ and } \\
	D &= \sum_{k=1}^6 Z_k + V.
\end{align*}
Regressors $Z$ and errors have the following distribution: 
\begin{equation*}
	(Z,U,V)\sim N\left(0, \begin{bmatrix}
		\operatorname{Id}_{6} & 0 & 0 \\
		0 & 1 & 1/2 \\
		0 & 1/2 & 1
	\end{bmatrix}\right),
\end{equation*}
where $\operatorname{Id}_6$ denotes the $6\times6$ Identity Matrix. Note that the fact that $Z \perp U$ and $Z \perp V$ guarantees that the Control Function Assumption is satisfied. In addition, $Z_6$ is excluded from the structural equation (i.e., it does not directly affect $Y$) and may be used as an instrument. In this case, $X = ( Z_1, \dots, Z_5, D)$. 

We estimate the CASF for the following counterfactual distribution $F^*$: (i) the distribution of $(Z_1,\dots, Z_5)$ remains unchanged and (ii) $D$ is normal with mean 1 (instead of 0) and the same variance as in the DGP. Therefore, the true parameter is $\theta_0=2$.

We evaluate the performance of three estimation procedures:
\begin{itemize}
    \item The the plug-in (PI) estimator $\hat{\theta}_{PI}$, given in equation~\eqref{eq:CASF_plugin}, which uses the original moment condition. The corresponding asymptotic variance estimator is also based on the original moment condition (see below).
    \item The DML estimator $\hat{\theta}_{DML}$, given in equation~\eqref{eq:CASF_DML}, which corrects for the second step. The corresponding estimator of the asymptotic variance only corrects for the second step (see below). 
    \item The three-step debiased (3SD) estimator $\hat{\theta}$, given in equation~\eqref{eq:CASF_D}. The corresponding estimator of the asymptotic variance corrects for both steps (see below).
    \item A Plug-in plus correct asymptotic variance (PI-CAV) procedure that uses the plug-in estimator to recover the parameter, but accounts for both steps when estimating the asymptotic variance (see below).
\end{itemize}

For all estimation procedures, we use Monte Carlo integration, with a sample of size $S=10^7$, to solve the integrals w.r.t. $F^*$. We set the number of cross-fitting partitions to $L=10$. The estimators for the nuisance parameters $g_0$ and $h_0$ are Lasso using the following dictionaries with linear terms: $\mathbf{c}_6(z) = (z_j)_{j=1}^6$ and $\mathbf{b}_7(x, v) = (z_1, \dots, z_5, d, v)$. The Lasso tuning parameter for both steps is chosen by cross-validation with 10 folds.

The asymptotic variance of each estimator is the usual GMM asymptotic variance. To be precise, recall that $\hat{V}_{i\ell}=D_i - \hat{g}_\ell(Z_i)$. The estimators of the asymptotic variances are
\begin{align*}
    \hat{\Psi}_{PI} &= \frac{1}{n} \sum_{\ell =1}^{L}\sum_{i\in I_{\ell }} \left( \frac1S \sum_{s=1}^S  \hat{h}_\ell(X_s^*, \hat{V}_{i\ell}) \right)^2 - \hat{\theta}_{PI}^2, \\
    \hat{\Psi}_{DML} &= \frac{1}{n} \sum_{\ell =1}^{L}\sum_{i\in I_{\ell }} \left( \frac1S \sum_{s=1}^S  \hat{h}_\ell(X_s^*, \hat{V}_{i\ell}) +  \hat{\alpha}_{2\ell}(X_i, \hat{V}_{i\ell}) (Y_i - \hat{h}_\ell(X_i, \hat{V}_{i\ell})) \right)^2 - \hat{\theta}_{DML}^2, \text{ and} \\
    \hat{\Psi} &= \frac{1}{n} \sum_{\ell =1}^{L}\sum_{i\in I_{\ell }} \left( \frac1S \sum_{s=1}^S  \hat{h}_\ell(X_s^*, \hat{V}_{i\ell}) + \hat{\alpha}_{1\ell}(Z_i) (D_i - \hat{g}_\ell(Z_i)) +  \hat{\alpha}_{2\ell}(X_i, \hat{V}_{i\ell}) (Y_i - \hat{h}_\ell(X_i, \hat{V}_{i\ell})) \right)^2 - \hat{\theta}^2,
\end{align*}
for the plug-in, DML, and three-step debiased estimators, respectively.  The estimator of the asymptotic variance in the PI-CAV procedure is $\hat{\Psi}$.

\subsubsection{Results}
Table~\ref{tab:MC_CASF} reports the results for a series of sample sizes ($n\in \{100, 500, 1000\}$). For each sample size, we conducted $2000$ replications. We present the mean bias of each estimator (Mean Bias), the mean estimated asymptotic standard deviation (Asymp. SD), the standard error of the estimators (Std. Error), and the coverage (Coverage) of a 95\% confidence interval.

\begin{table}[!htb]
	\centering
    \resizebox{\textwidth}{!}{
	\begin{tabular}{@{}lccccccccccccc@{}}
    \toprule
         & \multicolumn{3}{c}{Mean Bias} & \multicolumn{3}{c}{Asymp. SD} & \multicolumn{3}{c}{Std. Error} & \multicolumn{4}{c}{Coverage (\%)} \\ \cmidrule(lr{0.5em}){2-4} \cmidrule(lr{0.5em}){5-7} \cmidrule(lr{0.5em}){8-10} \cmidrule(lr{0.5em}){11-14}
         & PI       & DML      & 3SD     & PI       & DML      & 3SD     & PI       & DML      & 3SD      & PI     & PI-CAV     & DML    & 3SD    \\ \midrule
100  & 0.214   & 0.065    & 0.089    & 0.028       & 0.128        & 0.131        & 0.153 & 0.165  & 0.152  & 15.4   & 59.4               & 82.4    & 82.9    \\
500  & 0.141   & -0.019   & 0.013    & 0.015       & 0.059        & 0.061        & 0.063 & 0.068  & 0.063  & 4.3    & 37.2               & 90.6    & 92.4    \\
1000 & 0.116   & -0.027   & 0.003    & 0.011       & 0.042        & 0.043        & 0.045 & 0.048  & 0.045  & 2.4    & 24.5               & 86.5    & 94.1  \\ \bottomrule
\end{tabular}
}
	\caption{Simulation results for the CASF estimators with a Control-Function Approach. Number of replications is $2000$. \textit{PI} = Plug-in estimator, \textit{DML} = Double/Debiased machine learning estimator, \textit{3SD} = Three-Step debiased estimator, and  \textit{PI-CAV} = Plug-in plus correct asymptotic variance procedure (Mean Bias and Std. Error equal to PI, Asymp. SD equal to 3SD).}
	\label{tab:MC_CASF}
\end{table}

%%%%%%%%%%%%%%%%%%%%%%%%%%%%%%%%%%%%%%%%%%%%%%%%%%%%%%%%%%%%%%%%%%%%%%%%%%%%%%%%%%%%%%%%%%%%%%%
%%%%%%%%%%%%%%%%%%%%%%%%%%%%%%%%%%%%%%%%%%%%%%%%%%%%%%%%%%%%%%%%%%%%%%%%%%%%%%%%%%%%%%%%%%%%%%%%%%%%%%
\clearpage
\section{Regularity conditions for Assumption~\ref{ass:inclusion}}
\label{sec:app_inclusion} 

When the second step is
nonparametric or partialy linear, Assumption~\ref{ass:inclusion} reduces to
square-integrability conditions. For the nonparametric case ($%
\Delta_2(g)=L_2(X, V(g))$), the assumption requires that $\alpha_{02} \in
L_2(X, V(g_\tau))$ and $h(F_0, g_\tau) \in L_2(X, V)$ for small $\tau$. That
is, for all $0\leq\tau<\varepsilon$,
\begin{equation*}
\int \alpha_{02}(x, \varphi(d,z,g_\tau))^2 dF_0(w) <\infty \text{ and } \int
h(F_0, g_\tau)(x,\varphi(d,z,g_0))^2 dF_0(w)<\infty.
\end{equation*}

Assumption~\ref{ass:inclusion} also leads to similar requirements in the partially linear model, where $\Delta_2(g) = \{\beta^{\prime }x+\kappa(v)\colon
\beta\in\mathbb{R}^{p}, \kappa\in L_2(V(g))\}$. Note that, since $%
\alpha_{02}\in\Delta_2(g_0)$, we have that $\alpha_{02}(x,v)=\beta_0^{\prime
}x+\kappa_0(v)$ for $\beta_0\in\mathbb{R}^p$ and $\kappa_0\in L_2(V)$. Then $%
\alpha_{02}\in \Delta_2(g_\tau) \iff \kappa_0 \in L_2(V(g_\tau))$. In turn,
since $h(F_0,g_\tau)\in \Delta_2(g_\tau)$, we have that $h(F_0,g_\tau)(x,v)=%
\beta_\tau^{\prime }x+\kappa_\tau(v)$ for $\beta_\tau\in\mathbb{R}^p$ and $%
\kappa_\tau\in L_2(V(g_\tau))$. Therefore, Assumption~\ref{ass:inclusion:h}
imposes $\kappa_\tau \in L_2(V)$.

The next proposition gives primitive sufficient conditions for Assumption~%
\ref{ass:inclusion:alpha}:

\begin{prop}
{\label{prop:conditions_inclusion_alpha}} Under Assumption~\ref%
{ass:diff_h_alpha_phi}, if $\partial\alpha_{02}/\partial v$ is almost surely bounded,
then Assumption~\ref{ass:inclusion:alpha} is satisfied in the nonparametric
and partialy linear cases.
\end{prop}

We can give sufficient conditions for Assumption~\ref{ass:inclusion:h} in
terms of smoothness of the distribution of the data as $g_\tau$ approaches $%
g_0$. Let $F^{xv}_\tau$ and $F^{v}_\tau$ be the distributions of $(X,
V(g_\tau))$ and $V(g_\tau)$, respectively. Note that $F^{xv}_0$ and $F^v_0$
denote the distributions of $(X,V)$ and $V$, respectively. Then:

\begin{prop}
{\label{prop:conditions_inclusion_h}} Assumption~\ref{ass:inclusion:h} is
satisfied in the nonparametric case under the following conditions: \mbox{}
\\[-20pt]

\begin{enumerate}
\item $\mathbb{E}[Y^4]<\infty$,

\item there exists an $\varepsilon>0$ such that, for $\tau<\varepsilon$, $%
F^{xv}_\tau$ and $F^{xv}_0$ are equivalent measures (absolutely continuous
between each other), and

\item $\mathbb{E}[\nu_\tau(X,V)]<\infty$, being $\nu_\tau$ the Radon-Nikodym
density of $F^{xv}_0$ w.r.t. $F^{xv}_{\tau}$.
\end{enumerate}

Moreover, Assumption~\ref{ass:inclusion:h} is satisfied in the partialy linear
case if Condition~1 is replaced by: Condition~1$^{\ast }$. $\E[Y^rX^s]%
<\infty $, for every $s,r\in \mathbb{N}$ satisfying $s+r=4$, and
Conditions~2-3 hold with $F_{\tau }^{v}$ replacing $F_{\tau }^{xv}$.
\end{prop}

\clearpage

%%%%%%%%%%%%%%%%%%%%%%%%%%%%%%%%%%%%%%%%%%%%%%%%%%%%%%%%%%%%%%%%%%%%%%%%%%%%%%%%%%%%%%%%%%%%%%%%%%%%%%
%%%%%%%%%%%%%%%%%%%%%%%%%%%%%%%%%%%%%%%%%%%%%%%%%%%%%%%%%%%%%%%%%%%%%%%%%%%%%%%%%%%%%%%%%%%%%%%%%%%%%%
%%%%%%%%%%%%%%%%%%%%%%%%%%%%%%%%%%%%%%%%%%%%%%%%%%%%%%%%%%%%%%%%%%%%%%%%%%%%%%%%%%%%%%%%%%%%%%%%%%%%%%
\section{Verification of rates on derivatives for Lasso estimators}
\label{sec:lasso_derivatives}

This section provides primitive conditions under which
\begin{equation} \label{eq:deriv_rates}
    \norm{\partial\hat{h}_\ell/\partial v - \partial h_0 / \partial v}_2 \norm{\hat{g}_\ell - g_0}_2=o_p(n^{-1/2}) 
\text{and} 
\norm{\partial\hat{\alpha}_{2\ell}/\partial v - \partial \alpha_{02} /\partial v}_2 
\norm{\hat{g}_\ell - g_0}_2=o_p(n^{-1/2})
\end{equation}
hold when the second-step nuisance functions are estimated by Lasso on the dictionary
$\mathbf b_J(x,v)$. Note that $\alpha_{02}$ is also estimated by a Lasso on the same dictionary.

Recall that the estimators in Section~\ref{sec:CASF_details} take the form
\[
\hat h_\ell(x,v)=\mathbf b_J(x,v)'\widehat{\boldsymbol{\eta}}_\ell
 \text{ and }
\hat\alpha_{2\ell}(x,v)=\mathbf b_J(x,v)'\widehat{\boldsymbol{\rho}}_\ell.
\]
Therefore, if the atoms in $\mathbf{b}_J$ are differentiable,
\[
\frac{\partial \hat h_\ell}{\partial v}(x,v)
=
\left(\frac{\partial \mathbf b_J}{\partial v}(x,v)\right)'\widehat{\boldsymbol{\eta}}_\ell
\text{ and }
\frac{\partial \hat\alpha_{2\ell}}{\partial v}(x,v)
=
\left(\frac{\partial \mathbf b_J}{\partial v}(x,v)\right)'\widehat{\boldsymbol{\rho}}_\ell.
\]

Let $\boldsymbol\eta_{0J}$ and $\boldsymbol\rho_{0J}$ denote sparse coefficient vectors approximating
$h_0$ and $\alpha_{02}$, respectively, on the dictionary $\mathbf b_J$. Then
\begin{align*}
\frac{\partial \hat h_\ell}{\partial v}
-
\frac{\partial h_0}{\partial v}
&=
\frac{\partial \mathbf b_J'}{\partial v}(\widehat{\boldsymbol\eta}_\ell-\boldsymbol\eta_{0J})
+
\left(
\frac{\partial \mathbf b_J'}{\partial v}\boldsymbol\eta_{0J}
-
\frac{\partial h_0}{\partial v}
\right).
\end{align*}

The second term corresponds to the derivative approximation error
\[
a_h^\partial \equiv
\inf_{\boldsymbol\eta\in \mathbb{R}^J}
\left\|
\frac{\partial h_0}{\partial v}
-
\frac{\partial \mathbf b_J'}{\partial v}\boldsymbol\eta
\right\|_2.
\]

For the first term, we assume that the Gram matrices
\[
Q_J\equiv\E[\mathbf b_J(X,V)\mathbf b_J(X,V)'] 
\text{ and }
Q_J^\partial\equiv\E\!\left[
\left(\frac{\partial \mathbf b_J}{\partial v}(X,V)\right)\left(\frac{\partial \mathbf b_J}{\partial v}(X,V)\right)'
\right]
\]
have eigenvalues bounded away from zero and infinity uniformly in $J$.
Assume also that the dictionary satisfies the derivative bound
\[
\sup_{x,v}\left\|
\frac{\partial \mathbf b_J}{\partial v}(x,v)
\right\|_\infty \leq C \sqrt{J}.
\]

Given these conditions, under the standard Lasso coefficient rates in \citet{bickel2009simultaneous}, that is
\[
\|\widehat{\boldsymbol\eta}_\ell-\boldsymbol\eta_{0J}\|_1
=
O_p\!\left(s_h\sqrt{\frac{\log J}{n}}\right),
\]
it follows that
\[
\left\|
\frac{\partial \hat h_\ell}{\partial v}
-
\frac{\partial h_0}{\partial v}
\right\|_2
=
O_p\!\left(
a_h^\partial+s_h\sqrt{\frac{\log J}{n}}
\right).
\]
An identical argument yields
\[
\left\|
\frac{\partial \hat\alpha_{2\ell}}{\partial v}
-
\frac{\partial \alpha_{02}}{\partial v}
\right\|_2
=
O_p\!\left(
a_\alpha^\partial+s_\alpha\sqrt{\frac{\log J}{n}}
\right).
\]
Therefore the rates in equation~\eqref{eq:deriv_rates} hold provided that
\[
\left(
a_h^\partial+s_h\sqrt{\frac{\log J}{n}}
\right)
\|\hat g_\ell-g_0\|_2=o_p(n^{-1/2}),
\]
and
\[
\left(
a_\alpha^\partial+s_\alpha\sqrt{\frac{\log J}{n}}
\right)
\|\hat g_\ell-g_0\|_2=o_p(n^{-1/2}).
\]
Assumption~\ref{ass:gen_reg:continuous_deriv} follows from equation~\eqref{eq:deriv_rates} and some boundedness conditions (see the proof for the CASF in Proposition~\ref{prop:CASF_conditions}). 

\subsection{Derivative approximation errors}

We now give primitive conditions under which the derivative approximation errors
$a_h^\partial$ and $a_\alpha^\partial$ vanish sufficiently fast. We consider different dictionaries $\mathbf{b}_J$ for the second step.

\paragraph{Power series.}
Let $\mathbf b_J(x,v)$ consist of tensor products of power series,
\[
\mathbf b_J(x,v)=\{x^{k_1}v^{k_2}:0\le k_1+k_2\le M\},
\]
so that $J$ is of the order of $M^2$. If $h_0$ and $\alpha_{02}$ belong to a H\"older
class $C^s$, with $s>1$, classical approximation
results imply
\[
\inf_{\boldsymbol\eta}\|h_0-\mathbf b_J'\boldsymbol\eta\|_2=O(J^{-s/2})
\text{ and }
a_h^\partial=O(J^{-(s-1)/2}).
\]
The same rate holds for $\alpha_{02}$. These bounds follow from standard
approximation and inverse inequalities for polynomial series \citep[see][Ch.~6]{schumaker2007spline}.

\paragraph{Spline dictionaries.}

Let $\mathbf b_J(x,v)$ consist of tensor products of B-splines of order $k$
with $J_x$ and $J_v$ knots. If $h_0$ and $\alpha_{02}$ belong to a H\"older class
$C^s$, with $1<s\le k$, spline approximation
theory implies
\[
\|h_0-\mathbf b_J'\boldsymbol\eta\|_2=O(J^{-s/2})
\text{ and }
a_h^\partial=O(J^{-(s-1)/2}),
\]
with analogous bounds for $\alpha_{02}$. Moreover, B-spline bases satisfy
\[
\sup_{x,v}
\left\|
\frac{\partial \mathbf b_J}{\partial v}(x,v)
\right\|_\infty
\leq C \sqrt{J},
\]
which ensures the derivative stability conditions assumed above \citep[see][Th.~4.22 and 6.25]{schumaker2007spline}.

\subsection{Examples of first-step estimators}

The rate condition also depends on the convergence rate of the first-step
estimator $\hat g_\ell$.

\paragraph{Lasso or sparse series estimators.}

Suppose $g_0(z)$ admits a sparse representation on a dictionary
$c_K(z)$ with sparsity $s_g$. Under restricted eigenvalue conditions, following \citet{bickel2009simultaneous},

\[
\|\hat g_\ell-g_0\|_2
=
O_p\!\left(\sqrt{\frac{s_g\log K}{n}}\right).
\]

\paragraph{Deep neural networks.}

If $g_0$ belongs to a compositional H\"older class with smoothness $s_g$
and intrinsic dimension $d_g$, deep ReLU networks achieve
\[
\|\hat g_\ell-g_0\|_2
=
O_p\!\left(n^{-\kappa}\right),
\text{ with }
\kappa=\frac{s_g}{2s_g+d_g},
\]
under standard regularity conditions \citep{farrell2021deep}.

\paragraph{Random forests and boosted trees.}

Under Lipschitz smoothness and sparsity of the relevant covariates,
forest and boosting estimators satisfy
\[
\|\hat g_\ell-g_0\|_2
=
O_p\!\left(n^{-\kappa}\right),
\text{ with }
\kappa=\frac{1}{2+d_g},
\]
for effective dimension $d_g$ \citep{wager2018estimation}.

\subsection{Primitive sufficient conditions}

Combining the bounds above, equation~\eqref{eq:deriv_rates}
holds whenever
\[
\left(
J^{-(s-1)/2}
+
s_h\sqrt{\frac{\log J}{n}}
\right)
\|\hat g_\ell-g_0\|_2=o_p(n^{-1/2}),
\]
and similarly for $\alpha_{02}$. For example, if $\|\hat g_\ell-g_0\|_2=O_p(n^{-1/3})$, which is the slowest rate allowed by Assumption~\ref{ass:est_convergence:estimators}, the condition holds for
sieve dimensions satisfying
\[
J = O\left(n^{\frac{1}{3(s-1)}}\right).
\]

%%%%%%%%%%%%%%%%%%%%%%%%%%%%%%%%%%%%%%%%%%%%%%%%%%%%%%%%%%%%%%%%%%%%%%%%%%%%%%%%%%%%%%%%%%%%%%%%%%%%%%
%%%%%%%%%%%%%%%%%%%%%%%%%%%%%%%%%%%%%%%%%%%%%%%%%%%%%%%%%%%%%%%%%%%%%%%%%%%%%%%%%%%%%%%%%%%%%%%%%%%%%%
\section{Proofs of the results}
\label{sec:Proofs}
%%%%%%%%%%%%%%%%%%%%%% Commented
\begin{comment}
\begin{proofc}[Lemma~\ref{lma:two_effects}]
	Applying the chain rule several times to $d \bar{m}(g(F_\tau), h(F_\tau, g(F_\tau)),\theta)/d\tau$, we have that:
	\begin{equation*}
		\frac{d}{d\tau}\bar{m}(g(F_\tau), h(F_\tau, g(F_\tau)),\theta) = \frac{d}{d\tau}\bar{m}(g(F_\tau), h_0,\theta)+\frac{d}{d\tau}\bar{m}(g_0, h(F_\tau, g(F_\tau)),\theta).
	\end{equation*}
	Then, using the chain rule again:
	\begin{equation*}
		\begin{aligned}
			\frac{d}{d\tau}\bar{m}(g_0, h(F_\tau, g(F_\tau)),\theta) &=\frac{d}{d\tau}\bar{m}(g_0, h(F_0, g(F_\tau)),\theta) \\
			&+\frac{d}{d\tau}\bar{m}(g_0, h(F_\tau, g_0),\theta).
		\end{aligned}
	\end{equation*}
	Combining the above equations leads to:
	\begin{equation}{\label{eq:app_three_effects}}
		\begin{aligned}
			\frac{d}{d\tau}\bar{m}(g(F_\tau), h(F_\tau, g(F_\tau)),\theta) &=\frac{d}{d\tau}\bar{m}(g(F_\tau), h_0,\theta) \\
			&+\frac{d}{d\tau}\bar{m}(g_0, h(F_0, g(F_\tau)),\theta)\\
			&+	\frac{d}{d\tau}\bar{m}(g_0, h(F_\tau, g_0),\theta).		
		\end{aligned}
	\end{equation}
	
	Now, note that by the chain rule we have that:
	\begin{equation*}
		\begin{aligned}
			\frac{d}{d\tau}\bar{m}(g(F_\tau), h_0,\theta)&+\frac{d}{d\tau}\bar{m}(g_0, h(F_0, g(F_\tau)),\theta) \\
			&=\frac{d}{d\tau}\bar{m}(g(F_\tau), h(F_0, g(F_\tau)),\theta).
		\end{aligned}
	\end{equation*}
	Hence the first two terms in equation~\eqref{eq:app_three_effects} equal the derivative of $\bar{m}(g(F_\tau), h(F_0, g(F_\tau)),\theta)$.
\end{proofc}
\end{comment}

\begin{proofc}[Proposition~\ref{prop:zero_indirect}]
    The proposition is an application of the functional chain rule \citep[see][Property~1.2.8]{yamamuro2006differential}. If $h \mapsto \bar{m}(g_0, h)$ and $g \mapsto h(F_0, g)$ are Hadamard differentiable, then $g \mapsto \bar{m}(g_0, h(F_0, g))$ is Hadamard differentiable with derivative $D_{02} \circ D_h$, where $\circ$ denotes composition of functions. The functions $D_{02}$ and $D_h$ are the derivatives of $h \mapsto\bar{m}(g_0, h)$ at $h_0$ and $g \mapsto h(F_0, g)$ at $g_0$, respectively. We have that $D_{02} \circ D_h = 0$ since $D_{02}=0$ by hypothesis. Thus, the result follows from the characterization of Hadamard differentiation in \citet[Property~ 1.2.7]{yamamuro2006differential}. 
\end{proofc}

\begin{proofc}[Proposition~\ref{prop:second_step}]
	We derive the result for a one-dimensional $S$. Note that $\E[D_{02}(W,h)]$ is a linear and continuous functional in the Hilbert space $L_2(X,V)$, with $V\equiv\varphi(D,Z,g_0)$. Thus, by the Riesz Representation Theorem, there exists a $r_{02}$ such that $\E[D_{02}(W,h)]=\E[r_{02}(X,V)h(X,V)]$.  Then, for the (differentiable) path $\tau \mapsto h(F_\tau, g_0)$, by Assumption~\ref{ass:Diff_m_h},
	\begin{equation} \label{eq:Rr_{02}ndstep}
		\frac{d}{d\tau} \bar{m}(g_0, h(F_\tau, g_0)) = \frac{d}{d\tau}\E[r_{02}(X,V)h(F_\tau, g_0)(X,V)],
	\end{equation}
	where $h(F,g)(x,v)$ denotes $h(F,g)$ evaluated at $(x,v)$. This is Assumption~1 in \citet{ichimura2022influence}. Since Assumption~2 in that paper is satisfied in our setup, Proposition~1 in \citet{ichimura2022influence} gives: $\phi_2(w,g_0,h_0,\alpha_{02})=\alpha_{02}(x,\varphi(d,z,g_0))\{s-h_0(x,\varphi(d,z,g_0))\}$. The nuisance $\alpha_{02}$ is the $L_2$-projection of $r_{02}$ onto $\Delta_{2}(g_0)$:
	\begin{equation*}
		\alpha_{02} = \argmin_{\alpha\in\Delta_{2}(g_0)} \E[(r_{02}(X,V)-\alpha(X,V))^2].
	\end{equation*}

    In case $\func{dim}(S)>1$, the Influence Function $\phi_2$ is the sum of each individual Influence Function (which may be derived by the above procedure). Thus:
    \begin{equation*}
        \phi_2(w,g_0,h_0,\alpha_{02})=\alpha_{02}(x,\varphi(d,z,g_0))'\{s-h_0(x,\varphi(d,z,g_0))\},
    \end{equation*}
    where $\alpha_{02}$, $s$, and $h_0$ are $\func{dim}(S) \times 1$ vectors.

    %	We now show that, necessarily, $\alpha_{02}\in L_2(g_0)\equiv \{(d,z)\mapsto \delta(x,\varphi(d,z, g)) \colon \delta\in L_2(X,V)\}$. Note that $r_{02}\in L_2(g_0)$. Moreover, since $L_2(g_0)$  is a linear and closed subspace of $L_2(D,Z)$, by \citet[Th.~1 in Sec.~3.4]{luenberger1997optimization}, for every $\alpha\in\Delta_{2}(g_0)$ we have the decomposition $\alpha=m+m^\perp$, with $m\in L_2(g_0)$ and $m^\perp\in L_2(g_0)^\perp$, the orthogonal complement of $L_2(g_0)$. Therefore, for every $\alpha\in\Delta_{2}(g_0)$,
    %	\begin{equation*}
%		\lVert r_{02} - \alpha \rVert^2 = \lVert r_{02} - m - m^\perp \rVert^2 = \lVert r_{02} - m \rVert^2 + \lVert m^\perp \rVert^2 \geq \lVert r_{02} - m \rVert^2.
%	\end{equation*}
%	Note that $\lVert \delta \rVert^2=\E[\delta(D,Z)^2]$ for every $\delta\in L_2(D,Z)$. The above result uses that $r_{02}-m \in L_2(g_0)$ and Pitagoras' Theorem \citep[Lemma~1 in Sec.~3.3]{luenberger1997optimization}. Since equality is achieved when $m^\perp=0$, we have that $\lVert r_{02} - \alpha \rVert^2$ is minimized for an $\alpha\in\Delta_{2}(g_0)\cap L_2(g_0)$.
\end{proofc}

%%%%%%%%%%%%%%%%%%%%%%%%%%%%%%%%%%%%%%%%%%%%%%%%%%%%%%%%%%%%%%%%%%%%%%%%%%%%%%%%%%%%%%%%%%%%%%%%%%%%%%%%%%%%%%%%
\begin{proofc}[Lemma~\ref{lma:generalized_HR}]
    We proceed as in \citet[Lma.~1]{hahn2013asymptotic}. Let $\tau\mapsto g_\tau$ be a differentiable path. For any function $\delta_2\in\Delta_{2}(g_\tau)$,  we have that
	\begin{equation*}
		\E[\delta_{2}(X,V(g_\tau))\cdot \{S - h(F_0, g_\tau)(X,V(g_\tau))\}]=0.
	\end{equation*}
	This is the orthogonality condition that defines $h(F_0, g_\tau)$ (it is equation~\eqref{eq:ortho_genFg} for $(F,g)=(F_0,g_\tau)$). If $\delta_2 \in \Delta_2(g_\tau)$ when $\tau<\varepsilon$, we can take derivatives in the above equation. Thus, applying the chain rule, we get
	\begin{equation*}
		\begin{aligned}
			\frac{d}{d\tau}\E[\delta_2(X, V)\cdot h(F_0, g_\tau)(X,V)]&= -\frac{d}{d\tau}\E[\delta_{2}(X,V)\cdot h_0(X,V(g_\tau))]\\
			&+ \frac{d}{d\tau}\E\left[\delta_{2}(X,V(g_\tau))\cdot(S - h_0(X,V))\right].
		\end{aligned}
	\end{equation*}
\end{proofc}

%%%%%%%%%%%%%%%%%%%%%%%%%%%%%%%%%%%%%%%%%%%%%%%%%%%%%%%%%%%%%%%%%%%%%%%%%%%%%%%%%%%%%%%%%%%%%%%%%%%%%%%%%%%%%%%%
\begin{proofc}[Theorem~\ref{thm:first_step}]	
	We compute $d\bar{m}(g(F_\tau), h_0)/d\tau$ and $d\bar{m}(g_0, h(F_0, g(F_\tau)))/d\tau$ separately and then add them according to equation~\eqref{eq:1st_step_decomposition}. By Assumptions~\ref{ass:Diff_m_h}, \ref{ass:Diff_m_g}, and \ref{ass:inclusion:h}, we have that for the differentiable paths $\tau\mapsto g(F_\tau)$ and $\tau\mapsto h(F_0,g(F_\tau))$:
	\begin{equation}{\label{eq:lin_g_proof}}
		\frac{d}{d\tau} \bar{m}(g(F_\tau), h_0) = \frac{d}{d\tau}\E[D_{dir}(W,g(F_\tau))]=\frac{d}{d\tau}\E[r_{dir}(Z)g(F_\tau)(Z)],
	\end{equation}
	where $r_{dir}$ is the Riesz representer of $\E[D_{dir}(W,g)]$, and for $r_{02}$ as in equation~\eqref{eq:Rr_{02}ndstep},
	\begin{equation*}
		\frac{d}{d\tau} \bar{m}(g_0, h(F_0, g(F_\tau))) = \frac{d}{d\tau}\E[D_{02}(W,h(F_0, g(F_\tau)))]=\frac{d}{d\tau}\E[r_{02}(X, V)'h(F_0, g(F_\tau))(X,V)].
	\end{equation*}
	In these equations, $g(F)(z)$ means $g(F)$ evaluated at $z$, and $h(F,g)(x,v)$ means $h(F,g)$ evaluated at $(x,v)$. Note that we have linearized the moment condition with respect to each component of $h(F_0, g(F_\tau))$, in case $\func{dim}(S)>1$. Note that by Assumption~\ref{ass:inclusion:h}, $h(F_0, g(F_\tau))\in \Delta_2(g_0)$. This means that $h(F_0, g(F_\tau))$ is orthogonal to $r_{02}-\alpha_{02}$ (since $\alpha_{02}$ is the $L_2$-projection of $r_{02}$ onto $\Delta_2(g_0)$). Then, we can write:
    \begin{equation*}
        \frac{d}{d\tau} \bar{m}(g_0, h(F_0, g(F_\tau))) = \frac{d}{d\tau}\E[\alpha_{02}(X, V)'h(F_0, g(F_\tau))(X,V)].
    \end{equation*}
    
    By Assumption~\ref{ass:inclusion:alpha} we can apply Lemma~\ref{lma:generalized_HR} to the RHS in the above equation to get:
    \begin{equation*}
		\begin{aligned}
			\frac{d}{d\tau} \bar{m}(g_0, h(F_0, g(F_\tau))) = \frac{d}{d\tau}\E\left[\alpha_{02}(X,\varphi(D,Z,g(F_\tau)))'\{S-h_0(X,\varphi(D,Z,g(F_\tau)))\}\right].
		\end{aligned}
	\end{equation*}
	Under Assumption~\ref{ass:diff_h_alpha_phi}, the RHS of the above display can be linearized in $g(F_\tau)$ to get
	\begin{equation}{\label{eq:lin_indirect_proof}}
		\begin{aligned}
			\frac{d}{d\tau} \bar{m}(g_0, h(F_0, g(F_\tau)))
			=\E\left[\left.\frac{\partial}{\partial v}\left\{\alpha_{02}(X,v)' (S-h_0(X,v))\right\}\right\rvert_{v=V} D_\varphi g(F_\tau)(D,Z)\right],
		\end{aligned}	
	\end{equation}
    where $D_\varphi g(d,z)$ denotes $D_\varphi g$ evaluated at $(d,z)$. We have assumed that derivatives and expectations can be interchanged (we may impose some regularity conditions on $H$ such that this is possible). Since $D_\varphi$ is linear in $g$, the function inside the expectation in the RHS is linear in $g$.
    
	 We now use equation~\eqref{eq:1st_step_decomposition} to combine the results in equations \eqref{eq:lin_g_proof} and \eqref{eq:lin_indirect_proof}. This gives:
	\begin{equation*}
		\begin{aligned}
			\frac{d}{d\tau} \bar{m}(g(F_\tau), h(F_0, g(F_\tau)))=\frac{d}{d\tau}  \E & \biggl[  D_{dir}(W,g(F_\tau)) \\
			&\left. \left. +\frac{\partial}{\partial v}\left\{\alpha_{02}(X,v)' (S-h_0(X,v))\right\}\right\rvert_{v=V} D_\varphi g(F_\tau)(D,Z) \right],
		\end{aligned}
	\end{equation*}
	which gives the linearization result of the Theorem \textsc{(LIN)}.
	
	To find the shape of the IF, note that the adjoint $D_\varphi^*$ of $D_\varphi$ is defined by the equation $\E[\delta(D,Z)D_\varphi g(D,Z)]=\E[D_\varphi^*\delta(Z)g(Z)]$. Therefore, by the Law of Iterated Expectations in equation~\eqref{eq:lin_indirect_proof}, noting that $V\equiv\varphi(D,Z,g_0)$ is a function of $(D,Z)$:
	\begin{equation*}
		\begin{aligned}
			\frac{d}{d\tau} \bar{m}(g_0, h(F_0, g(F_\tau)))&=\frac{d}{d\tau}  \E\left[ \E\left[ \left. \left.\frac{\partial}{\partial v}\left\{\alpha_{02}(X,v)' (S-h_0(X,v))\right\}\right\rvert_{v=V} D_\varphi g(F_\tau)(D,Z) \right\lvert D, Z\right]\right] \\
			&=\E[\nu(D,Z)D_\varphi g(F_\tau)(D,Z)]=\E[D_\varphi^* \nu (Z)g(F_\tau)(Z)],
		\end{aligned}
	\end{equation*}
	with
	\begin{equation*}
		\nu(d,z)\equiv \left.\frac{\partial}{\partial v}\left\{\alpha_{02}(x,v)'(\E[S|D=d,Z=z]-h_0(x,v))\right\}\right\rvert_{v=\varphi(d,z,g_0)}.
	\end{equation*}
	
	Again, we can use equation~\eqref{eq:1st_step_decomposition} to combine this last result with that in equation~\eqref{eq:lin_g_proof}:
	\begin{equation*}
		\frac{d}{d\tau} \bar{m}(g(F_\tau), h(F_0, g(F_\tau)))=\frac{d}{d\tau}\E[\{r_{dir}(Z)+D_\varphi^* \nu (Z)\}g(F_\tau)(Z)].
	\end{equation*}
	This is Assumption~1 in \citet{ichimura2022influence}. Since Assumption~2 in that paper is guaranteed by our Assumption~\ref{rho}, Proposition~1 in \citet{ichimura2022influence} gives the shape of the IF: $\phi_1(w,g_0,\alpha_{01})=\alpha_{01}(z)\cdot \epsilon(w, g_0)$.
	The parameter $\alpha_{01}$ is a $L_2$-projection:
	\begin{equation}{\label{eq:alpha1_defi}}
		\alpha_{01} = \argmin_{\alpha\in\Delta_{1}} \E\left[- r_e(Z) \cdot \left(-\frac{r_{01}(Z)}{r_e(Z)}-\alpha(Z) \right)^2 \right],
	\end{equation}
	where $r_{01}=r_{dir}+D_\varphi^*\nu$.
\end{proofc}

%%%%%%%%%%%%%%%%%%%%%%%%%%%%%%%%%%%%%%%%%%%%%%%%%%%%%%%%%%%%%%%%%%%%%%%%%%%%%%%%%%%%%%%%%%%%%%%%%%%%%%%%%%%%%%%%%%%%%%%%%%%%%%%%
The asymptotic normality and consistent estimation of the asymptotic variance result in Theorem~\ref{thm:asymptotics} relies on the following lemmas:
%%%%%%%%%%%%%%%%
\begin{lma}
\label{lma:IF2v}
Assume that differentiation and integration are interchangeable. Moreover, $h, \alpha_2 \in \Delta_2(g_0)$ are almost surely twice continuously differentiable with respect to $v$ and the mapping $g \mapsto \varphi(d, z, g)$, from $\Delta_1$ to $L_2(D,Z)$, is twice Frechet differentiable with continuous second derivative. Then,
    \begin{equation*}
\resizebox{\textwidth}{!}{
$
       \left\lvert \E\left[ \phi_2(W, g, h, \alpha_2) - \phi_2(W, g_0, h, \alpha_2) - \frac{%
\partial }{\partial v}\left[ \alpha_{2}(x,v)' (s-h(x,v))%
\right] \cdot D_{\varphi }[g-g_0] \right] \right\rvert \leq C \cdot \norm{g-g_0}_2^2, $
}
    \end{equation*}
  where the derivative w.r.t. $v$ is evaluated at $v=\varphi(d,z, g_0)$ and $D_\varphi$ is the first derivative of $g \mapsto \varphi(d,z,g)$ at $g_0$.
\end{lma}

\begin{proofc}[Lemma~\ref{lma:IF2v}]
    Without loss of generality, we assume that $S$ has one dimension. To shorten notation, define $\varphi(g) \equiv \varphi(d,z, g)$ and $\bar{\phi}_2(g) \equiv \E[\phi_2(W, g, h, \alpha_2)]$. The Riesz representer of the derivative of $\bar{\phi}_2(g)$ at a point $g \in \Delta_1$ is:
    \begin{equation*}
        D_{\bar{\phi}_2}(g) = \left.\frac{\partial }{\partial v}\left[ \alpha_{2}(x,v)(s-h(x,v))\right] \right|_{v=\varphi(g)}  D_\varphi(g),
    \end{equation*}
    where $D_\varphi(g)$ is the derivative of $\varphi(g)$ at $g \in \Delta_1$. We can therefore find the second derivative of $\bar{\phi}_2(g)$, which is given by
    \begin{align*}
        D_{\bar{\phi}_2}^2(g)  =& \left. \frac{\partial }{\partial v}\left[ \alpha_{2}(x,v)(s-h(x,v))\right] \right|_{v=\varphi(g)}   D^2_\varphi(g) + \\
        &\left.\frac{\partial^2 }{\partial v^2}\left[ \alpha_{2}(x,v)(s-h(x,v))\right] \right|_{v=\varphi(g)}  (D_\varphi(g))^2,
    \end{align*}
    where $D^2_\varphi(g)$ is the second derivative of  $\varphi(g)$. The assumptions in the theorem guarantee continuity of $g \mapsto D_{\bar{\phi}_2}^2(g)$. Proposition~3 in \citet[p.~177]{luenberger1997optimization} yields the result.
\end{proofc}

%%%%%%%%%%%%%%%%
\begin{lma}
Under the conditions of Lemma~\ref{lma:IF2v}, being $\nu(h, \alpha_2) \equiv \partial/\partial v\{\alpha_2(x,v) \cdot (s - h(x,v)\}$  we have that
\label{lma:IF2lin}
    \begin{align*}
        \E\left[ \left(\phi_2(W, g, h, \alpha_2) - \phi_2(W, g_0, h, \alpha_2)\right) - \left(\phi_2(W, g, h_0, \alpha_{02}) - \phi_2(W, g_0, h_0, \alpha_{02})\right)\right] \leq \\
        C \cdot \norm{g-g_0}_2 \cdot \left(\norm{g-g_0}_2+\norm{\nu(h, \alpha_2)-\nu(h_0, \alpha_{02})}_2\right).
    \end{align*}
\end{lma}

\begin{proofc}[Lemma~\ref{lma:IF2lin}]
     Without loss of generality we assume that $S$ has one dimension. By the triangle inequality, we have that
     \begin{align*}
         \left|\E\left[ \left(\phi_2(W, g, h, \alpha_2) - \phi_2(W, g_0, h, \alpha_2)\right) - \left(\phi_2(W, g, h_0, \alpha_{02}) - \phi_2(W, g_0, h_0, \alpha_{02})\right)\right] \right| \leq \\
         \left|\E\left[ \phi_2(W, g, h, \alpha_2) - \phi_2(W, g_0, h, \alpha_2) - \frac{%
\partial }{\partial v}\left[ \alpha_{2}(x,v) (s-h(x,v))%
\right] \cdot D_{\varphi }[g-g_0] \right] \right| + \\
\left|\E\left[ \phi_2(W, g, h_0, \alpha_{02}) - \phi_2(W, g_0, h_0, \alpha_{02}) - \frac{%
\partial }{\partial v}\left[ \alpha_{02}(x,v) (s-h_2(x,v))%
\right] \cdot D_{\varphi }[g-g_0] \right] \right| + \\
\left|\E\left[ \left( \frac{%
\partial }{\partial v}\left[ \alpha_{2}(x,v) (s-h(x,v))%
\right] - \frac{%
\partial }{\partial v}\left[ \alpha_{02}(x,v) (s-h_0(x,v))%
\right] \right) \cdot D_{\varphi }[g-g_0] \right] \right|.
     \end{align*}
     The terms in the second and third row are bounded by $\norm{g-g_0}_2^2$ by Lemma~\ref{lma:IF2v} applied to $(h, \alpha_2)$ and $(h_0, \alpha_{02})$. By the Cauchy-Schwarz inequality and boundedness of $D_\varphi$, the term in the last row is bounded by $\norm{g-g_0}_2 \cdot \norm{\nu(h, \alpha_2)-\nu(h_0, \alpha_{02})}_2$.

\end{proofc}

%%%%%%%%%%%%%%%%
\begin{proofc}[Theorem~\ref{thm:asymptotics}]
We start with asymptotic normality of $	\sqrt{n}(\hat{\theta}-\theta_0)$. A relevant deviation from \citet{chernozhukov2022locally} is that the second-step influence function is evaluated at the generated regressor. The cornerstone of the asymptotic normality result is the asymptotic equivalence \citep[c.f.][Lemma~8]{chernozhukov2022locally}
    \begin{equation} \label{eq:asump_equiv}
		\sqrt{n}\hat{\psi}(\theta_0)=\frac{1}{\sqrt{n}}\sum_{i=1}^{n}\psi(W_i, g_0,h_0,\alpha_0,\theta_0)+o_p(1).
	\end{equation}
     To handle estimators evaluated at the generated regressor, we provide a novel expansion of $\sqrt{n}\hat{\psi}(\theta_0)-n^{-1/2}\sum_{i=1}^{n}\psi(W_i, g_0,h_0,\alpha_0,\theta_0) = n^{-1/2}\sum_{i=1}^n (\hat{\psi}_{i\ell}(\theta_0) - \psi(W_i, g_0,h_0,\alpha_{0},\theta_0))$. 
     
     As in \citet{chernozhukov2022locally}, we add and subtract $\phi_1(W_i, \hat{g}_\ell, \alpha_{01})$, $\phi_1(W_i, g_0, \hat{\alpha}_{1\ell})$, and $\phi_1(W_i, g_0, \alpha_{01})$. This allows to deal with the estimated first-step influence function. The second step influence function is more complex, as it is evaluated at the generated regressor. A novel expansion is needed to deal with it. In particular, we add and subtract: $\phi_2(W_i, g_0, \hat{h}_\ell, \alpha_{02})$,  $\phi_2(W_i, g_0, h_0, \hat{\alpha}_{2\ell})$, $\phi_2(W_i, \hat{g}_\ell, h_0, \alpha_{02})$, and $2\phi_2(W_i, g_0, h_0, \alpha_{02})$. This results in  
     \begin{equation*}
         \hat{\psi}_{i\ell}(\theta_0) - \psi(W_i, g_0,h_0,\alpha_{0},\theta_0)=\hat{R}_{1i\ell}+\hat{R}_{2i\ell}+\hat{R}_{3i\ell}+\hat{R}_{4i\ell}+ \hat{\Delta}_{1i\ell}+\hat{\Delta}_{2i\ell},
     \end{equation*}
     where
	\begin{equation}
    \label{eq:expansion}
		\begin{aligned}
			\hat{R}_{1i\ell} &\equiv m(W_i, \hat{g}_\ell, \hat{h}_\ell,\theta_0)-m(W_i,g_0,h_0,\theta_0), \\
			\hat{R}_{2i\ell} &\equiv \phi_1(W_i, \hat{g}_\ell, \alpha_{01}) + \phi_2(W_i, g_0, \hat{h}_\ell, \alpha_{02}) - \phi_1(W_i, g_0, \alpha_{01}) - \phi_2(W_i, g_0, h_0, \alpha_{02}), \\
            \hat{R}_{3i\ell} &\equiv \phi_2(W_i, \hat{g}_\ell, h_0, \alpha_{02}) - \phi_2(W_i, g_0, h_0, \alpha_{02}), \\
            \hat{R}_{4i\ell} &\equiv \phi_1(W_i, g_0, \hat{\alpha}_{1\ell}) + \phi_2(W_i, g_0, h_0, \hat{\alpha}_{2\ell}) - \phi_1(W_i, g_0, \alpha_{01}) - \phi_2(W_i, g_0, h_0, \alpha_{02}), \\
            \hat{\Delta}_{1i\ell} &\equiv \phi_1(W_i,\hat{g}_\ell, \hat{\alpha}_{1\ell}) - \phi_1(W_i, \hat{g}_\ell, \alpha_{01}) - \phi_1(W_i, g_0, \hat{\alpha}_{1\ell}) + \phi_1(W_i, g_0, \alpha_{01}), \text{ and} \\
            \hat{\Delta}_{2i\ell} &\equiv \phi_2(W_i, \hat{g}_\ell, \hat{h}_\ell, \hat{\alpha}_{2\ell}) - \phi_2(W_i, g_0, \hat{h}_\ell, \alpha_{02}) -  \phi_2(W_i, g_0, h_0, \hat{\alpha}_{2\ell}) - \phi_2(W_i, \hat{g}_\ell, h_0, \alpha_{02}) \\
            &+2\phi_2(W_i, g_0, h_0, \alpha_{02})
		\end{aligned}
	\end{equation}
    The goal is to show $n^{-1/2}\sum_{i\in I_\ell}(\hat{R}_{1i\ell}+\hat{R}_{2i\ell}+\hat{R}_{3i\ell}+\hat{R}_{4i\ell}+ \hat{\Delta}_{1i\ell}+\hat{\Delta}_{2i\ell}) \xrightarrow{P} 0$. To simplify notation, throughout the proof we consider that $m(W, g, h, \theta)$, $S$, $h(X, V)$, and $\alpha_{02}(V, V)$ are one dimensional. The proof for higher dimensions follows \textit{mutatis mutandis}, e.g., replacing $|m(w,\hat{g}_\ell,\hat{h}_\ell,\theta_0)-m(w,g_0,h_0,\theta_0)|^2$ by $|m_j(w,\hat{g}_\ell,\hat{h}_\ell,\theta_0)-m_j(w,g_0,h_0,\theta_0)|^2$ for each dimension $j=1,\dots,q$; or $\alpha_{02}(x,v)^2 \cdot [\hat{h}_\ell(x,v)-h_0(x,v)]^2$ by $(\alpha_{02}(x,v)'[\hat{h}_\ell(x,v)-h_0(x,v)])^2$.

    We begin by providing rates for estimation of $\alpha_{01}$ and $\alpha_{02}$. Assumption~\ref{ass:bounded_dicts} allows us to apply Lemma~A10 in \citet{chernozhukov2022automatic} to get $\norm{\hat{B}_\ell-\E[\mathbf{b}_J(X,V)\mathbf{b}_J(X,V)']}_\infty=O_p(\sqrt{\log(J)/n})$ and $\norm{\hat{C}_\ell-\E[\mathbf{c}_K(Z)\mathbf{c}_K(Z)']}_\infty=O_p(\sqrt{\log(K)/n})$. The fact that Assumption~\ref{ass:rates:numterms} imposes a polynomial rate on $J$ and $K$ then implies that $O_p(\sqrt{\log(J)/n})=O_p(n^{-r})$ and $O_p(\sqrt{\log(K)/n})=O_p(n^{-r})$, since $r<1/2$ (Assumption~\ref{ass:est_convergence}). This, on top of Assumptions~\ref{ass:span},  \ref{ass:abs_sumability}, \ref{ass:est_convergence:linearizations}, and \ref{ass:rates}, means that we can apply Theorem~2 in \citet{chernozhukov2022automatic} to get:
	\begin{equation*}
		\norm{\hat{\alpha}_{1\ell}-\alpha_{01}}_2=o_p(n^\xi n^{-r/2}) \text{ and } \norm{\hat{\alpha}_{2\ell}-\alpha_{02}}_2=o_p(n^\xi n^{-r/2})
	\end{equation*}
	for any $\xi>0$. We choose a $\xi$ satisfying $0<\xi<(3r-1)/2<r/2$, which is possible since, by Assumption~\ref{ass:est_convergence}, $r\in (1/3,1/2)$. This guarantees that
    \begin{equation}
        \norm{\hat{\alpha}_{1\ell}-\alpha_{01}}_2=o_p(1) \text{ and } \norm{\hat{\alpha}_{2\ell}-\alpha_{02}}_2=o_p(1) \label{eq:alpha_consist}
    \end{equation}
    and
	\begin{equation}
		 \label{eq:product_consist}
         \begin{aligned}
        \sqrt{n}\norm{\hat{\alpha}_{1\ell}-\alpha_{01}}_2\norm{\hat{g}-g_0}_2=o_p(1) \text{ and } \sqrt{n}\norm{\hat{\alpha}_{2\ell}-\alpha_{02}}_2\norm{\hat{h}_\ell-h_0}_2=o_p(1)
         \end{aligned}
	\end{equation}
	where the previous display follows from Assumption~\ref{ass:est_convergence:estimators}.

    In the following, recall that $(\hat{g}_\ell, \hat{h}_\ell, \hat{\alpha}_{1\ell}, \hat{\alpha}_{2\ell})$ are estimated using observations in $I_\ell^c$. First, under Assumption~\ref{ass:sq_convergence:m_conv}, we have that
    \begin{equation*}
        \E[\hat{R}_{1i\ell}^2 | I_\ell^c] = \int |m(w,\hat{g}_\ell,\hat{h}_\ell,\theta_0)-m(w,g_0,h_0,\theta_0)|^2 dF_0(w)\xrightarrow{P} 0.
    \end{equation*}
    
    Second, by Assumptions \ref{ass:bounded_dicts} and \ref{ass:abs_sumability:main}:
	\begin{equation*}
		\begin{aligned}
					|\alpha_{01}(z)| &\leq \sum_{k=1}^{\infty} |\beta_{k}| |c_k(z)| \leq \sup_{k\in \mathbb{N}}|c_k(z)| \cdot \sum_{k=1}^{\infty} |\beta_{k}| \equiv \kappa_1 < \infty \text{ and } \\
					|\alpha_{02}(x,v)| &\leq \sum_{j=1}^{\infty} |\rho_{j}| |b_j(x,v)| \leq \sup_{j\in \mathbb{N}}|b_j(x,v)| \cdot \sum_{j=1}^{\infty} |\rho_{j}| \equiv \kappa_2 < \infty.
		\end{aligned}
	\end{equation*}
	Thus, by the triangle inequality, being $v\equiv \varphi(d,z,g_0)$:
	\begin{equation*}
		\begin{aligned}
			 \E[\hat{R}_{2i\ell}^2 | I_\ell^c]  &\leq \int \alpha_{01}(z)^2\left[\epsilon(w,\hat{g}_\ell)-\epsilon(w, g_0)\right]^2 dF_0(w)
			+ \int \alpha_{02}(x,v)^2\left[\hat{h}_\ell(x,v)-h_0(x,v)\right]^2 dF_0(w) \\
			&\leq \kappa_1^2 \lVert \epsilon(\cdot, \hat{g}_\ell) - \epsilon(\cdot, g_0) \rVert_2^2 + \kappa_2^2 \lVert \hat{h}_\ell - h_0 \rVert_2^2.
		\end{aligned}
	\end{equation*}
    Moreover, by Assumption~\ref{ass:general_erro}, $\lVert \epsilon(\cdot, \hat{g}_\ell) - \epsilon(\cdot, g_0) \rVert_2 = \lVert \epsilon(\cdot, \hat{g}_\ell) - \epsilon(\cdot, g_0) +D_e(\hat{g}_\ell-g_0)-D_e(\hat{g}_\ell-g_0)\rVert_2 \leq \lVert D_e \rVert \lVert \hat{g}_\ell-g_0 \rVert_2 + o(\lVert \hat{g}_\ell-g_0 \rVert_2)$. Then, $\E[\hat{R}_{2i\ell}^2 | I_\ell^c] \xrightarrow{P} 0$ follows from the above display and Assumption~\ref{ass:est_convergence:estimators}.

    Third, with $v(g) \equiv \varphi(d,z,g)$, adding and subtracting $\alpha_{02}(x, v(\hat{g}_\ell)) (s - h_0(x, v))$ and by the triangle inequality:
    \begin{equation*}
        \begin{aligned}
             \E[\hat{R}_{3i\ell}^2 | I_\ell^c]  &\leq \int \alpha_{02}(x, v(\hat{g}_\ell))^2 (h_0(x, v(\hat{g}_\ell)) - h_0(x, v))^2 dF_0(w) \\
             &+ \int (s-h_0(x,v))^2 (\alpha_{02}(x, v(\hat{g}_\ell))-\alpha_{02}(x, v))^2dF_0(w). 
        \end{aligned}
    \end{equation*}
     Each term can be dealt with using the Mean Value Theorem. For the first term, let $\kappa_3$ be the bound of $\partial h_0/\partial v$ given by Assumption~\ref{ass:gen_reg:bounded_deriv} and recall that $\alpha_{02}$ is bounded by $\kappa_2$. Then, with $\bar{g}_\ell = g_0 + \bar{\tau} (\hat{g}_\ell - g_0)$ for some $\tau\in[0, 1]$:
     \begin{equation*}
         \begin{aligned}
             \int \alpha_{02}(x, v(\hat{g}_\ell))^2 (h_0(x, v(\hat{g}_\ell)) - h_0(x, v))^2 dF_0(w) &\leq \kappa_2^2 \int \frac{\partial h_0}{\partial v}(x, v(\bar{g}_\ell))^2 (v(\hat{g}_\ell) - v)^2 dF_0(w) \\ 
             &\leq  \kappa_2^2 \kappa_3^2 \norm{\varphi(\cdot,\cdot,\hat{g}_\ell) - \varphi(\cdot,\cdot,g_0)}_2^2.
         \end{aligned}
     \end{equation*}
     By Assumption~\ref{ass:gen_reg:phi_Fderiv}, the term in the second row satisfies $\norm{\varphi(\cdot,\cdot,\hat{g}_\ell) - \varphi(\cdot,\cdot,g_0)}_2 \leq \norm{D_\varphi}\norm{\hat{g}_\ell-g_0}_2 + o(\norm{\hat{g}_\ell-g_0}_2)$. The same procedure applies to the second term in the bound of $ \E[\hat{R}_{3i\ell}^2 | I_\ell^c]$, since $\E[(S-h_0(X,V)^2|D,Z]$ is bounded by Assumption~\ref{ass:sq_convergence:var_bound} and $\partial\alpha_{02}/\partial v$ is bounded by Assumption~\ref{ass:gen_reg:bounded_deriv}. Hence, $\E[\hat{R}_{3i\ell}^2 | I_\ell^c] \xrightarrow{P} 0$ by Assumption~\ref{ass:est_convergence:estimators}.

     Fourth, calling $\kappa_4,\kappa_5<\infty$ to the bounds given by Assumption~\ref{ass:sq_convergence:var_bound}:
     \begin{equation*}
         \begin{aligned}
             \E[\hat{R}_{4i\ell}^2 | I_\ell^c] &\leq \int \epsilon(w, g_0)^2 (\hat{\alpha}_{1\ell}(z) - \alpha_{01}(z))^2 dF_0(w) + \int (s - h_0(x,v))^2 (\hat{\alpha}_{2\ell}(x,v) - \alpha_{02}(x,v))^2 dF_0(w) \\
             &\leq \kappa_4^2 \norm{\hat{\alpha}_{1\ell} - \alpha_{01}}_2^2 + \kappa_5^2 \norm{\hat{\alpha}_{2\ell} - \alpha_{02}}_2^2.
         \end{aligned}
     \end{equation*}
     Hence, $\E[\hat{R}_{4i\ell}^2 | I_\ell^c] \xrightarrow{P} 0$ by equation~\eqref{eq:alpha_consist}.

     Since the data are iid, the above three results guarantee that
     \begin{equation*}
        \E\left[\left. \left( \frac{1}{\sqrt{n}} \sum_{i\in I_\ell} (\hat{R}_{ji\ell} -\E[\hat{R}_{ji\ell} | I_\ell^c]) \right)^2 \right| I_\ell^c \right] \leq \E[\hat{R}_{ji\ell}^2 | I_\ell^c] \xrightarrow{P} 0 \text{ for } j \in \{1, 2, 3, 4\}.
    \end{equation*}   
    This, on top of the triangle and conditional Markov inequalities, leads to
    \begin{equation*}
        \frac{1}{\sqrt{n}} \sum_{i\in I_\ell} \left(\hat{R}_{1i\ell}+\hat{R}_{2i\ell}+\hat{R}_{3i\ell} +\hat{R}_{4i\ell}- \E[\hat{R}_{1i\ell}+\hat{R}_{2i\ell}+\hat{R}_{3i\ell}+\hat{R}_{4i\ell}| I_\ell^c] \right) \xrightarrow{P} 0.
    \end{equation*}

    We now show that $n^{-1/2} \sum_{i\in I_\ell}\E[\hat{R}_{1i\ell}+\hat{R}_{2i\ell}+\hat{R}_{3i\ell}+\hat{R}_{4i\ell}| I_\ell^c] \xrightarrow{P} 0$. Orthogonality of $\epsilon(W,g_0)$ and $S-h_0(X,V)$ to $\Delta_1$ and $\Delta_2(g_0)$, respectively, combined with $\hat{\alpha}_{1\ell} \in \Delta_1$ and $\hat{\alpha}_{2\ell} \in \Delta(g_0)$, gives $\E[\hat{R}_{4i\ell} | I_\ell^c] = 0$. Now, as in the proof of Theorem~\ref{thm:first_step}, let
    \begin{equation*}
        \nu(d,z)\equiv \left.\frac{\partial}{\partial v}\left\{\alpha_{02}(x,v)(\E[S|D=d,Z=z]-h_0(x,v))\right\}\right\rvert_{v=\varphi(d,z,g_0)}.
    \end{equation*}
    To linearize the generalized error and the indirect effect, we add and subtract $\alpha_{01}(z)r_e(z)(\hat{g}_\ell(z)-g_0(z))$ and $\nu(d,z)D_\varphi(\hat{g}_\ell-g_0)$ inside the expectation to get
    \begin{equation*}
        \begin{aligned}
            \E[\hat{R}_{1i\ell}+\hat{R}_{2i\ell}+\hat{R}_{3i\ell}| I_\ell^c] = \\
            \E\left[\left. m(W_i, \hat{g}_\ell, \hat{h}_\ell, \theta_0) - m(W_i, g_0, h_0, \theta_0) - \alpha_{02}(X_i, V_i) (\hat{h}_\ell(X_i, V_i) - h_0(X_i, V_i)) \right\lvert I_\ell^c \right] \\
            + \E\left[\left. \alpha_{01}(Z_i)r_e(Z_i)(\hat{g}_\ell(Z_i) - g_0(Z_i)) + \nu(D_i, Z_i)D_\varphi(\hat{g}_\ell - g_0)  \right\lvert I_\ell^c \right] \\
            + \E\left[\left. \alpha_{01}(Z_i) \cdot \left[\epsilon(W_i, \hat{g}_\ell) - \epsilon(W_i, g_0) - r_e(Z_i)(\hat{g}_\ell(Z_i) - g_0(Z_i)) \right]  \right\lvert I_\ell^c \right] \\
            + \E\left[\left. \phi_2(W_i, \hat{g}_\ell, h_0, \alpha_{02}) - \phi_2(W_i, g_0, h_0, \alpha_{02}) - \nu(D_i, Z_i)D_\varphi(\hat{g}_\ell - g_0)  \right\lvert I_\ell^c \right].
        \end{aligned}
    \end{equation*}
    We deal with the terms in the third to fifth rows separately.
    
    To bound the term in the fourth row, we use that $\alpha_{01}$ is bounded by $\kappa_1$, that $\E[r_e(Z)(g(Z) - g_0(Z)] = \E[D_e(g-g_0)]$, and Assumption~\ref{ass:general_erro}:
    \begin{equation*}
        \begin{aligned}
            |\E\left[\left. \alpha_{01}(Z_i) \cdot \left[\epsilon(W_i, \hat{g}_\ell) - \epsilon(W_i, g_0) - r_e(Z_i)(\hat{g}_\ell(Z_i) - g_0(Z_i)) \right]  \right\lvert I_\ell^c \right] | \\
            \leq \kappa_1 |\E\left[\left. \epsilon(W_i, \hat{g}_\ell) - \epsilon(W_i, g_0) - D_e(\hat{g}_\ell - g_0)   \right\lvert I_\ell^c \right] | \leq \kappa_1 C \norm{\hat{g}_\ell - g_0}_2^2.
        \end{aligned}
    \end{equation*}
    The term in the fifth row is also bounded by $\norm{\hat{g}_\ell-g_0}_2^2$ by Lemma~\ref{lma:IF2v}, which can be applied to $h_0, \alpha_{02}\in \Delta_2(g_0)$ by Assumptions~\ref{ass:gen_reg:phi_Fderiv} and \ref{ass:gen_reg:twice_deriv}.

    Regarding the term in the third row, note that the projection result in equation~\eqref{eq:alpha1_defi} implies that, for every $\delta_1\in\Delta_1$,
    \begin{equation*}
        \E\left[ r_e(Z_i) \delta_1(Z_i) \left( - \frac{r_{01}(Z_i)}{r_e(Z_i)} - \alpha_{01}(Z_i) \right) \right] = 0,
    \end{equation*}
    where $r_{01} \equiv r_{dir} + D_\varphi^*\nu$, $r_{dir}$ is the linearization of the direct effect and $D_\varphi^*$ is the adjoint of $D_\varphi$ (see the proof of Theorem~\ref{thm:first_step}). Taking $\delta_1 = \hat{g}_\ell - g_0$, we can develop the term in the third row as follows:
    \begin{equation*}
        \begin{aligned}
            \E\left[\left. \alpha_{01}(Z_i)r_e(Z_i)(\hat{g}_\ell(Z_i) - g_0(Z_i)) + \nu(D_i, Z_i)D_\varphi(\hat{g}_\ell - g_0)  \right\lvert I_\ell^c \right] = \\
            \E\left[\left. -r_{01}(Z_i)(\hat{g}_\ell(Z_i) - g_0(Z_i)) + \nu(D_i, Z_i)D_\varphi(\hat{g}_\ell - g_0)  \right\lvert I_\ell^c \right] = \\
            \E\left[\left. -r_{dir}(Z_i)(\hat{g}_\ell(Z_i) - g_0(Z_i)) -D_\varphi^*\nu(Z_i)(\hat{g}_\ell(Z_i) - g_0(Z_i)) + \nu(D_i, Z_i)D_\varphi(\hat{g}_\ell - g_0)  \right\lvert I_\ell^c \right] = \\
            \E\left[\left. -r_{dir}(Z_i)(\hat{g}_\ell(Z_i) - g_0(Z_i))\right\lvert I_\ell^c \right],
        \end{aligned}
    \end{equation*}
    since, by the definition of the adjoint operator, $\E[D_\varphi^*\nu \delta_1] = \E[\nu D_\varphi\delta_1]$ for every $\nu\in L_2(D,Z)$ and every $\delta_1\in\Delta_1$.

    Combining all these results and the triangle inequality leads to
    \small{
    \begin{equation*}
        \begin{aligned}
            |\E[\hat{R}_{1i\ell}+\hat{R}_{2i\ell}+\hat{R}_{3i\ell}| I_\ell^c] | \leq \\
            \left|\E\left[\left. m(W_i, \hat{g}_\ell, \hat{h}_\ell, \theta_0) - m(W_i, g_0, h_0, \theta_0) -r_{dir}(Z_i)(\hat{g}_\ell(Z_i) - g_0(Z_i)) - \alpha_{02}(X_i, V_i) (\hat{h}_\ell(X_i, V_i) - h_0(X_i, V_i)) \right\lvert I_\ell^c \right]\right| \\
            + C \norm{\hat{g}_\ell-g_0}_2^2 \\
            = \left|\E\left[\left. m(W_i, \hat{g}_\ell, \hat{h}_\ell, \theta_0) - m(W_i, g_0, h_0, \theta_0) -D_{dir}(W_i,\hat{g}_\ell - g_0) - D_2(W_i, \hat{h}_\ell - h_0) \right\lvert I_\ell^c \right]\right| + C \norm{\hat{g}_\ell-g_0}_2^2 \\
           = \bar{\psi}(\hat{g}_\ell, \hat{h}_\ell) + C \norm{\hat{g}_\ell-g_0}_2^2,
        \end{aligned}
    \end{equation*}
    }
    where we have used Assumption~\ref{ass:gen_reg:bounded_alpha2} for the equality. Therefore,
    \begin{equation*}
        \frac{1}{\sqrt{n}}\sum_{i\in I_\ell}\E[\hat{R}_{1i\ell}+\hat{R}_{2i\ell}+\hat{R}_{3i\ell}+\hat{R}_{4i\ell}| I_\ell^c] \leq \sqrt{n}\bar{\psi}(\hat{g}_\ell, \hat{h}_\ell) + C \sqrt{n}\norm{\hat{g}_\ell-g_0}_2^2.
    \end{equation*}
    The above converges to zero in probability by Assumptions~\ref{ass:est_convergence:estimators} and \ref{ass:linear_approx}.

    To conclude with the proof of equation~\eqref{eq:asump_equiv}, we show that $n^{-1/2} \sum_{i\in I_\ell} (\hat{\Delta}_{1i\ell} + \hat{\Delta}_{2i\ell}) \xrightarrow{P} 0$. Regarding
    \begin{equation*}
    \begin{aligned}
        \hat{\Delta}_{1i\ell} = [\hat{\alpha}_{1\ell}(Z_i)-\alpha_{01}(Z_i)]\cdot [\epsilon(W_i,\hat{g}_\ell)-\epsilon(W_i, g_0)],
    \end{aligned}
    \end{equation*}
    an application of the Cauchy-Schwarz and conditional Markov inequalities leads to:
	 \begin{equation*}
	 	\left|\frac{1}{\sqrt{n}}\sum_{i\in I_\ell} \hat{\Delta}_{1i\ell} \right|= O_p(\sqrt{n}\norm{\hat{\alpha}_{1\ell}-\alpha_{01}}_2 \norm{\epsilon(\cdot, \hat{g}_\ell)-\epsilon(\cdot, g_0)}_2).
	 \end{equation*}
	 Recall that, by Assumption~\ref{ass:general_erro}, $\norm{\epsilon(\cdot, \hat{g}_\ell)-\epsilon(\cdot, g_0)}_2 = O(\norm{\hat{g}_\ell - g_0}_2)$. Thus, equation~\eqref{eq:product_consist} gives the result for $\hat{\Delta}_{1i\ell}$.

     Regarding $\hat{\Delta}_{2i\ell}$, adding and subtracting $\hat{\alpha}_{2\ell}(X_i, V_i)(S_i - \hat{h}_\ell(X_i,V_i))$, where $V_i = \varphi(D_i, Z_i, g_0)$, we get
     \begin{equation*}
         \begin{aligned}
              \hat{\Delta}_{2i\ell} = -[\hat{\alpha}_{2\ell}(X_i, V_i)-\alpha_{02}(X_i, V_i)]\cdot [\hat{h}_\ell(X_i, V_i) - h_0(X_i, V_i)] \\
              +\left[\phi_2(W,\hat{g}_\ell, \hat{h}_\ell, \hat{\alpha}_{2\ell}) - \phi_2(W, g_0,\hat{h}_\ell, \hat{\alpha}_{2\ell})\right] - \left[\phi_2(W, \hat{g}_\ell, h_0, \alpha_{02}) - \phi_2(W, g_0, h_0, \alpha_{02})\right]
         \end{aligned}
     \end{equation*}
    Therefore, by the combination of the Cauchy-Schwarz, conditional Markov, and the triangle inequalities; on top of Lemma~\ref{lma:IF2lin} (which is valid under Assumption~\ref{ass:gen_reg}) gives: 
    \begin{equation*}
        \left|\frac{1}{\sqrt{n}}\sum_{i\in I_\ell} \hat{\Delta}_{2i\ell} \right|= O_p(\sqrt{n}\norm{\hat{\alpha}_{2\ell}-\alpha_{02}}_2 \norm{\hat{h}_\ell-g_0}_2) + O_p(\sqrt{n}\norm{\hat{g}_\ell-g_0}_2 \cdot(\norm{\hat{g}_\ell-g_0}_2+\norm{\hat{\nu}_\ell-\nu_0}_2 )),
    \end{equation*}
    where recall that $\hat{\nu}_\ell=\nu(\hat{h}_\ell, \hat{\alpha}_{2\ell})$, $\nu_0=\nu(h_0, \alpha_{02})$, and $\nu(h, \alpha_2) = \partial/\partial v\{\alpha_2(x,v) \cdot (s - h(x,v)\}$ gives the derivative of the second-step IF understood as a functional of $(h, \alpha_2)$. Thus, equation \eqref{eq:product_consist} and Assumptions~\ref{ass:est_convergence:estimators} and \ref{ass:gen_reg:continuous_deriv} give the result. Finally, the asymptotic equivalence in equation~\eqref{eq:asump_equiv} follows from the above discussion and the triangle inequality.
    
	When equation~\eqref{eq:asump_equiv} holds for each component of $\hat\psi$, consistency of $\hat{\theta}$ follows under standard conditions that guarantee uniform convergence of $\hat{\psi}(\theta)'\hat{\Upsilon}\hat{\psi}(\theta)$ in $\Theta$ \citep[c.f.][Th.~14.1]{wooldridge2010econometric}. These conditions will follow from Assumption~\ref{ass:jacobian} if $\Theta$ is compact. Moreover, by Assumptions \ref{ass:est_convergence:estimators}  and \ref{ass:jacobian:diff} we can apply the Mean Value Theorem to get
	\begin{equation*}
		\sqrt{n}\left(\hat{\psi}(\hat{\theta})-\hat{\psi}\left(\theta_0\right)\right)=\sqrt{n}\frac{\partial\hat\psi}{\partial\theta}(\bar\theta)\cdot (\hat{\theta}-\theta_0)
	\end{equation*}
	for $\bar{\theta}$ a point between $\theta_0$ and $\hat{\theta}$ (that is $\bar{\theta}\xrightarrow{P}\theta_0$). Then, if equation~\eqref{eq:asump_equiv} holds:
	\begin{equation} \label{eq:MVT}
		\sqrt{n}\frac{\partial\hat\psi}{\partial\theta}(\bar\theta)\cdot (\hat{\theta}-\theta_0) = \frac{1}{\sqrt{n}}\sum_{i=1}^{n}\psi(W_i, g_0,h_0,\alpha_0,\theta_0) + \sqrt{n}\hat{\psi}(\hat{\theta})+o_p(1).
	\end{equation}
	
	Now, note that
	\begin{equation*}
		\begin{aligned}
				\frac{\partial\hat\psi}{\partial\theta}(\theta) &=\frac{\partial}{\partial\theta}\left( \frac{1}{n}\sum_{\ell=1}^{L}\sum_{i\in I_{\ell}} \left[ m(W_{i},\hat{g}_{\ell},\hat{h}_\ell, \theta) +\phi_1(W_{i},\hat{g}_{\ell}, \hat{\alpha}_{1\ell})+\phi_2(W_{i},\hat{g}_{\ell}, \hat{h}_\ell, \hat{\alpha}_{2\ell}) \right]  \right) \\
				&= \frac{1}{n}\sum_{\ell=1}^{L}\sum_{i\in I_{\ell}} \frac{\partial m}{\partial\theta}(W_i, \hat{g}_\ell,\hat{h}_\ell, \theta)
		\end{aligned}
	\end{equation*}
	Then, since Assumptions~\ref{ass:est_convergence:estimators} and \ref{ass:jacobian}, on top of $\hat{\theta}\xrightarrow{P}\theta_0$, guarantee that we can apply Lemma~E2 in \cite{chernozhukov2022locally}, we have that $\partial\hat{\psi}(\hat\theta)/\partial\theta \xrightarrow{P} M$. Thus, this term is bounded in probability. Therefore, since $\hat{\Upsilon}$ is also $O_p(1)$, equation~\eqref{eq:MVT} implies
	\begin{equation*}
		\begin{aligned}
			\frac{\partial\hat\psi}{\partial\theta}(\hat\theta)'\hat \Upsilon \frac{\partial\hat\psi}{\partial\theta}(\bar\theta) \cdot \sqrt{n} (\hat{\theta}-\theta_0) &= \frac{\partial\hat\psi}{\partial\theta}(\hat\theta)'\hat \Upsilon \cdot \frac{1}{\sqrt{n}}\sum_{i=1}^{n}\psi(W_i, g_0,h_0,\alpha_0,\theta_0) \\
			&+ \sqrt{n} \frac{\partial\hat\psi}{\partial\theta}(\hat\theta)'\hat \Upsilon\hat\psi(\hat\theta)+o_p(1).
		\end{aligned}
	\end{equation*}
	
	Thus, since  $(\partial\hat\psi(\hat\theta)/\partial\theta)'\hat \Upsilon\hat\psi(\hat\theta)=0$ is the first-order condition for the minimization problem in equation~\eqref{dgmm}, $\partial\hat{\psi}(\bar\theta)/\partial\theta \xrightarrow{P} M$ \citep[by Lemma~E2 in][]{chernozhukov2022locally}, and $M'\Upsilon M$ is non-singular:
	\begin{equation*}
		\sqrt{n} (\hat{\theta}-\theta_0) = (M'\Upsilon M)^{-1} \frac{1}{\sqrt{n}}\sum_{i=1}^{n}\psi(W_i, g_0,h_0,\alpha_0,\theta_0) + o_p(1).
	\end{equation*}
	Then, the asymptotic normality result follows from $n^{-1/2}\sum_{i=1}^{n}\psi(W_i, g_0,h_0,\alpha_0,\theta_0) \xrightarrow{D} N(0, \Psi)$.

    We conclude the proof of Theorem~\ref{thm:asymptotics} by providing consistency of $\hat{\Xi}$. Call $\psi_i\equiv \psi(W_i, g_0,h_0,\alpha_{0},\theta_0)$ and $\bar\Psi \equiv n^{-1}\sum_{i=1}^n \psi_i\psi_i'$.  We have that
	 \begin{equation*}
	 	\norm{\hat{\Psi}-\bar{\Psi}}_\infty \leq \sum_{\ell=1}^L \frac1n \sum_{i\in I_\ell} \left(\norm{\hat\psi_{i\ell}-\psi_i}_\infty^2 +2 \norm{\hat\psi_{i\ell}-\psi_i}_\infty\norm{\psi_i}_\infty\right)
	 \end{equation*}
	
	 We now expand  $\hat{\psi}_{i\ell}(\tilde{\theta}_\ell) - \psi(W_i, g_0,h_0,\alpha_{0},\theta_0)=\hat{R}_{1i\ell}+\hat{R}_{2i\ell}+\hat{R}_{3i\ell}+\hat{R}_{4i\ell}+\hat{R}_{5i\ell}+\hat{\Delta}_{1i\ell}+\hat{\Delta}_{2i\ell}$, with
	 	\begin{equation*}
	 		\hat{R}_{5i\ell} \equiv m(W_i, \hat{g}_\ell, \hat{h}_\ell,\tilde{\theta}_\ell)-m(W_i,\hat{g}_\ell, \hat{h}_\ell,\theta_0)
	 	\end{equation*}
	 and the remaining terms are given in equation~\eqref{eq:expansion}. Then
	 \begin{eqnarray*}
	 	\frac1n \sum_{i\in I_{\ell}} \norm{\hat\psi_{i\ell}-\psi_i}_\infty^2 \leq C\frac1n \sum_{i\in I_{\ell}} \left(\norm{\hat{R}_{1i\ell}}_\infty^2+\norm{\hat{R}_{2i\ell}}_\infty^2+\norm{\hat{R}_{3i\ell}}_\infty^2+\norm{\hat{R}_{4i\ell}}_\infty^2+\norm{\hat{R}_{5i\ell}}_\infty^2+\norm{\hat{\Delta}_{1i\ell}}_\infty^2+\norm{\hat{\Delta}_{2i\ell}}_\infty^2\right)
	 \end{eqnarray*}
	 by the triangle inequality. The constant $C$ comes from the presence of the interation terms: for instance, we have that $2\norm{\hat{R}_{1i\ell}}_\infty\norm{\hat{R}_{2i\ell}}_\infty \leq 2\max\{\norm{\hat{R}_{1i\ell}}_\infty,\norm{\hat{R}_{2i\ell}}_\infty\}$.
	
	 We apply Assumptions~\ref{ass:sq_convergence:m_conv} and \ref{ass:sq_convergence:m_conv_theta} to each component of $\hat{R}_{1i\ell}$ and $\hat{R}_{5i\ell}$, respectively. This yields $\E[\norm{\hat{R}_{1i\ell}}_\infty^2|I_\ell^c]\xrightarrow{P}0$ and  $\E[\norm{\hat{R}_{5i\ell}}_\infty^2|I_\ell^c]\xrightarrow{P}0$. Moreover, by the arguments in our asymptotic normality proof, $\E[\norm{\hat{R}_{2i\ell}}_\infty^2|I_\ell^c]\xrightarrow{P}0$, $\E[\norm{\hat{R}_{3i\ell}}_\infty^2|I_\ell^c]\xrightarrow{P}0$, and $\E[\norm{\hat{R}_{4i\ell}}_\infty^2|I_\ell^c]\xrightarrow{P}0$. Also, by the Cauchy-Schwarz inequality, equation~\eqref{eq:product_consist}, Assumptions~\ref{ass:general_erro} and \ref{ass:gen_reg}, and applying Lemma~\ref{lma:IF2lin}:
	 \begin{equation*}
     \begin{aligned}
	 	\E[\norm{\hat{\Delta}_{1i\ell}}_\infty^2|I_\ell^c] &\leq C \left(\norm{\hat{\alpha}_{1\ell}-\alpha_{01}}_2\norm{\epsilon(\cdot,\hat{g}_\ell)-\epsilon(\cdot, g_0)}_2\right)=o_p(1) \text{ and} \\
 \E[\norm{\hat{\Delta}_{2i\ell}}_\infty^2|I_\ell^c] &\leq C (\norm{\hat{\alpha}_{2\ell}-\alpha_{02}}_2 \norm{\hat{h}_\ell-g_0}_2) \\ 
 &+ C(\norm{\hat{g}_\ell-g_0}_2 \cdot(\norm{\hat{g}_\ell-g_0}_2+\norm{\hat{h}_\ell-h_0}_2 + \norm{\hat{\alpha}_{2\ell}-\alpha_{02}}_2)) = o_p(1).        
     \end{aligned}
	 \end{equation*}
	 Thus, collecting the above results:
	 \begin{equation*}
     \begin{aligned}
	 	\E\left[ \left. \frac1n \sum_{i\in I_{\ell}} \norm{\hat\psi_{i\ell}-\psi_i}_\infty^2 \right\rvert I_\ell^c \right] \leq \\
        C \E\left[ \left. \norm{\hat{R}_{1i\ell}}_\infty^2+\norm{\hat{R}_{2i\ell}}_\infty^2+\norm{\hat{R}_{3i\ell}}_\infty^2+\norm{\hat{R}_{4i\ell}}_\infty^2+\norm{\hat{R}_{5i\ell}}_\infty^2+\norm{\hat{\Delta}_{1i\ell}}_\infty^2+\norm{\hat{\Delta}_{2i\ell}}_\infty^2 \right\rvert I_\ell^c\right]=o_p(1).         
     \end{aligned}
	 \end{equation*}
	
	 An application of the conditional Markov inequality gives then $n^{-1} \sum_{i\in I_{\ell}} \norm{\hat\psi_{i\ell}-\psi_i}_\infty^2 =o_p(1)$. Also, by Assumptions~\ref{ass:bounded_dicts}, \ref{ass:abs_sumability:main}, \ref{ass:sq_convergence:m_bound}, and \ref{ass:sq_convergence:var_bound}: $\E[\psi_i\psi_i']<\infty$. So, by the Law of Large Numbers, $\bar\Psi\xrightarrow{P}\E[\psi_i\psi_i']$. Therefore, by Cauchy-Schwarz:
	 \begin{equation*}
	 	\begin{aligned}
	 		\norm{\hat\Psi-\bar\Psi}_\infty &\leq  \sum_{\ell=1}^L \left[ \frac1n\sum_{i\in I_\ell} \norm{\hat\psi_{i\ell}-\psi_i}_\infty^2 +2 \sqrt{ \frac1n \sum_{i\in I_\ell} \norm{\hat\psi_{i\ell}-\psi_i}^2_\infty } \sqrt{ \frac1n \sum_{i\in I_\ell} \norm{\psi_i}^2_\infty } \right] \\
	 		&= o_p(1) + o_p(1)\cdot O_p(1) = o_p(1).
	 	\end{aligned}
	 \end{equation*}
	 This leads to $\hat\Psi=\bar\Psi+o_p(1)\xrightarrow{P} \E[\psi_i\psi_i']$.
	
\end{proofc}

%%%%%%%%%%%%%%%%%%%%%%%%%%%%%%%%%%%%%
\begin{proofc}[Proposition~\ref{prop:ATE_conditions}] \mbox{} \\[-25pt]

    \paragraph{Assumptions~\ref{ass:Diff_m_h} and \ref{ass:Diff_m_g}}
     Since $D_2(W,h)=0$, we just need to ensure that $\E[D_{dir}(W,g)]=\E[-(Y-h_0(V))V(1-V)g(Z)]$ is $L_2$-continuous. By the Cauchy-Schwarz inequality
    \begin{equation*}
        |\E[D_{dir}(W,g)]| \leq \sqrt{\E[(Y-h_0(V))^2V^2(1-V)^2]} \norm{g}_2 \leq \sqrt{\E[(Y-h_0(V))^2]} \norm{g}_2.
    \end{equation*}
    We have used that $V \in [0, 1] \Rightarrow V(1-V) \leq 1 \Rightarrow V^2(1-V)^2 \leq 1$. Since  $\E[Y^2] = \E[\E[Y^2|D,Z]]<\infty$ (Assumption~\ref{ass:regularity_ATE_expectation}) implies $\E[(Y-h_0(V))^2] < \infty$, this gives Assumption~\ref{ass:Diff_m_g}. 

    \paragraph{Assumption~\ref{ass:sq_convergence}} We note that $|D-V| \leq 1$ almost surely. Therefore, $|Y-h_0(V) -\theta_0(D-V)| \leq |Y-h_0(V)| + |\theta_0|$. Using these result we get
    \begin{equation*}
        \E[|m(W, g_0, h_0, \theta_0)|^2] = \E[|Y-h_0(V) -\theta_0(D-V)|^2 |D-V|^2] \leq \E[(Y-h_0(V))^2] + 2|\theta_0|\E[|Y-h_0(V)|] + \theta_0^2.
    \end{equation*}
    The right hand side is bounded by Assumption~\ref{ass:regularity_ATE_expectation}.

    Let $\hat{V}_\ell \equiv \Lambda(\hat{g}_\ell(Z))$. In the following expectations, data $W=(Y, D, Z)$ is independent of the observations in $I_\ell^c$. For  Assumption~\ref{ass:sq_convergence:m_conv}, we have that:
    \begin{equation} \label{eq:bound_diff_m_gh}
    \begin{aligned}
        \int [m(w,\hat{g}_\ell,\hat{h}_\ell,\theta_0)-m(w,g_0,h_0,\theta_0)]^2 dF_0(w) \\
        = \E\left[ \left. \left( (Y - \hat{h}_\ell(\hat{V}_\ell) -\theta_0 (D - \hat{V}_\ell))(D-\hat{V}_\ell) - (Y-h_0(V)-\theta_0(D-V))(D-V) \right)^2 \right| I_\ell^c \right] \\
        \leq \E\left[ \left. \left( (Y - \hat{h}_\ell(\hat{V}_\ell) -\theta_0 (D - \hat{V}_\ell))(V-\hat{V}_\ell) \right)^2 \right| I_\ell^c \right] \\ +
        \E\left[ \left. \left( (h_0(V) - \hat{h}_\ell(\hat{V}_\ell) -\theta_0 (V - \hat{V}_\ell))(D-\hat{V}_\ell) \right)^2 \right| I_\ell^c \right]
    \end{aligned}
    \end{equation}
    
    Also, the following bounds hold almost surely: $|\hat{V}_\ell| \leq 1$, $|D-\hat{V}_\ell| \leq 1$, and $|Y - \hat{h}_\ell(\hat{V}_\ell) -\theta_0 (D - \hat{V}_\ell)| \leq |Y| + |\hat{h}_\ell(\hat{V}_\ell)| + |\theta_0|$. Thus, almost surely,
    \begin{equation*}
        (Y - \hat{h}_\ell(\hat{V}_\ell) -\theta_0 (D - \hat{V}_\ell))^2 \leq Y^2 + \hat{h}_\ell(\hat{V}_\ell)^2 + \theta_0^2 + 2|Y||\hat{h}_\ell(\hat{V}_\ell)| +2|\theta_0| |Y| + 2|\theta_0||\hat{h}_\ell(\hat{V}_\ell)|. 
    \end{equation*}
    Assumption~\ref{ass:regularity_ATE_dictionary} provides that the dictionary $\mathbf{b}_J(v)$ used to estimate $h_0$ has continuous atoms, hence $|\hat{h}_\ell(\hat{V}_\ell)| < C_h$ almost surely. Therefore, $ (Y - \hat{h}_\ell(\hat{V}_\ell) -\theta_0 (D - \hat{V}_\ell))^2 \leq C_1 + 2C_h|Y| +Y^2$ almost surely, where $C_1 = \theta_0^2 + C_h^2 + 2|\theta_0|C_h$. We can use this to bound the term in the third row from equation~\eqref{eq:bound_diff_m_gh}.
    \begin{align*}
         \E\left[ \left. \left( (Y - \hat{h}_\ell(\hat{V}_\ell) -\theta_0 (D - \hat{V}_\ell))(V-\hat{V}_\ell) \right)^2 \right| I_\ell^c \right] &\leq  \E\left[ \left. (C_1 + 2C_h \E[|Y||Z] + \E[Y^2|Z])(V-\hat{V}_\ell)^2 \right| I_\ell^c \right] \\
         & \leq C \cdot \norm{\varphi(\cdot, \cdot, \hat{g}_\ell) - \varphi(\cdot, \cdot, g_0)}_2^2,
    \end{align*}
    since $C_1 + 2C_h \E[|Y||D,Z] + \E[Y^2|D,Z]$ is bounded by Assumption~\ref{ass:regularity_ATE_expectation}. Also, recall that $D_\varphi g = V(1-V)g$. Therefore, it is continuous as $V(1-V) \leq 1/2$ almost surely, and, by the Cauchy-Schwarz inequality, $\norm{D_\varphi g}_2 \leq 1/4 \norm{g}_2$. Hence, by  Assumption~\ref{ass:est_convergence:estimators},
    \begin{equation} \label{eq:hdps_phi_conv}
        \norm{\varphi(\cdot, \cdot, \hat{g}_\ell) - \varphi(\cdot, \cdot, g_0)}_2 \leq \norm{D_\varphi} \norm{\hat{g}_\ell - g_0}_2 + o_p(\norm{\hat{g}_\ell - g_0}_2) \xrightarrow{P} 0.
    \end{equation}

    Adding and subtracting $\hat{h}_\ell(V)$ inside the square, the term in the fourth row of equation~\eqref{eq:bound_diff_m_gh} satisfies, recalling that $|D-\hat{V}_\ell|\leq 1$,
    \begin{equation*}
    \begin{aligned}
            \E\left[ \left. \left( (h_0(V) - \hat{h}_\ell(\hat{V}_\ell) -\theta_0 (V - \hat{V}_\ell))(D-\hat{V}_\ell) \right)^2 \right| I_\ell^c \right] \leq  \E\left[ \left. \left( h_0(V) - \hat{h}_\ell(V)\right)^2 \right| I_\ell^c \right] \\
            + \E\left[ \left. \left( \hat{h}_\ell(V) - \hat{h}_\ell(\hat{V}_\ell)\right)^2 \right| I_\ell^c \right] \\ + \theta_0^2 \E\left[ \left. \left( V - \hat{V}_\ell\right)^2 \right| I_\ell^c \right].
    \end{aligned}
    \end{equation*}
    The terms in the first and third row converge to 0 in probability by Assumption~\ref{ass:est_convergence:estimators} and equation~\eqref{eq:hdps_phi_conv}. For the term in the second row,  Assumption~\ref{ass:regularity_ATE_dictionary} guarantees that $\sup_{v\in [0, 1]} \partial \hat{h}_\ell/\partial v (v) \leq C$. Hence, it converges to 0 in probability by the Mean Value Theorem, Assumption~\ref{ass:est_convergence:estimators}, and equation~\eqref{eq:hdps_phi_conv}.

    Regarding Assumption~\ref{ass:sq_convergence:m_conv_theta}, we have that
    \begin{equation*}
        \int |m(w, \hat{g}_\ell, \hat{h}_\ell, \tilde{\theta}_\ell) - m(w, \hat{g}_\ell, \hat{h}_\ell, \theta_0)|^2 dF_0(w) = \E\left[ \left. \left( (\tilde{\theta}_\ell - \theta_0) (D - \hat{V}_\ell) \right)^2 \right| I_\ell^c \right] \leq (\tilde{\theta}_\ell - \theta_0)^2.
    \end{equation*}
    Hence, it suffices to have a consistent preliminary estimator of $\theta_0$ under the partially linear model.

    Finally, Assumption~\ref{ass:regularity_ATE_expectation} combined with the conditional Jensen's inequality gives the first part of Assumption~\ref{ass:sq_convergence:var_bound}. Just note that $h_0(V)^2 \leq \E[Y^2|V] = \E[ \E[Y^2|D,Z]|V]$. For the second part, we have that $\E[\epsilon(W, g_0)^2|Z]=V(1-V) \leq 1/2$.

    \paragraph{Assumption~\ref{ass:general_erro}} The derivative of the generalized error at an arbitrary $g\in\Delta_1$, evaluated at $\delta_1\in\Delta_1$, is $D_e(g)[\delta_1] = \E[-\Lambda(g(Z)) \cdot (1-\Lambda(g(Z)) \delta_1(Z)]$. We have that $D_e \equiv D_e(g_0)$ with Riesz representer $r_e(Z)=-V(1-V)$, which is non-positive. Assumption~\ref{ass:regularity_ATE_compact} leads to $r_e$ bounded away from 0. Note also that $|r_e| \leq 1/2$ almost surely.

    To ensure a quadratic remainder in the linear expansion of the generalized error, we take the second derivative at $g\in \Delta_1$, evaluated at $\delta_1, \delta_1'\in\Delta_1$, $D_{e}^2(g)[\delta_1][\delta_1'] = \E[-\Lambda(g(Z)) \cdot (1-\Lambda(g(Z))\cdot(1-2\Lambda(g(Z))\delta_1(Z)\delta_1'(Z)]$. Since $\Lambda(g(z)) \in [0, 1]$ almost surely, we have that $|D_{e}^2(g)[\delta_1][\delta_1']
    | \leq C \norm{\delta_1}_2\norm{\delta_1'}_2$, with constant $C$ independent of $g$. Hence, $\sup_{g\in\Delta_1} \norm{D_{e}^2(g)} \leq C$ and Proposition~3 in \citet[p.~177]{luenberger1997optimization} gives the result. 

     \paragraph{Assumption~\ref{ass:linear_approx}} We check variant (i) of the assumption. Recall that, with $v = \Lambda(g_0(z))$,
     \begin{equation*}
        D_{dir}(w, g) = -(y-h_0(v))v(1-v)g(z)  \text{ and } D_2(w, h) = 0
     \end{equation*}
     Adding and subtracting $(Y-h_0(V)-\theta_0(D-V))(D-\hat{V}_\ell)$ to $m(W, \hat{g}_\ell, \hat{h}_\ell, \theta_0) - m(W, g_0, h_0, \theta_0) - D_{dir}(W, \hat{g}_\ell-g_0)$, we get:
     \begin{equation} \label{eq:hd_ps_Ass8_expansion}
     \begin{aligned}
          (\hat{h}_\ell(\hat{V}_\ell) - h_0(V) -\theta_0(\hat{V}_\ell - V))(\hat{V}_\ell - D) \\
         + (Y - h_0(V) - \theta_0 (D-V))(D - \hat{V}_\ell) - (Y - h_0(V) - \theta_0 (D-V))(D - V) \\
         + (Y-h_0(V))V(1-V)(\hat{g}_\ell(Z) - g_0(Z)).
     \end{aligned}
     \end{equation}
     We separately deal with the term in the first row, and the terms in the second and third rows.

     For the term in the first row, after conditioning on $Z$, adding and subtracting $\hat{h}_\ell(V)(\hat{V}_\ell - V)$, and using the triangle inequality:
     \begin{align*}
         |\E[(\hat{h}_\ell(\hat{V}_\ell) - h_0(V) -\theta_0(\hat{V}_\ell - V))(\hat{V}_\ell - D) | I_\ell^c] | \leq |\E[(\hat{h}_\ell(\hat{V}_\ell) - \hat{h}_\ell(V))(\hat{V}_\ell - V)| I_\ell^c] | \\
         + |\E[(\hat{h}_\ell(V) - h_0(V))(\hat{V}_\ell - V)| I_\ell^c] | \\
         + |\theta_0| \E[(\hat{V}_\ell - V)^2|I_\ell^c]|.
     \end{align*}
     Recall that Assumption~\ref{ass:regularity_ATE_dictionary} guarantees that $\sup_{v\in [0, 1]} \partial \hat{h}_\ell/\partial v (v) \leq C$. Therefore, combining the Mean Value Theorem and the Cauchy-Schwarz inequality gives $|\E[(\hat{h}_\ell(\hat{V}_\ell) - \hat{h}_\ell(V))(\hat{V}_\ell - V)| I_\ell^c] | \leq C  \norm{\varphi(\cdot, \cdot, \hat{g}_\ell) - \varphi(\cdot, \cdot, g_0)}_2^2$. By equation~\eqref{eq:hdps_phi_conv}, this term is bounded by $\norm{\hat{g}_\ell - g_0}_2^2$ with probability tending to one. For the second term, by the Cauchy-Schwarz inequality, $ |\E[(\hat{h}_\ell(V) - h_0(V))(\hat{V}_\ell - V)| I_\ell^c] | \leq \norm{\hat{h}_\ell - h_0}_2 \norm{\hat{g}_\ell - g_0}_2$ with probability tending to one. Finally, the third term is bounded by $\norm{\hat{g}_\ell - g_0}_2^2$.

     We now move to bound the terms in the second and third rows of equation~\eqref{eq:hd_ps_Ass8_expansion}. If we define the mapping $\xi(g) \equiv \E[(Y-h_0(V) - \theta_0 (D-V))(D - \Lambda(g(Z)))]$, with derivative at $g_0$ given by $D_\xi g =\E[-(Y-h_0(V))V(1-V)g(Z)]$, we have that
     \begin{align*}
          \E[ (Y - h_0(V) - \theta_0 (D-V))(D - \hat{V}_\ell) - (Y - h_0(V) - \theta_0 (D-V))(D - V) \\
         + (Y-h_0(V))V(1-V)(\hat{g}_\ell(Z) - g_0(Z)) | I_\ell^c] = |\xi(\hat{g}_\ell) - \xi(g_0) - D_\xi (\hat{g}_\ell - g_0) |.
     \end{align*}
     Therefore, it suffices to show that the linear expansion of $\xi$ around $g_0$ has quadratic remainder.

     Call $D_\xi(g)$ to the derivative of of $\xi$ at an arbitrary $g \in \Delta_1$, so that $D_\xi = D_\xi(g_0)$. Let $V(g) \equiv \Lambda(g(Z))$. The first derivative evaluated at $\delta_1 \in \Delta_1$ is
     \begin{equation*}
         D_\xi(g)[\delta_1] = \E[-(Y-h_0(V) -\theta_0(D-V)) \cdot \Lambda(g(Z)) \cdot (1-\Lambda(g(Z))\delta_1(Z)].
     \end{equation*}
    For the second derivative, with $\delta_1, \delta_1' \in \Delta_1$,
    \begin{equation*}
        D^2_\xi(g)[\delta_1][\delta_1'] = \E[-(Y-h_0(V) -\theta_0(D-V)) \cdot \Lambda(g(Z)) \cdot (1-\Lambda(g(Z))\cdot (1-2\Lambda(g(Z))\delta_1(Z)\delta_1'(Z)].
    \end{equation*}
     By Assumption~\ref{ass:regularity_ATE_expectation} and the conditional Jensen's inequality, $\E[Y|D,Z]$ and $h_0(V)$ are bounded. Moreover, $|D-V| \leq 1$ and $\Lambda(g(Z)) \cdot (1-\Lambda(g(Z))\cdot (1-2\Lambda(g(Z)) \leq C$, with $C$ independent of $g$, since $\Lambda(g(Z)) \in [0, 1]$. Hence, by conditioning on $(D,Z)$ and applying the Cauchy-Schwarz inequality, we get $|D_\xi^2(g)[\delta_1][\delta_2] | \leq C \norm{\delta_1}_2\norm{\delta_1'}_2$. Thus, Proposition~3 in \citet[p.~177]{luenberger1997optimization} gives that $|\xi(\hat{g}_\ell) - \xi(g_0) - D_\xi (\hat{g}_\ell - g_0) | \leq C \norm{\hat{g}_\ell - g_0}_2^2$.

    \paragraph{Assumption~\ref{ass:gen_reg}} It is satisfied since $\alpha_{02}= 0$.
    
    \paragraph{Assumption~\ref{ass:jacobian}} We have that
    \begin{equation*}
        \frac{\partial m}{\partial\theta}(W,g,h,\theta)=-(D-V(g))^2.
    \end{equation*}
    Assumptions~\ref{ass:jacobian:diff}-\ref{ass:jacobian:existence} are trivial to check. For Assumption~\ref{ass:jacobian:convergence}, we have that:
    \begin{align*}
        \int \left| \frac{\partial m}{\partial\theta}(w,\hat{g}_\ell,\hat{h}_\ell,\theta_0) -\frac{\partial m}{\partial\theta}(w,g_0,h_0,\theta_0) \right| dF_0(w)  &= \E\left[ \left. \left| (D - \hat{V}_\ell)^2 - (D - V)^2\right| \right| I_\ell^c \right] \\
        &\leq \E\left[ \left. |2D + V + \hat{V}_\ell| \cdot |V - \hat{V}_\ell| \right| I_\ell^c \right] \\
        &\leq C \cdot \norm{\varphi(\cdot, \cdot, g_0) - \varphi(\cdot, \cdot, \hat{g}_\ell)}_2,
    \end{align*}
    since $|2D + V + \hat{V}_\ell|$ is bounded almost surely. The RHS converges in probability to zero by equation~\eqref{eq:hdps_phi_conv}, so the assumption is satisfied.
\end{proofc}

\begin{proofc}[Proposition~\ref{prop:CASF_conditions}] \mbox{} \\[-25pt]

    \paragraph{Assumptions~\ref{ass:Diff_m_h} and \ref{ass:Diff_m_g}}
    Let $V \equiv D - g_0(Z)$. We start by formally checking that
        \begin{align*}
        \E[D_{dir}(W,g)]&=\E\left[-\int \frac{\partial h_0}{\partial v}(x^*, V)dF^*(x^*) \cdot g(Z)\right], \\
        \E[D_2(W,h)] &= \E\left[ \int h(x^*, V)dF^*(x^*) \right] =\E\left[ \frac{f^{\ast }(X)f_{v}(V)}{f_{xv}(X,V)}\cdot h(X,V)\right]
    \end{align*}
     are continuous. Note that the second equality in the second row comes from Assumption~\ref{ass:CASF_conditions_alpha2}. This assumption, on top of the Cauchy-Schwarz inequality, also guarantees that $|\E[D_2(W,h)]| \leq C \norm{h}_2$. On the other hand, Assumption~\ref{ass:CASF_conditions_h} and the Cauchy-Schwarz inequality give that $|\E[D_{dir}(W,g))]| \leq C \norm{g}_2$.
     
    \paragraph{Assumption~\ref{ass:sq_convergence}} If the CASF is well defined (finite), Assumption~\ref{ass:sq_convergence:m_bound} is equivalent to
    \begin{equation*}
        \E\left[\left(\int h_0(x^*,V)dF^*(x^*\right)^2\right]<\infty.
    \end{equation*}
    By Jensen's inequality:
    \begin{align*}
        \E\left[\left(\int h_0(x^*,V)dF^*(x^*)\right)^2\right] \leq \E\left[\frac{f^*(X)f_{v}(V)}{f_{xv}(X,V)}h_0(X,V)^2\right].
    \end{align*}
    This is finite by Assumptions~\ref{ass:CASF_conditions_alpha2} and \ref{ass:CASF_conditions_expectation}. Indeed, note that $h_0(X,V)^2 \leq \E[Y^2 | X, V] = \E[ \E[ Y^2 | D, Z] | X, V]$, so it is bounded.

    To check Assumption~\ref{ass:sq_convergence:m_conv}, again by Jensen's inequality, with $\hat{V}_\ell \equiv D - \hat{g}_\ell(Z)$,
    \begin{align*}
        \int [m(w,\hat{g}_\ell,\hat{h}_\ell,\theta_0)-m(w,g_0,h_0,\theta_0)]^2 dF_0(w) \leq \E\left[ \left.\int\left(\hat{h}_\ell(x^*, \hat{V}_\ell)-h_0(x^*,V) \right)^2 dF^*(x^*)\right| I_\ell^c\right] \\
        \leq \E\left[ \left.\int\left(\hat{h}_\ell(x^*, \hat{V}_\ell)-\hat{h}_\ell(x^*,V) \right)^2 dF^*(x^*)\right| I_\ell^c\right] + \E\left[ \left.\int\left(\hat{h}_\ell(x^*,V)-h_0(x^*,V) \right)^2 dF^*(x^*)\right| I_\ell^c\right]
    \end{align*}
    For the first term, by the Mean Value Theorem and $\partial\hat{h}_\ell/\partial v(x,v) \leq C$ (Assumption~\ref{ass:CASF_conditions_dictionary}):
    \begin{equation*}
        \E\left[ \left.\int\left(\hat{h}_\ell(x^*, \hat{V}_\ell)-\hat{h}_\ell(x^*,V) \right)^2 dF^*(x^*)\right| I_\ell^c\right] \leq \E[C^2(\hat{V}_\ell -V)^2| I_\ell^c] = C^2\norm{\hat{g}_\ell - g_0}_2^2
    \end{equation*}
    This converges to zero by Assumption~\ref{ass:est_convergence:estimators}. For the second term, by Assumption~\ref{ass:CASF_conditions_alpha2}:
    \begin{align*}
        \E\left[ \left.\int\left(\hat{h}_\ell(x^*,V)-h_0(x^*,V) \right)^2 dF^*(x^*)\right| I_\ell^c\right] = \\
        \E\left[ \left. \frac{f^*(X)f_{v}(V)}{f_{xv}(X,V)}\left(\hat{h}_\ell(X,V)-h_0(X,V) \right)^2\right| I_\ell^c\right] \leq C \norm{\hat{h}_\ell - h_0}_2^2,
    \end{align*}
    which also tends to zero Assumption~\ref{ass:est_convergence:estimators}.
    
    Regarding Assumption~\ref{ass:sq_convergence:m_conv_theta}, since $m(w,\hat{g}_\ell,\hat{h}_\ell,\tilde{\theta}_\ell)-m(w,\hat{g}_\ell,\hat{h}_\ell,\theta_0)=\tilde{\theta}_\ell-\theta_0$, it suffices to have a consistent preliminary estimator of the CASF. Finally, Assumptions~\ref{ass:CASF_conditions_expectation}, on top the the conditional Jensen's inequality, gives Assumption~\ref{ass:sq_convergence:var_bound}.

    \paragraph{Assumption~\ref{ass:general_erro}} For $\varepsilon(W, g) = D - g(Z)$, we have that $r_e(z) = -1$, which is independent of $g$. Indeed, the generalized error is linear in this case. Therefore, $\E[\varepsilon(W,g) - \varepsilon(W, g_0) - D_e(g-g_0)] =0$ and the assumption is satisfied.

    \paragraph{Assumption~\ref{ass:linear_approx}} We check variant (ii) of the assumption. Recall that, with $v = d - g_0(z)$,
        \begin{equation*}
        \begin{aligned}
            D_{dir}(w,g)&= -\int \frac{\partial h_0}{\partial v}(x^*, v)dF^*(x^*) g(z), \text{ and } \\
            D_2(w, h)&= \int h(x^*, v)dF^*(x^*).
        \end{aligned}
    \end{equation*}
    Adding and subtracting $\int [h_0(x^*,\hat{V}_\ell) - h_0(x^*, V)] dF(x^*)$ to $m(W, \hat{g}_\ell, \hat{h}_\ell, \theta_0) - m(W, g_0, h_0, \theta_0) - D_{dir}(W, \hat{g}_\ell-g_0) - D_2(W, \hat{h}_\ell - h_0)$, we get:
    \begin{equation} \label{eq:CASF_Ass8_expansion}
        \begin{aligned}
            \int \left[ \hat{h}_\ell(x^*, \hat{V}_\ell) - \hat{h}_\ell(x^*, V) \right]dF(x^*) - \int \left[ h_0(x^*, \hat{V}_\ell) - h_0(x^*, V) \right]dF(x^*)  \\
            + \int h_0(x, \hat{V}_\ell)dF^*(x^*) - \int h_0(x, V)dF^*(x^*) + \int \frac{\partial h_0}{\partial v}(x^*, V)dF^*(x^*) (\hat{g}_\ell(Z) - g_0(Z)).
        \end{aligned}
    \end{equation}

    We show that the term in the first row is quadratic. A second order Taylor expansion of $\hat{h}_\ell$ and $h_0$ gives
    \begin{align*}
         \hat{h}_\ell(x^*, \hat{V}_\ell) - \hat{h}_\ell(x^*, V) = \frac{\partial \hat{h}_\ell}{\partial v}(x^*, V)(\hat{V}_\ell - V) + \frac{\partial^2 \hat{h}_\ell}{\partial v^2}(x^*, \tilde{V}_\ell) (\hat{V}_\ell - V)^2 \text{ and } \\
         h_0(x^*, \hat{V}_\ell) - h_0(x^*, V) = \frac{\partial h_0}{\partial v}(x^*, V)(\hat{V}_\ell - V) + \frac{\partial^2 h_0}{\partial v^2}(x^*, \tilde{V}_0) (\hat{V}_\ell - V)^2,
    \end{align*}
    with both $\tilde{V}_\ell$ and $\tilde{V}_0$ between $\hat{V}_\ell$ and $V$. Hence, since $\hat{V}_\ell - V = -(\hat{g}_\ell(Z) - g_0(Z))$,
    \begin{align*}
        \left|\E\left[\left.  \int \left[ \hat{h}_\ell(x^*, \hat{V}_\ell) - \hat{h}_\ell(x^*, V) \right]dF(x^*) - \int \left[ h_0(x^*, \hat{V}_\ell) - h_0(x^*, V) \right]dF(x^*)  \right| I_\ell^c \right] \right|= \\
        \left|\E\left[\left.  \int \left( \frac{\partial \hat{h}_\ell}{\partial v}(x^*, V) - \frac{\partial h_0}{\partial v}(x^*, V) \right) \left(\hat{g}_\ell(Z) - g_0(Z)\right)  \right| I_\ell^c \right] \right| \\
        + \left|\E\left[\left.  \int \left( \frac{\partial^2 \hat{h}_\ell}{\partial v^2}(x^*, \tilde{V}_\ell) - \frac{\partial^2 h_0}{\partial v^2}(x^*, \tilde{V}_0) \right) \left(\hat{g}_\ell(Z) - g_0(Z)\right)^2  \right| I_\ell^c \right] \right|.
    \end{align*}
    The term in the third row is bounded by $\norm{\hat{g}_\ell - g_0}_2^2$ by Assumptions~\ref{ass:CASF_conditions_h} and \ref{ass:CASF_conditions_dictionary}. Recall that $\sqrt{n}\norm{\hat{g}_\ell - g_0}_2^2 \xrightarrow{P} 0$ by Assumption~\ref{ass:est_convergence:estimators}.
    
    For the term in the second row, by Assumption~\ref{ass:CASF_conditions_alpha2} and the Cauchy-Schwarz inequality:
    \begin{align*}
        \left|\E\left[\left.  \int \left( \frac{\partial \hat{h}_\ell}{\partial v}(x^*, V) - \frac{\partial h_0}{\partial v}(x^*, V) \right) \left(\hat{g}_\ell(Z) - g_0(Z)\right)  \right| I_\ell^c \right] \right| \leq \\
        \sqrt{\E\left[\left. \frac{f^*(X)f_{v}(V)}{f_{xv}(X,V)} \left( \frac{\partial \hat{h}_\ell}{\partial v}(X, V) - \frac{\partial h_0}{\partial v}(X, V) \right)^2  \right| I_\ell^c \right]} \cdot \norm{\hat{g}_\ell - g_0}_2 \leq \\
        C \norm{\partial\hat{h}_\ell/\partial v - \partial h_0 / \partial v}_2 \norm{\hat{g}_\ell - g_0}_2.
    \end{align*}
    The term in the last row is $o_p(n^{-1/2})$ by Assumptions~\ref{ass:CASF_conditions_deriv}.
    
    To deal with the term in the second row of equation~\eqref{eq:CASF_Ass8_expansion}, we introduce the mapping $\xi(g) \equiv \bar{m}(g, h_0) + \theta_0 = \E[\int h_0(x^*, D - g(Z))dF^*(x^*)]$. Hence, its derivative is $D_\xi g = \E[D_{dir}(W, g)]$. That is, we have that
    \begin{align*}
        \left|\E\left[\left. \int h_0(x, \hat{V}_\ell)dF^*(x^*) - \int h_0(x, V)dF^*(x^*) + \int \frac{\partial h_0}{\partial v}(x^*, V)dF^*(x^*) (\hat{g}_\ell(Z) - g_0(Z)) \right| I_\ell^c \right] \right|= \\
        |\xi(\hat{g}_\ell) - \xi(g_0) - D_\xi(\hat{g}_\ell - g_0) |.
    \end{align*}
     Therefore, it suffices to show that the linear expansion of $\xi$ around $g_0$ has quadratic remainder.

     Call $D_\xi(g)$ to the derivative of of $\xi$ at an arbitrary $g \in \Delta_1$, so that $D_\xi = D_\xi(g_0)$. The first derivative evaluated at $\delta_1 \in \Delta_1$ is
     \begin{equation*}
         D_\xi(g)[\delta_1] = \E\left[ -\int \frac{\partial h_0}{\partial v}(x^*, D-g(Z))dF^*(x^*)\delta_1(Z)\right].
     \end{equation*}
    For the second derivative, with $\delta_1, \delta_1' \in \Delta_1$,
    \begin{equation*}
        D^2_\xi(g)[\delta_1][\delta_1'] = \E\left[ \int \frac{\partial^2 h_0}{\partial v^2}(x^*, D-g(Z))dF^*(x^*)\delta_1(Z)\delta_1'(Z)\right].
    \end{equation*}
    Assumption~\ref{ass:CASF_conditions_h} and the Cauchy-Schwarz inequality give $|D_\xi^2(g)[\delta_1][\delta_2] | \leq C \norm{\delta_1}_2\norm{\delta_1'}_2$, with the constant $C$ independent of $g$. Thus, Proposition~3 in \citet[p.~177]{luenberger1997optimization} leads to $|\xi(\hat{g}_\ell) - \xi(g_0) - D_\xi (\hat{g}_\ell - g_0) | \leq C \norm{\hat{g}_\ell - g_0}_2^2$. Hence, it is $o_p(n^{-1/2})$.

    \paragraph{Assumption~\ref{ass:gen_reg}} Since the derivative of $\varphi$ at an arbitrary $g\in\Delta_1$ is $D_\varphi(g) = -1$, the second derivative is zero so Assumption~\ref{ass:gen_reg:phi_Fderiv} is satisfied. For the CASF, $\alpha_{02} = f^*f_v/f_{xv}$, so Assumptions~\ref{ass:CASF_conditions_alpha2} and \ref{ass:CASF_conditions_h} give Assumptions~\ref{ass:gen_reg:twice_deriv} and \ref{ass:gen_reg:bounded_deriv}. Assumption~\ref{ass:gen_reg:bounded_alpha2} is satisfied since, in this case, $\hat{h}_\ell$ is a linear combination of the atoms in $\mathbf{b}_J$, which are all in $\Delta_2(g_0)$.
    
   Regarding Assumption~\ref{ass:gen_reg:continuous_deriv}, we use the decomposition
\begin{align*}
\frac{\partial}{\partial v}\left[\hat{\alpha}_{2\ell} \cdot (y-\hat{h}_\ell)\right]
-
\frac{\partial}{\partial v}\left[\alpha_{02} \cdot (y-h_0)\right]
&=
-\frac{\partial \hat{\alpha}_{2\ell}}{\partial v} \cdot (\hat{h}_\ell-h_0)
+\left(\frac{\partial\hat{\alpha}_{2\ell}}{\partial v}-\frac{\partial\alpha_{02}}{\partial v} \right) \cdot (y-h_0) \\
&\quad
-(\hat{\alpha}_{2\ell}-\alpha_{02}) \cdot\frac{\partial h_0}{\partial v}
-\hat{\alpha}_{2\ell} \cdot \left(\frac{\partial\hat{h}_\ell}{\partial v}-\frac{\partial h_0}{\partial v} \right).
\end{align*}
We have that $\hat{\alpha}_{2\ell}$ and $\partial\hat{\alpha}_{2\ell}/\partial v$ are bounded by Assumption~\ref{ass:CASF_conditions_dictionary}, $(Y-h_0(X,V))^2$ is bounded (conditionally on $D, Z$) by Assumption~\ref{ass:CASF_conditions_expectation}, and $\partial h_0/\partial v$ is bounded by Assumption~\ref{ass:CASF_conditions_h}. Hence, using these bounds and the Cauchy-Schwarz inequality:
\begin{equation*}
    \|\hat\nu_\ell-\nu_0\|_2 \leq C \left( \norm{\hat{h}_\ell - h_0}_2 + \norm{\hat{\alpha}_{2\ell} - \alpha_{02}}_2 + \norm{\partial\hat{h}_\ell/\partial v - \partial h_0 / \partial v}_2 + \norm{\partial\hat{\alpha}_{2\ell}/\partial v - \partial \alpha_{02} /\partial v}_2\right).
\end{equation*}
Then, Assumption~\ref{ass:est_convergence:estimators} guarantees that $\norm{\hat{h}_\ell - h_0}_2 \norm{\hat{g}_\ell - g_0}_2 = o_p(n^{-1/2})$,  Assumptions~\ref{ass:span}-\ref{ass:rates} guarantee that $\norm{\hat{\alpha}_{2\ell} - \alpha_{02}}_2 \norm{\hat{g}_\ell - g_0}_2 = o_p(n^{-1/2})$ (see the arguments leading to equation~\eqref{eq:product_consist}), and Assumption~\ref{ass:CASF_conditions_deriv} gives that $\norm{\hat{g}_\ell - g_0}_2 \cdot (\norm{\partial\hat{h}_\ell/\partial v - \partial h_0 / \partial v}_2 +\norm{\partial\hat{\alpha}_{2\ell}/\partial v - \partial \alpha_{02} /\partial v}_2) = o_p(n^{-1/2})$.

    \paragraph{Assumption~\ref{ass:jacobian}} This is trivial in case of the moment condition identifying the CASF. Simply note that
    \begin{equation*}
        \frac{\partial m}{\partial\theta}(W,g,h,\theta)=-1.
    \end{equation*}

    %%%%% Previous attempt with Song (2012)
    %We now check Assumtion~\ref{ass:linear_approx}. Set $V(g)\equiv D-g(Z)$ and recall that $V\equiv D-g_0(Z)$. We have that:
    %\begin{align*}
    %    \E\left[m(W, g,h,\theta_0)-m(W,g_0,h_0,\theta_0) - D_1(W,g-g_0)-D_2(W,h-h_0)\right] \\
    %    = \E\left[ \int h(x^*, V(g))dF^*(x^*)\right] - \E\left[ \int h_0(x^*, V)dF^*(x^*)\right]  \\
    %    - \E\left[\alpha_{01}(Z)\cdot (g(Z)-g_0(Z))\right]-\E\left[\int (h(x^*, V)-h_0(x^*,V))dF^*(x^*)\right] \\
    %    =\E\left[\int (h(x^*, V(g)-h(x^*,V))dF^*(x^*)\right]-\E\left[\alpha_{01}(Z)\cdot (g(Z)-g_0(Z))\right].
    %\end{align*}
    %By Assumption~\ref{ass:gen_reg_dicts:inclusion}:
    %\begin{align*}
    %    \E\left[\int (h(x^*, V(g)-h(x^*,V))dF^*(x^*)\right]=\E[\alpha_{02}(X,V)\cdot(h(X, V(g))-h(X,V))] \\
    %    =\E[\alpha_{02}(X,V)\cdot(\E[Y|X,V(g)]-\E[Y|X, V(g_0)])]
    %\end{align*}
    %Under Assumption~\ref{ass:CASF_conditions}, this term is bounded by $C_1 \cdot \norm{g-g_0}_2^2$ by Theorem~1 in \citet{song2012smoothness}.

    %We also need o bound $\E[\alpha_{01}(Z)\cdot (g(Z)-g_0(Z))]$, the derivative of $g\mapsto \bar{m}(g,h(F_0, g),\theta)$ at $g_0$. For %$g$ in a neighborhood of $g_0$, the derivative of $\bar{m}(g,h(F_0, g),\theta)$ at $g$ is given by
    %\begin{equation*}
    %    D_g\bar{m}=\E\left[-\int \frac{\partial h_0}{\partial v}(x^*, V(g))dF^*(x^*) + \frac{\partial h_0}{\partial v}(X, V(g))\alpha_{02}(X, V(g)) \right]
    %\end{equation*}
\end{proofc}

%%%%%%%%%%%%%%%%%%%%%%%%%%%%%%%%%%%%%%%%%%%%%%%%%%%%%%%%%%%%%%%%%%%%%%%%%%%%%%%%%%%%%%%%%%%%%%%%%%%%%%%%%%%%%%%%%%%%%%%%%%%%%%%%
\begin{proofc}[Proposition~\ref{prop:conditions_inclusion_alpha}]
    Start with the nonparametric case. Let $v(g)\equiv\varphi(d,z,g)$ By the triangle inequality (applied to the $L_2(W)$-norm):
    \begin{equation*}
    \begin{aligned}
        \left( \int \alpha_{02}(x, v(g_\tau))^2dF_0(w) \right)^{1/2} &\leq  \left( \int \alpha_{02}(x, v(g_0))^2dF_0(w) \right)^{1/2} \\
        &+  \left( \int \left[\alpha_{02}(x, v(g_\tau))-\alpha_{02}(x, v(g_0))\right]^2dF_0(w) \right)^{1/2}
    \end{aligned}
    \end{equation*}
    The fist quantity in the RHS is finite since $\alpha_{02}\in L_2(X,V)$. For quantity in the second row, by the Mean Value Theorem and Hadamard differentiability of $\varphi$:
\begin{equation*}
	\begin{aligned}
		\left(\int \left(\alpha_{02}(x,v(g_\tau))-\alpha_{02}(x,v)\right)^2dF_0(w)\right)^{1/2} \leq C \norm{\varphi(\cdot,\cdot,g_\tau)-\varphi(\cdot,\cdot,g_0)}_2  \\ =C\norm{\varphi(\cdot,\cdot,g_\tau)-\varphi(\cdot,\cdot,g_0)-D\varphi(g_\tau-g_0)+D\varphi(g-g_0)}_2 \\
        \leq C\left(\norm{D\varphi}\cdot\norm{g_\tau-g_0}_2+o(\norm{g_\tau-g_0}_2)\right),
	\end{aligned}
\end{equation*}
where $C$ is the bound of $\partial\alpha_{02}/\partial v$. Let $\varepsilon$ be such that $o(\norm{g_\tau-g_0}_2)\leq \norm{g_\tau-g_0}_2$. For $\tau<\varepsilon$:
\begin{equation*}
		\left(\int \left(\alpha_{02}(x,v(g_\tau))-\alpha_{02}(x,v)\right)^2dF_0(w)\right)^{1/2} \leq C(\norm{D\varphi}+1)\norm{g_\tau-g_0}_2,
\end{equation*}
which is finite since $\tau\mapsto g_\tau$ is a differentiable path in $\Delta_1$.

For the partialy linear case, where $\alpha_{02}(x,v)=\beta_0'x+\kappa_0(v)$, simply note that
\begin{equation*}
    \frac{\partial\alpha_{02}}{\partial v}(x,v)= \frac{\partial\kappa_{0}}{\partial v}(x,v).
\end{equation*}
Thus, $\partial\kappa_0/\partial v$ is bounded and we can proceed as above to show that $\int \kappa_0(v(g_\tau))^2dF_0(w)$ is finite.
\end{proofc}

%%%%%%%%%%%%%%%%%%%%%%%%%%%%%%%%%%%%%%%%%%%%%%%%%%%%%%%%%%%%%%%%%%%%%%%%%%%%%%%%%%%%%%%%%%%%%%%%%%%%%%%%%%%%%%%%%%%%%%%%%%%%%%%%
\begin{proofc}[Proposition~\ref{prop:conditions_inclusion_h}]
    We start with the nonparametric case. Throughout the proof, we call $h_\tau \equiv h(F_0, g_\tau)$. We note that $h_\tau \in \Delta_2(g_0)=L_2(X,V)$ if and only if $\int h_\tau(x,v)^2dF_{xv}^0(x,v)<\infty$. Under the conditions in the statement of the proposition, by Cauchy-Schwarz' inequality:
    \begin{equation*}
	\int h_\tau(x,v)^2 dF_{xv}^0(x,v) = \int h_\tau(x,v)^2\nu_\tau(x,v)dF_{xv}^\tau(x,v) \leq \int  h_\tau(x,v)^4 dF_{xv}^\tau(x,v)+ \int \nu_\tau(x,v)^2 dF_{xv}^\tau(x,v).
\end{equation*}
Furthermore, since $h_\tau(X, V(g_\tau))=\E[Y|X, V(g_\tau)]$, by conditional Jensen's inequality and the Law of Iterated Expectations
\begin{equation*}
	\int  h_\tau(x,v)^4 dF_{xv}^\tau(x,v)=\E[\E[Y|X,V(g_\tau)]^4]\leq \E[\E[Y^4|X,V(g_\tau)]]=\E[Y^4]<\infty.
\end{equation*}
Also, since the Radon-Nikodym density of $F_{xv}^\tau$ w.r.t. $F_{xv}^{0}$ is $\nu_\tau(x,v)^{-1}$ \citep[Prop.~1.7]{shao2003mathematical}:
\begin{equation*}
	\int \nu_\tau(x,v)^2dF_{xv}^\tau(x,v)=\int \nu_\tau(x,v)dF_{xv}^{0}(x,v)=\E[\nu_\tau(X,V)]<\infty.
\end{equation*}

For the partialy linear case we need to show that $\int \kappa_\tau(v)^2 dF_v^0(v)<\infty$. We can proceed as above to get:
\begin{equation*}
    \int \kappa_\tau(v)^2 dF_v^0(v) \leq \int \kappa_\tau(v)^4 dF_v^\tau(v) + \E[\nu_\tau(V)],
\end{equation*}
where the second quantity in the RHS is finite by assumption. In the partialy linear model we have that $\E[Y|V(g_\tau)]=\beta_\tau'\E[X|V(g_\tau)]+\kappa_\tau(V(g_\tau))$. Therefore,
\begin{equation*}
    \int \kappa_\tau(v)^4 dF_v^\tau(v)=\E[\kappa_\tau(V(g_\tau))^4]=\E[\E[Y-\beta_\tau'X|V(g_\tau)]^4] \leq \E[(Y-\beta_\tau'X)^4].
\end{equation*}
This is finite if the expectation of the fourth-order cross-products between $Y$ and $X$ is finite.
\end{proofc}

\end{appendices}

\end{document}